\documentclass[prd,nofootinbib,superscriptaddress,twocolumn]{revtex4}
\usepackage{amsfonts,amssymb,amsthm,bbm,amsmath}
\usepackage{color,psfrag}
\usepackage[dvips]{graphicx}
\usepackage{hyperref}

\newcommand{\nocontentsline}[3]{}
\let\origcontentsline\addcontentsline
\newcommand\stoptoc{\let\addcontentsline\nocontentsline}
\newcommand\resumetoc{\let\addcontentsline\origcontentsline}

\makeatletter
\def\l@subsubsection#1#2{}
\makeatother
\def\nn{\nonumber}

\def\be{\begin{equation}}
\def\ee{\end{equation}}

\newcommand{\p}{\partial}

\newcommand{\N}{\nabla}

\def\cH{ {\cal H}}
\def\b{\beta}

\def\d{\delta}

\def\G{\Gamma}
\def\e{\epsilon}

\def\lam{\lambda}

\def\l{\left}
\def\r{\right}

\def\MO{\mathcal {O}}

\def\MH{\mathcal {H}}
\def\MA{\mathcal {A}}
\def\MM{\mathcal {M}}
\def\MN{\mathcal {N}}
\def\MG{\mathcal {G}}
\def\MV{\mathcal {V}}
\def\MQ{\mathcal {Q}}

\def\tX{\widetilde{X}}
\def\tP{\widetilde{P}}
\def\tB{\widetilde{B}}
\def\tY{\widetilde{Y}}
\def\tPi{\widetilde{\Pi}}
\def\tQ{\widetilde{Q}}
\def\tJ{\widetilde{J}}
\begin{document}
\title{Schrödinger Symmetry in\\ Spherically-symmetric Static Mini-superspaces with Matter Fields} 
\author{Taishi Sano}
\email{t.sano@ruri.waseda.jp}
\affiliation{Department of Physics, Waseda University, 3-4-1 Okubo, Shinjuku, Tokyo 169-8555, Japan}

\author{Yuki Yokokura}
\email{yokokura.yuki@kochi-tech.ac.jp}
%\email{yuki.yokokura@kek.jp}
\affiliation{School of Engineering Science, Kochi University of Technology, 185 Miyanokuchi, Tosayamada, Kami, Kochi 782-8502, Japan}
\affiliation{Theory Center, KEK, Tsukuba, Ibaraki 305-0801, Japan}
\affiliation{RIKEN iTHEMS, Wako, Saitama 351-0198, Japan}

%%%%%%%%%%%%%%%%%%
\begin{abstract}
Schr\"{o}dinger symmetry has been shown to emerge in a ``fluid limit" from the full superspace to several mini-superspace models. To investigate one aspect of the robustness of this emergent symmetry, we consider two spherically-symmetric static mini-superspace models with matter fields at the classical level: (i) a Maxwell field with a cosmological constant and (ii) $n$ massless scalar fields. By developing a method based on canonical transformations, we demonstrate that for model (i), 3D Schrödinger symmetry emerges, and the solution is the (anti-)de Sitter Reissner-Nordström spacetime, and for model (ii), $(2+n)$D Schrödinger symmetry appears, and the solution is a generalized Janis-Newman-Winicour spacetime and its ``interior", a Kantowski-Sachs type closed universe. Furthermore, for the vacuum model, we find that 2D Schrödinger symmetry holds with different lapse functions and mini-superspace coordinates, suggesting the potential, yet unconfirmed, covariance of the symmetry. Finally, we propose a physical interpretation of the symmetry under the Hamiltonian constraint $H$: symmetry generators commuting with $H$ map a solution to another one, while those non-commuting with $H$ generate a new theory with the Schrödinger symmetry and the transformed configuration is a solution to the new theory. These results reinforce the robustness of the emergent Schrödinger symmetry and open new frontiers for exploring dynamics of matter and gravity. 
\end{abstract}

\maketitle

\setcounter{secnumdepth}{2}
\setcounter{tocdepth}{2}
\tableofcontents

%\documentclass[aps,pra,reprint,superscriptaddress,amsmath,amssymb,floatfix]{revtex4-2}
%\usepackage{amsmath}
%\usepackage{lipsum}
%\usepackage{txfonts}
%\usepackage{comment}
%\usepackage{xcolor}
%\usepackage{graphicx}
%\usepackage{dcolumn}
%\usepackage{braket}
%\usepackage{booktabs}
%\usepackage{bm}
%\newcommand{\colr}[1]{{\color{magenta}#1}}
%\newcommand{\colb}[1]{{\color{blue}#1}}
%\newcommand{\colg}[1]{{\color{gray}#1}}
%\newcommand{\pos}{\mathrm{e}^+}
%\bibliographystyle{apsrev4-2}

%%%%%%%%%%%%%%%%%%%%%%%%%%%%%%%%%%%%%%%%%%%%%%%%%%%%%%%%%%%%
\section{Introduction}\label{sec:intro}
Symmetry has played a significant role in physics. In the quest for a theory of quantum gravity, where the fundamental principles are still unknown, symmetry should serve as a guiding principle. The characteristic symmetry of gravity is general covariance, the gauge symmetry of spacetime, which leads to background independence and the dynamics of gravity. Here, the Hamiltonian constraint $\MH=0$ is the master key; it generates diffeomorphism \cite{Dirac}, the dynamics of gravity ultimately reduce to it \cite{Peres1962,DeWitt:1967yk,osti_4124259,Gerlach:1969ph, Moncrief:1972cx}, and it plays a central role in various crucial problems, such as the so-called ``problem of time" \cite{Kuchar:1991qf,Isham:1992ms,Hoehn:2020epv} and the quantization of gravity \cite{DeWitt:1967yk,osti_4124259}. Thus, the Hamiltonian constraint should be the essential ingredient in exploring the symmetry of (quantum) gravity. 

Then, what is the \textit{physical symmetry} in the configuration space of gravity? That is, what is the \textit{dynamical symmetry}, such as transforming a physical configuration into another? One idea is corner symmetry; the existence of a spacetime boundary transforms the gauge symmetries into physical ones, and charges like energy evaluated on that boundary form the symmetry algebra \cite{Ciambelli:2022cfr,Freidel:2023bnj}. The crucial point here is that the vanishing of the bulk contribution is a consequence of $\MH=0$. Therefore, exploring the physical symmetries associated with $\MH=0$ from the bulk perspective is highly intriguing. The present paper studies this and focuses particularly on symmetries that emerge in a certain limit.

One framework to describe the configuration space of the gravitational field and matter fields is \textit{superspace}, an abstract manifold with infinite dimensions determined by the Hamiltonian constraint $\MH$.
$\MH=0$ as the dynamical equation in general relativity is the Hamilton-Jacobi equation that determines a trajectory in the configuration space (superspace) of geometries and matter fields on a hypersurface in spacetime \cite{Peres1962,DeWitt:1967yk,osti_4124259,Gerlach:1969ph}. 

When geometric symmetries such as homogeneity  or spherical symmetry are imposed on spacetime, the number of configuration modes becomes finite, resulting in a finite-dimensional superspace (\textit{mini-superspace}). This corresponds to restricting the full superspace to collective modes or taking a fluid limit, and thus universal structures independent of the details of the system can be expected to emerge \cite{Oriti:2024elx,BenAchour:2024gir,Calcinari:2026hbm}. Furthermore, typical examples of mini-superspace models such as black holes and homogeneous universe exhibit thermodynamic properties. This suggests that underlying microscopic degrees of freedom may exist while their statistical behavior is governed by universal mechanisms.

In fact, Schrödinger symmetry appears in both the homogeneous isotropic mini-superspace and the spherically-symmetric static vacuum mini-superspace \cite{PhysRevLett.90.031302,Achour2019,Achour2020,Achour_2021,1021468SciPostPhys101022,Achour:2021dtj,BenAchour:2023dgj,Christodoulakis:2013xha,BenAchour:2022fif,Geiller:2022baq}. It is a conformal extension of Galilean invariance \cite{Niederer:1972zz}, and the generators consist of momenta $p_i$, angular momenta $j_{ij}$, Galilean boost $b^i$, Hamiltonian $H=q_+$, dilatation $q_0$, and special conformal transformation $q_-$ (see Sec.\ref{sec:Sch_sym} or Ref.\cite{Duval:2024eod} for a review). It is known not only as the maximal symmetry group of the free Schrödinger equation~\cite{Niederer:1972zz} but also as a symmetry inherent in the Gross-Pitaevskii equation, which is the effective equation of quantum liquids \cite{Kolomeisky:2000zz}. Note that the symmetry algebra for the mini-superspaces contains the Hamiltonian constraint $q_+\propto \MH$, %$q_+=H(=0)$ 
but the algebra is not a constraint algebra like the hypersurface deformation algebra \cite{Dirac}. Interpreting this symmetry as a universal symmetry of quantum gravity in the fluid limit, we can study models and quantization based on the symmetry; the symmetry can be useful for restricting the type of interactions in effective models of quantum gravity \cite{BenAchour:2023dgj} or for controlling the ambiguity that arises when quantizing mini-superspace models to resolve singularities \cite{Sartini:2021ktb}. 

Currently, the Schrödinger symmetry has been observed only in certain mini-superspace models, where the symmetry is 2-dimensional. Therefore, to establish it as an emergent universal symmetry in (quantum) gravity, it is important to examine its robustness from various perspectives. Among these, one essential aspect is to test whether the symmetry is preserved when interactions with matter fields are taken into account. Introducing new fields increases the dimensions of the mini-superspace, and it is not trivial whether Schrödinger symmetry still holds.
%Therefore, it is crucial to verify the robustness of the emergence of the Schrödinger symmetry from various perspectives. In particular, it is essential to check whether the symmetry is preserved when interaction with matter is considered. 
In this paper, as a first step towards this, we consider spherically-symmetric static mini-superspace models with matter fields.\footnote{Spherically-symmetric static mini-superspace models can describe black holes. In quantum theory, matter fields play non-trivial roles. Hawking radiation originates from quantum fluctuations of matter fields, and interaction between matter fields and gravity should be considered in order to understand the time evolution of the information of a collapsing matter during the evaporation \cite{Kawai:2015uya}. The origin of the entropy-area law, which should be consistent with the mechanism of information recovery, is also likely to be relevant to effects of matter \cite{Yokokura:2022phq}. Furthermore, high curvatures, say, around the central region, induce large quantum fluctuations of matter fields,  generating a large backreaction to the geometry \cite{Kawai:2020rmt}. Thus, incorporating matter fields is crucial not only for examining the robustness of the symmetry but also for gaining a deeper understanding of quantum black holes.}% and examine the robustness.
 
More precisely, we analyze two mini-superspace models at the \textit{classical} level and show explicitly that Schrödinger symmetry emerges in both cases: 
\begin{itemize}
\item[(I)] Spherically-symmetric static mini-superspace with a Maxwell field and a cosmological constant,
\item[(II)] Spherically-symmetric static mini-superspace with $n$ massless scalar fields.
\end{itemize}
Model I describes (anti-)de Sitter Reissner-Nordström ((A)dS-RN) solutions, which reduce to the Schwarzschild solutions in the limit of vanishing electric charge and cosmological constant. It thus serves as a natural first step toward incorporating matter into black hole mini-superspace models. Model II leads to the Janis-Newman-Winicour (JNW) solution~\cite{Janis:1968zz}, which asymptotes to the Schwarzschild geometry at $r \to \infty$ but lacks an event horizon and instead possesses a naked singularity, and it also provides a  basic setup for relational formalism using a scalar field as a ``clock" \cite{Blyth:1975is,Halliwell:1984eu,Ashtekar:2011ni,Domagala:2010bm}.

In a spherically symmetric spacetime, the areal radius $r$ is well-defined, and it should be natural to consider foliations of spacetime in terms of $r$ or an associated parameter $\tau$. Then, integrating over time (due to the static nature) and over spherical directions (due to spherical symmetry) leaves only $\tau$ as a non-trivial coordinate. This parameter serves as ``time", and we obtain the Hamiltonian $H=N\MH$. Here, $N$ is the lapse function, and $\MH$ is the Hamiltonian constraint that guarantees the reparametrization invariance for $\tau$. Thus, the two mini-superspace models are a kind of one-dimensional mechanical system with Hamiltonian constraints $H=0$ (\textit{``black-hole mechanics"} \cite{Cavaglia:1994yc,Bodendorfer:2019cyv,Achour:2021dtj,BenAchour:2022fif,BenAchour:2023dgj,Perez:2023jrq,Livine:2025soz}). We study the symmetry of such systems. 

In Sec.\ref{sec:Sch_sym}, we first review how Schrödinger symmetry can be easily obtained for a free-particle system by calculating the algebra of the Noether charges. Here, symmetry means that the Hamiltonian remains invariant under the transformation (see below for a more physical meaning). Then, we propose a simple strategy for obtaining the Schrödinger symmetry in the mini-superspace models: one performs canonical transformations and chooses the lapse function $N$ so that the Hamiltonian $H$ takes the same form as that of the free particle system.\footnote{In Ref.\cite{Paliathanasis:2025nmj}, a similar method is applied to an extended mini-superspace metric by the ED lift method.} By following this procedure, the Schrödinger algebra can be automatically constructed, and the Schrödinger symmetry appears in terms of the new canonical variables. Here, the solution configuration is represented as a uniform linear motion in the mini-superspace with the momentum constrained by $H=0$, which can be translated into the spacetime geometry through the canonical transformation. Note that this canonical transformation method can be applied to models where it is practically not so useful to identify Schrödinger symmetry by using the Eisenhart–Duval (ED) lift method \cite{cccc1220-d9a2-3162-a3c5-a8782bbc8c8b,BenAchour:2022fif,Dantas:2026hvz} (see Appendix \ref{A:ED} for the ED lift method).

In Sec.\ref{sec:Schwarzschild}, we implement this procedure explicitly in the spherically-symmetric static vacuum mini-superspace model with cosmological constant and demonstrate the usefulness and consistency by reproducing the 2D-Schrödinger symmetry \cite{BenAchour:2023dgj} in a systematic manner. %\YY{Here, the notation of this paper is also explained.}

In Sec.\ref{sec:EM}, we analyze Model I, where the presence of the Maxwell field $A_0(r)$ increases the dimension of the mini-superspace by one. By applying our strategy, we show that the 3D Schrödinger symmetry appears. The 2D Schrödinger symmetry is embedded as a subalgebra, and in the limit of vanishing the electric charge, the 3D symmetry reduces to the 2D one. (A)dS-RN metrics can be obtained as the solution in a consistent manner. %In Appendix \ref{A:M_Q}, the formulae of mass and electric charge are discussed.

In Sec.\ref{sec:scalar}, we study Model II, where the dimension of the mini-superspace is $n+2$. By performing a similar analysis but using a different choice of $N$ and a different coordinate in the mini-superspace, we prove that $(n+2)$D Schrödinger symmetry is obtained. 
Solving Model II leads to the JNW solution (generalized to $n$ scalar fields) and its ``interior" solution (a Kantowski-Sachs-type closed universe) in a unified manner.

This Schrödinger symmetry is obtained by choosing the lapse properly in each model, and it is not clear yet whether the symmetry is gauge-dependent or not. While further research is needed to fully answer this important question (see also Sec.\ref{sec:discussion}), the analysis in this paper already offers an interesting observation. If the scalar field contribution is dropped in Model II, the 2D Schrödinger symmetry is reproduced again. Note that in the matter decoupling limit, both Model I (without cosmological constant) and Model II reduce to the same model, the vacuum spherically symmetric static model. They lead to the 2D Schrödinger symmetry in terms of different lapse functions and different mini-superspace coordinates. This is expected to support (at least, partially) the covariance of the 2D Schrödinger symmetry in the mini-superspace.

In studying these models, we find one interesting result. In both models, the ADM energy $M$, represented in different canonical coordinates, commutes with the Hamiltonian constraints and thus is gauge invariant. We can construct a gauge-invariant observable $T$ that is associated with the time coordinate at infinity and  canonically conjugate to $M$, $\{T,M\}=1$. Here, $\{f,g\}$ is the Poisson bracket.

In Sec.\ref{sec:interpretation}, we discuss the physical meaning of the obtained symmetry under the Hamiltonian constraint $H=0$. First, the generators $G$ of the Schrödinger symmetry obtained for a certain choice of $N$ satisfy $\{G,H\}+\partial_\tau G=0$. Momenta $p_i$ and angular momenta $j_{ij}$ commute with $H$, and we can show that such generators transform a solution to another one satisfying $H=0$ (dynamical symmetry). As examples, we construct  symmetry transformations that map the RN solution to the Schwarzschild one and those that map the JNW solution to the Schwarzschild one. On the other hand, Galilean boost $b^i$, dilatation $q_0$, and special conformal transformation $q_-$ do not commute with $H$, which prevents interpretation as a dynamical symmetry consistent with $H=0$. Therefore, understanding the transformations by such generators as a kind of symmetry presents a challenge, and there may be several possibilities. Here, we suggest one possible interpretation: such a non-commutative symmetry transforms the theory itself,  $H\to H_{\rm New}$, and the transformed configuration is a physical solution of the new theory with $H_{\rm New}=0$. We will demonstrate this concretely: in Model I, as the effect of adding a two-dimensional cosmological term, and in Model II, as the effect of adding a new scalar field. This provides one interpretation of Noether's theorem in a theory with the Hamiltonian constraint $H= 0$ and symmetry generators $G$ that do not commute with $H$. In Appendix \ref{A:Action}, the symmetry is studied in the action principle.

In Sec.\ref{sec:discussion}, we conclude with future prospects.

Our main result, that Schrödinger symmetry appears in mini-superspace models with different matter contents and different dimensions, should support one aspect of the robustness of the emergence of Schrödinger symmetry in a fluid limit of (quantum) gravity and contribute to future studies for quantum dynamics of gravity and matter.

%%%%%%%%%%%%%%%%%%%%%%%%%%%%%%%%%%%%%%%%%%%%%%%%%%%%%%%%%%%%%%
\section{Classical Schrödinger symmetry}\label{sec:Sch_sym}
%%%%%%%%%%%%%%%%%%%%%%%%%%
%\subsection{Schrödinger symmetry in free particle system}
\subsection{Free particle}\label{sec:particle}
We review  Schrödinger symmetry \cite{Niederer:1972zz,Duval:2024eod} in a classical free particle system and see how to find  Schrödinger symmetry in a heuristic manner. We consider 
\begin{equation}
   S[x^i(t)]=\int dt \frac{1}{2} g_{ij}\dot{x}^i\dot{x}^j =\int dt L,
   \label{action_free}
\end{equation}
where $\dot x^i \equiv \frac{dx^i}{dt}$ and $g_{ij}$ is an invertible symmetric constant metric on the configuration space spanned by $\{x^i\}_{i=1}^d$, say, $g_{ij}=\delta_{ij}$. Defining the canonical momentum $p_i\equiv \frac{\p L}{\p \dot x^i}=g_{ij}\dot{x}^j$, the Hamiltonian is given by 
\begin{equation}
   H=\frac{1}{2}g^{ij}p_ip_j,
   \label{hamiltonian_free}
\end{equation}
and the canonical relation in the Poisson bracket holds:
\begin{equation}
\left\{x^i,p_j\right\}=\delta^i_j,~~\left\{x^i,x^j\right\}=0=\left\{p_i,p_j\right\}
\label{canonical relation_free}
\end{equation}
Solving the canonical equations 
\begin{equation}\label{xp_eq}
    \dot x^i=\{x,H\}=g^{ij}p_j,~~\dot p_i =\{p_i,H\}=0,
\end{equation}
we can obtain the solution,  a \textit{uniform linear motion in the configuration space}: 
\begin{equation}
   x^i(t)=g^{ij} \alpha_jt+\beta^i~~{\rm with}~~ p_i(t)=\alpha_i,
   \label{solution_free}
\end{equation}
where $\alpha_i$ and $\beta^i$ are $2d$ integration constants fixed at $t=0$ that characterize the solution.

Generally, in the canonical form of Noether's theorem, a symmetry transformation is generated by a function  $G(q,p,t)$ that satisfies, in off-shell,
\begin{equation}\label{symmetry_cond}
    \{G,H\}+\frac{\p G}{\p t}=0,
\end{equation}
which is the condition for the Hamiltonian to be invariant. That is, a trajectory generated by such a function $G$ from a solution $(x^i(t), p_i(t))$,
\begin{align}
    x'^i(t)&=x^i(t)+\e \{x^i,G\}|_{x(t),p(t)},\nn\\
    p'_i(t)&=p_i(t)+\e \{p_i,G\}|_{x(t),p(t)},
\end{align}
also satisfies the canonical equation as another solution with different values of integration constants, where $\epsilon$ is a small parameter.\footnote{This is a dynamical symmetry if no Hamiltonian constraint exists. %If no Hamiltonian constraint exists, this leads to dynamical symmetries. 
See Sec.\ref{sec:interpretation} for more details.}  
On the other hand, such a generator $G$ is conserved in on-shell:
\begin{equation}
   \l.\frac{dG}{dt}\r|_{x(t),p(t)} = \l[ \left\{ G, H \right\} + \frac{\partial G}{\partial t}\r]_{x(t),p(t)} = 0.
   \label{conservation}
\end{equation}

%Generally, 
In the case of the free-particle Hamiltonian \eqref{hamiltonian_free}, symmetry generators $G$ transform a uniform linear motion \eqref{solution_free} with certain values of $(\alpha_i,\beta^i)$ to another  one with different ones $(\alpha'_i,\beta'^i)$. (We will provide a more explicit argument in Sec.\ref{sec:interpretation}.)
In particular, the momentum $p_i$ and the Galilean boost generator $b^i$ defined by 
\begin{equation}
   b^i := x^i - g^{ij} p_j t
   \label{b_free}
\end{equation}
satisfy the symmetry condition \eqref{symmetry_cond} and play roles of the symmetry generators.
Note that using the solution trajectory \eqref{solution_free}, we have a on-shell relation:
\begin{equation}\label{pb_initial}
    p_i|_{x(t),p(t)}=\alpha_i,~~b^i|_{x(t),p(t)}=\b^i,
\end{equation}
meaning that the initial value of the position, $\beta^i$, corresponds to the boost generator $b^i$ in on-shell.

Now, quantities consisting of $(b^i,p_i)$ also satisfy the symmetry condition \eqref{symmetry_cond} through the Leibniz rule. In particular, we can construct the quadratic ones: 
\begin{subequations} \label{jq_free}
\begin{gather}
j_{ij} := g_{ik} b^k p_j - g_{jk} b^k p_i = g_{ik} x^k p_j - g_{jk} x^k p_i, \\
q_+ := \frac{1}{2} g^{ij} p_i p_j = H, \\
q_0 := \frac{1}{2} b^i p_i=\frac{1}{2} x^i p_i -Ht, \\
q_- := \frac{1}{2} g_{ij} b^i b^j=\frac{1}{2} g_{ij} x^i x^j-x^ip_it+Ht^2.
\end{gather}
\end{subequations}
Noting that $b^i$ and $p_i$ satisfy
\begin{equation}
\left\{ b^i, p_j \right\} = \delta^i_j,
\label{bp_canonical_relation_free}
\end{equation}
we can check the following algebra: 
\begin{subequations} \label{jalg_free}
\begin{gather}
\{b^i,b^j\}=0=\{p_i,p_j\},\\
\left\{ j_{ij}, b^k \right\} = \delta^k_i g_{jl} b^l - \delta^k_j g_{il} b^l, \label{jb}\\
\left\{ j_{ij}, p_k \right\} = g_{ik} p_j - g_{jk} p_i, \label{jp}\\
\left\{ j_{ij}, j_{kl} \right\} = g_{ik} j_{jl} - g_{jk} j_{il} - g_{il} j_{jk} + g_{jl} j_{ik}.
\end{gather}
\end{subequations}
These together with \eqref{bp_canonical_relation_free} form the Galilean algebra for $g_{ij} = \delta_{ij}$. On the other hand, we can show from \eqref{bp_canonical_relation_free} that $q_+, q_0$ and $q_-$ satisfy $\mathfrak{sl}(2,\mathbb{R})$ algebra:
\begin{subequations} \label{qalg_free}
\begin{gather}
\left\{ q_0, q_\pm \right\} = \pm q_\pm, \\
\left\{ q_+, q_- \right\} = -2 q_0.
\end{gather}
\end{subequations} 
Specifically, \( q_+ \) generates time translations, \( q_0 \) dilatations (scaling transformations), and \( q_- \) special conformal transformations. % (see e.g. Ref.\cite{XX} for details \YY{Add a good Ref.}).
%\YY{Notation $q_\pm$ would be confusing because it looks like $q_{a=\pm}$. It is better to use $q_+=H,~q_0=D,~q_-=W$?}
These two algebras are intertwined as  
\begin{subequations} \label{qpbalg_free}
\begin{gather}
\left\{ q_0, b^i \right\} = -\frac{1}{2} b^i, \quad \left\{ q_0, p_i \right\} = \frac{1}{2} p_i, \\
\left\{ q_+, b^i \right\} = -g^{ij} p_j, \quad \left\{ q_-, p_i \right\} = g_{ij} b^j.
\end{gather}
\end{subequations}

Thus, the full algebra consisting of 
\eqref{bp_canonical_relation_free}, \eqref{jalg_free}, \eqref{qalg_free}, and \eqref{qpbalg_free} forms the $d$-dimensional Schrödinger algebra: 
\begin{equation}
\mathfrak{sh}(d) = \left( \mathfrak{sl}(2,\mathbb{R}) \oplus \mathfrak{so}(d) \right) \oplus_s \left( \mathbb{R}^d \oplus \mathbb{R}^d \right),
\label{sh_algebra_euclidean}
\end{equation}
for $g_{ij}=\d_{ij}$, where \( \oplus_s \) denotes a semidirect sum defined by the relations \eqref{jb},\eqref{jp} and \eqref{qpbalg_free}. 
The corresponding symmetry is the d-dimensional  Schrödinger symmetry: 
\begin{equation}
\mathrm{Sh}(d) = \left( \mathrm{SL}(2,\mathbb{R}) \times \mathrm{SO}(d) \right) \ltimes \left( \mathbb{R}^d \times \mathbb{R}^d \right).
\label{sh_group_euclidean}
\end{equation}

When the configuration space has a Lorentzian signature, for example,
\begin{equation}
g_{ij} = 
\begin{pmatrix}
\begin{matrix}
0 & 1 \\
1 & 0
\end{matrix} & 0 \\
0 & I_{d-2}
\end{pmatrix}
\quad \text{or} \quad
\begin{pmatrix}
\begin{matrix}
-1 & 0 \\
0 & 1
\end{matrix} & 0 \\
0 & I_{d-2}
\end{pmatrix},
\label{eta_lorentzian}
\end{equation}
the generators \( j_{ij} \) form the Lie algebra \( \mathfrak{so}(1,d-1) \), and the Schrödinger algebra becomes
\begin{equation}
\mathfrak{sh}(d) = \left( \mathfrak{sl}(2,\mathbb{R}) \oplus \mathfrak{so}(1,d-1) \right) \oplus_s \left( \mathbb{R}^d \oplus \mathbb{R}^d \right),
\label{sh_algebra_lorentzian}
\end{equation}
corresponding to the symmetry group:
\begin{equation}
\mathrm{Sh}(d) = \left( \mathrm{SL}(2,\mathbb{R}) \times \mathrm{SO}(1,d-1) \right) \ltimes \left( \mathbb{R}^d \times \mathbb{R}^d \right).
\label{sh_group_lorentzian}
\end{equation}
Generically, gravitational (mini-)superspaces have a Lorentzian signature \cite{DeWitt:1967yk}. Indeed, a Lorentzian version of Schrödinger algebra will appear in our models below.
%Since the gravitational minisuperspace is a relativistic system whose configuration space has a Lorentzian signature, we will focus on this Lorentzian version of the Schrödinger algebra.

Before proceeding to the next subsection, we emphasize two important points. 
The first one is that, in the classical theory, not all generators of the Schrödinger symmetry are independent \cite{BenAchour:2023dgj}. The classical phase space has $2d$ degrees of freedom corresponding to the integration constants $(\alpha_i,\beta^i)$, while the number of the generators $b^i,p_i,j_{ij},q_0,q_\pm$ is % in the \( d \)-dimensional Schrödinger algebra is
$2d + \frac{1}{2}d(d-1) + 3$, with the Casimirs of the symmetry algebra vanishing. 
In quantum theory, on the other hand, the ``shape" of wave functions, corresponding to an infinite number of degrees of freedom, has the physical information of the system and leads to non-vanishing Casimirs of the algebra of the Schrödinger symmetry, which gives the maximal symmetry of the free Schrödinger equation \cite{Niederer:1972zz}. Therefore, this symmetry, which is classically redundant, would provide, once quantized, the necessary and sufficient symmetry to characterize the system. %a system that can be equivalent to the free Schrödinger equation. 

%In fact, we started with the \( 2d \) Noether charges \( b^i \) and \( p_i \), which correspond to \( c_b^i \) and \( c^p_i \), and the conservation laws for the remaining Noether charges—\( j_{ij} \), \( q_{\pm} \), and \( q_0 \)—can be derived solely from the conservation of \( b^i \) and \( p_i \). In the classical system, one can recover the original mechanics from the Schrödinger algebra by setting its Casimir operators to particular value. However, this situation changes in the quantum theory, where the system generally has an infinite number of degrees of freedom. (See Ref.~\cite{BenAchour:2023dgj}.) In fact, the maximal symmetry of the free Schrödinger equation is precisely the Schrödinger symmetry ~\cite{Niederer:1972zz}.

The second point is that, the (classical) Schrödinger algebra can be constructed solely from quantities like $(b^i,p_i)$ that satisfy both the symmetry condition \eqref{symmetry_cond} and the canonical relation \eqref{bp_canonical_relation_free}. When such ones are obtained and the Hamiltonian takes the same form as the free one \eqref{hamiltonian_free}, the remaining generators of the Schrödinger algebra can be obtained in the same way as in \eqref{jq_free}. This observation provides a powerful guiding principle for finding the Schrödinger algebra in a more generic setup, which will be explored below.% in the next subsection.

%The second point is that the entire \( d \)-dimensional Schrödinger algebra can be reconstructed solely from the conserved charges \( b^i \) and \( p_i \), which satisfy both the conservation law \eqref{conservation} and the canonical Poisson bracket relation \eqref{bp_canonical_relation_free}. Once such \( b^i \) and \( p_i \) are obtained, the remaining Noether charges that form the Schrödinger algebra can be constructed in the same way as in \eqref{jq_free}, since all Poisson bracket relations involving \( j_{ij} \), \( q_{\pm} \), and \( q_0 \) are completely determined by the Poisson bracket relations among \( b^i \) and \( p_i \) thanks to Leibniz rule. 

%This provides a powerful guiding principle for identifying the Schrödinger algebra, which we will explore further in the next subsection.

%%%%%%%%%%%%%%%%%%%%%%%%%%%%%%%%%%%%%%%%%%%%%%%%%%%%%%%%%%%%%%%%%%%%%
\subsection{Canonical-transformation Method}\label{sec:strategy}
An interesting and important question for understanding the robustness of the emergent symmetry of gravitational systems is: under which condition can Schrödinger symmetry appear in general? Although finding a complete answer would be non-trivial, there is one related result based on Eisenhart–Duval (ED) lift method; one considers a ``lifted" configuration space endowed with a lifted metric and can show that when the lifted space is conformally flat, the Schrödinger symmetry is obtained  \cite{cccc1220-d9a2-3162-a3c5-a8782bbc8c8b,BenAchour:2022fif} (see Appendix \ref{A:ED} for a review).
Here, we propose another method to obtain Schrödinger symmetries, which can be applied to cases where the ED-lift method is not very effective for finding Schrödinger symmetries.
%Instead of using this method, we will consider more general (but not completely general yet) cases in which the ED method may not be useful, and find a sufficient condition for the Schrödinger symmetry to appear. %This will become our strategy for analyzing mini-superspace models in the following sections. 

Our idea follows from the observation discussed at the end of the previous subsection. If the Hamiltonian takes the same form as that of a free particle, the Schrödinger symmetry is naturally obtained. Therefore, even for systems with non-trivial metrics $g_{ij}(x)$ or potentials $V(x)$, if a canonical transformation exists that transforms the Hamiltonian into the form of the free particle Hamiltonian, the Schrödinger symmetry is automatically derived for the new canonical variables by applying the method in the previous subsection. 

More explicitly, we consider a system with a $x$-dependent metric $g_{ij}(x)$ and a potential $V(x)$: 
\begin{equation}
   H = \frac{1}{2} g^{ij}(x) p_i p_j + V(x).
   \label{hamiltonian_potential}
\end{equation}
\textit{Assume} that we can find a canonical transformation $(x^i,p_i)\to(\tilde x^i, \tilde p_i)$ such that this Hamiltonian becomes 
\begin{equation}
   H = \frac{1}{2} \tilde{g}^{ij} \tilde{p}_i \tilde{p}_j,
   \label{hamiltonian_potential_ptilde}
\end{equation}
where $ \tilde{g}^{ij}$ is a constant metric. 
Note here that canonical transformations preserve the symplectic structure: $\{x^i,p_j\}=\delta^i_j\to \{\tilde x^i,\tilde p_j\}=\delta^i_j$.  
Therefore, we can apply the same procedure as in the previous subsection to construct the generators of Schrödinger algebra. Formally, they are given by replacing $ (x^i, p_i)$ with $(\tilde{x}^i, \tilde{p}_i)$ in the generators \eqref{b_free} and \eqref{jq_free}:  \begin{subequations} \label{Noethers_tilde}
\begin{gather}
%\tilde{p}_i = \tilde{p}_i, \\
\tilde{b}^i := \tilde{x}^i - \tilde g^{ij} \tilde{p}_j t, \\
\tilde{j}_{ij} := \tilde g_{ik} \tilde{x}^k \tilde{p}_j - \tilde g_{jk} \tilde{x}^k \tilde{p}_i = \tilde g_{ik} \tilde{b}^k \tilde{p}_j - \tilde g_{jk} \tilde{b}^k \tilde{p}_i, \\
\tilde{q}_+ := \frac{1}{2} \tilde g^{ij} \tilde{p}_i \tilde{p}_j = H, \\
\tilde{q}_0 := \frac{1}{2} \tilde{b}^i \tilde{p}_i, \\
\tilde{q}_- := \frac{1}{2} \tilde g_{ij} \tilde{b}^i \tilde{b}^j.
\end{gather}
\end{subequations}
By construction, these charges satisfy the symmetry condition \eqref{symmetry_cond} and obey the same algebra as in \eqref{bp_canonical_relation_free}, \eqref{jalg_free}, \eqref{qalg_free}, and \eqref{qpbalg_free}. The above is our strategy for identifying Schrödinger symmetries in several non-trivial models.

Let us now supplement the procedure for finding the canonical transformation from the original form \eqref{hamiltonian_potential} to the desired form \eqref{hamiltonian_potential_ptilde}. Given that $\tilde{p}_i$ commute with the Hamiltonian $H$, the first step is to find functions $f_i(x,p)$ that commute with $H$, identify those with new momenta $\tilde{p}_i=f_i(x,p)$ and express the Hamiltonian \eqref{hamiltonian_potential} as the free form \eqref{hamiltonian_potential_ptilde}. 
The next step is to complete the canonical transformation. The obtained functions  $\tilde{p}_i=f_i(x,p)$ determine the relations $p_i=p_i(x,\tilde p)$. Using these, we obtain a generating function $F(x,\tilde{p})$ such that 
\begin{equation}\label{Feq1}
    p_i=\frac{\p F}{\p x^i}. 
\end{equation}
Then, we gain $\tilde x^i$ by 
\begin{equation}\label{Feq2}
    \tilde{x}^i = \frac{\p F}{\p \tilde{p}_i}.
\end{equation}
Thus, the desired canonical transformation $(x,p)\to (\tilde x,\tilde p)$ is obtained. %As shown in, say, Sec.\ref{sec:AdS}, the relations $\tilde{x}^i(x,p),\tilde{p}^i(x,p)$ connect symmetries to physics.  

Before moving on to applying this method to models, we discuss its subtle points. This is a heuristic approach; whether the desired canonical transformation exists can only be checked by actually performing the analysis. In particular, the first step above is very non-trivial. So, under what conditions does such a canonical transformation exist? What does its existence physically imply?  %Understanding these questions would require further studies. 
Another aspect is that the existence of such a canonical transformation is merely one sufficient condition for the existence of Schrödinger symmetries. How much broader is its applicability compared to the ED-lift method? What is the relationship between these two?  We leave these as future challenges.

\section{Vacuum with Cosmological Constant}\label{sec:Schwarzschild}
To demonstrate the usefulness and consistency of our method, we consider a case where the ED method cannot be applied directly to obtain Schrödinger symmetries, and use our strategy to reproduce a known result. That is, we apply the canonical-transformation method to the spherically-symmetric static vacuum mini-superspace model with cosmological constant $\Lambda$ and rederive the 2D-Schrödinger symmetry \cite{BenAchour:2023dgj}.

%%%%%%%%%%%%%%%%%%%%%%%%%%%%
\subsection{X-coordinate in the mini-superspace}\label{sec:vacuum}
We first construct the Hamiltonian. 
We consider a radial foliation of a 4D spherically-symmetric static spacetime, $\MM=\mathbb{R}\times \Sigma$, and set the metric as 
\begin{equation}
   ds^2=\epsilon\left(-N(\tau)^2d\tau^2+\frac{4X^+(\tau)}{X^-(\tau)}dt^2\right)+\frac{l_s^2}{4}{X^-(\tau)}^2d\Omega^2.
   \label{metric_X}
\end{equation}
Here, $\mathbb{R}$ is a domain of the radial variable $\tau$; $\Sigma$ is a 3D hypersurface specified by fixing $\tau$, where $\Sigma$ is timelike for $\e=-1$ and spacelike for $\e=1$; $N(\tau)$ is the lapse function associated with the reparametrization invariance of the radial coordinate; $X^\pm(\tau)$ are metric functions of $\tau$;\footnote{Any spherically-symmetric static metric can be cast into the form \eqref{metric_X}. A similar one is used in Ref.\cite{BenAchour:2023dgj}.} and $l_s$ is a fiducial length scale that introduces the dimensionful unit of the 2-sphere radius.

By using this metric, we evaluate the Einstein-Hilbert action with cosmological constant $\Lambda$: 
\begin{equation}
  S_{g}[g_{\mu\nu}]=\frac{1}{16\pi G\hbar}\int_\mathcal{M}d^4x\sqrt{-g}(R-2\Lambda).
   \label{EHaction}
\end{equation}
Here, for convenience, we introduce the Planck constant $\hbar$ to make dimensional analysis easier. To make the hypersurface $\Sigma$ finite, we set an IR cutoff scale for the range of $t$ integral: $l_0\equiv\int^{t_1}_{t_0}dt=t_1-t_0$. We then integrate this action over time and 2-sphere, perform integration by parts, and drop boundary terms to obtain the reduced Lagrangian $L_g$ via $S_g=\int d\tau L_g$: 
\begin{align}\label{L_g}
    L_g&= N\sqrt{\frac{X^+}{X^-}}\left(-\epsilon cl_p \frac{X^-}{4X^+} \frac{{\dot X^+}{\dot X^-}}{N^2}+\frac{cl_p}{l_s^2}-\frac{cl_p}{l_\Lambda^2}(X^-)^2\right).
\end{align}
Here, $\dot X^+\equiv\frac{dX^+}{d\tau}$, $l_\Lambda\equiv \frac{2}{\sqrt{\Lambda}}$, and 
\begin{equation}
c:=\frac{l_0l_s^2}{l_p^3}
   \label{c}
\end{equation}
is the dimensionless constant, the ratio between the UV cutoff scale $l_p\equiv\sqrt{\hbar G}$ and the IR cutoff scale $l_0$ \cite{BenAchour:2022fif,BenAchour:2023dgj}
% \YY{To Taishi: Add some Refs.}
. %\YY{Add a physical comment on $c$ simply.} %\TS{Note that for negative $\Lambda$, $l_\Lambda$ is purely imaginary.}\YY{$\leftarrow$ Remove, since it is not important.}

We introduce the canonical momenta as 
\begin{align}\label{P_L}
P_\pm:=\frac{\partial L_g}{\partial {\dot X^\pm}}=-\frac{\epsilon cl_p}{4}\sqrt{\frac{X^-}{X^+}}\frac{{\dot X^\mp}}{N},
\end{align}
so that the canonical relation holds:
\begin{align}\label{canonical}
    \{X^a,P_b\}=\delta^a_b,~~~\{X^a,X^b\}=0=\{P_a,P_b\},
\end{align}
where $a,b=+,-$, and the Poisson bracket is defined as 
\begin{equation}\label{Poisson}
    \{f_1,f_2\}\equiv \frac{\p f_1}{\p X^a}\frac{\p f_2}{\p P_a}-\frac{\p f_2}{\p X^a}\frac{\p f_1}{\p P_a},
\end{equation}
for functions $f_1,f_2$ on the phase space. %: $f_1(X,P),~f_2(X,P)$. 
We then obtain the Hamiltonian:  
\begin{align}
    H_g:=&P_+ \dot X^+ +P_- \dot X^-- L_g\nn\\
    =&N\sqrt{\frac{X^+}{X^-}}\l(-\frac{4}{\e c l_p}P_+P_- -\frac{cl_p}{l_s^2}+\frac{cl_p}{l_\Lambda^2}(X^-)^2  \r)\nn\\
    \equiv& N \MH_g.
    \label{H_g}
\end{align}
$\MH_g$ is the Hamiltonian constraint, which must be set to zero in order to gain the physical solution: $\cH_g\approx 0$. Here, $\approx$ is the weak equality that holds after calculating Poisson brackets, as usual \cite{Dirac}. 

We now fix the lapse as 
\begin{equation}\label{N1}
    N(\tau)=\sqrt{\frac{X^-}{4X^+}}.
\end{equation}
Then, the Hamiltonian \eqref{H_g} becomes 
\begin{align}
    H_g=&-\frac{2}{\e c l_p}P_+P_- -\frac{cl_p}{2l_s^2}+\frac{cl_p}{2l_\Lambda^2}(X^-)^2.
    \label{H_g2}
\end{align}
This can be expressed as the form of \eqref{hamiltonian_potential}: 
\begin{align}
    H_g=\frac{1}{2}\MG ^{ab}P_a P_b +\MV_\Lambda(X)
    \label{H_g3}
\end{align}
with the quadratic potential 
$\MV_\Lambda(X)= -\frac{cl_p}{2l_s^2}+\frac{cl_p}{2l_\Lambda^2}(X^-)^2$ and the metric of the 2D mini-superspace (\textit{supermetric}) 
\begin{equation}
\mathcal{G}_{ab} = 
\begin{pmatrix}
0 & -\frac{\epsilon c l_p}{2} \\
-\frac{\epsilon c l_p}{2} & 0
\end{pmatrix}
\quad \text{and} \quad
\mathcal{G}^{ab} = 
\begin{pmatrix}
0 & -\frac{2}{\epsilon c l_p} \\
-\frac{2}{\epsilon c l_p} & 0
\end{pmatrix},
\label{metric_L}
\end{equation}
where $a,b=+,-$, and the mini-superspace is ``Lorentzian", $\rm{det}\MG_{ab}<0$, as consistent with the general result \cite{DeWitt:1967yk}. Note here that %the mini-superspace is flat, but 
the lifted space is not conformally flat due to $\Lambda\neq0$, which makes the ED-lift method not very effective in practice for finding Schrödinger symmetries \cite{BenAchour:2022fif} (see Appendix \ref{A:ED}). 

%%%%%%%%%%%%%%%%%%%%%%%%%%%%%%%%%%%%%%%
\subsection{2D Schrödinger Symmetry}\label{sec:2D_Sch}
To find the symmetry of the Hamiltonian \eqref{H_g2}, we perform a canonical transformation $(X,P)\to(\tX, \tP)$ that rewrites the original Hamiltonian $H_g$ \eqref{H_g2} as the free-particle form \eqref{hamiltonian_potential_ptilde}. 

According to the explanation in the previous section, our first step is to find %conserved 
quantities $f_a(X,P)$ commuting with $H_g$ and identify them with new momenta $\tP_a(X,P)$. Noting that $\{H_g,H_g\}=\{H_g,P_+\}=0$, we can write \eqref{H_g2} as 
\begin{align}\label{H_g4}
    H_g&=-\frac{2}{\e c l_p}P_+\l[P_--\frac{\e c^2l_p^2}{4l_\Lambda^2}\frac{(X^-)^2}{P_+} \r] -\frac{cl_p}{2l_s^2}
\end{align}
and notice that the bracket part $[\cdots]$ commutes with $H_g$. Therefore, we set 
\begin{align}\label{tP_L}
    \tP_+=P_+,~~~
    \tP_-=P_--\frac{\e c^2l_p^2}{4l_\Lambda^2}\frac{(X^-)^2}{P_+}.
\end{align}

The second step is to construct a generating function $F_g(X,\tP)$ satisfying the condition \eqref{Feq1} for the relation \eqref{tP_L}. We can find \footnote{More precisely, an arbitrary function of $\tP_a$ may be added in the right hand side of \eqref{F_L}, but it dose not change the following analysis. We drop it for simplicity. }
\begin{align}\label{F_L}
    F_g(X,\tP)&=\int dX^+ P_+(X,\tP)+\int dX^- P_-(X,\tP)\nn\\
    &=X^+ \tP^+ + X_- \tP^- + \frac{\e c^2l_p^2}{12l_\Lambda^2}\frac{(X^-)^3}{\tP_+}.
\end{align}
Then, the new position coordinates are given through $\tX^a=\frac{\p F_g}{\p \tP_a}$ \eqref{Feq2} by 
\begin{align}\label{tX_L}
    \tX^+=X^+-\frac{\e c^2l_p^2}{12l_\Lambda^2}\frac{(X^-)^3}{(P_+)^2},~~~
    \tX^-=X^-.
\end{align}
We can check by explicit calculation that these new variables satisfy
\begin{equation}\label{canonical_tilde}
    \{\tX^a,\tP_b\}=\delta^a_b,~~\{\tX^a,\tX^b\}=0=\{\tP_a,\tP_b\},
\end{equation}
as expected from the general property of canonical transformations. Therefore, the Poisson brackets \eqref{Poisson} can be calculated in terms of $(\tX^a,\tP_a)$, instead of $(X^a,P_a)$.

Thus, we complete the canonical transformation, and the Hamiltonian \eqref{H_g2} becomes, in the lapse choice \eqref{N1}, the desired form:
\begin{align}\label{H_g5}
    H_g&=-\frac{2}{\e c l_p}\tP_+ \tP_- -\frac{cl_p}{2l_s^2}\\
    &=\frac{1}{2}\MG ^{ab}\tP_a \tP_b-\frac{cl_p}{2l_s^2},\nn
\end{align}
where $\MG ^{ab}$ is still given by \eqref{metric_L}. 

Now that the Hamiltonian takes the same form as the free-particle Hamiltonian, we can follow the steps in Sec.\ref{sec:particle} to
%apply the strategy  in Sec.\ref{sec:strategy} and
obtain the Schrödinger symmetry. The set of the generators consists of $\tP_\pm$ and 
\begin{subequations} \label{Noethers_tilde_X}
\begin{gather}
%P_\pm, \\
\tB^\pm := \tX^\pm - \mathcal{G}^{\pm a} \tP_a \tau,\\ %= X^\pm - P^\pm \tau, \\
\widetilde J_{+-} := \mathcal{G}_{+a} \tX^a \tP_- - \mathcal{G}_{-a} \tX^a \tP_+ =\frac{\e c l_p}{2}\l(\tB^+ \tP_+ - \tB^- \tP_-\r) ,\\% \mathcal{G}_{+a} \tB^a \tP_- - \mathcal{G}_{-a} \tB^a \tP_+,\\
\widetilde Q_+ := \frac{1}{2} \mathcal{G}^{ab} \tP_a \tP_b = H_g + \frac{cl_p}{2l_s^2}, \\
\widetilde Q_0 := \frac{1}{2} \tB^a \tP_a, \\
\widetilde Q_- := \frac{1}{2} \mathcal{G}_{ab} \tB^a \tB^b,
\end{gather}
\end{subequations}
which satisfy the symmetry condition \eqref{symmetry_cond} for the Hamiltonian \eqref{H_g5}. These have the relations \eqref{bp_canonical_relation_free}, \eqref{jalg_free}, \eqref{qalg_free}, and \eqref{qpbalg_free}, establishing the 2D Schrödinger algebra: 
\begin{equation}
\mathfrak{sh}(2) = \left( \mathfrak{sl}(2,\mathbb{R}) \oplus \mathfrak{so}(1,1) \right) \oplus_s \left( \mathbb{R}^2 \oplus \mathbb{R}^2 \right).
\label{sh_algebra_Sch}
\end{equation}
Its exponentiation gives the 2D Schrödinger group,
\begin{equation}
\mathrm{Sh}(2) = \left( \mathrm{SL}(2,\mathbb{R}) \times \mathrm{SO}(1,1) \right) \ltimes \left( \mathbb{R}^2 \times \mathbb{R}^2 \right).
\label{sh_group_lorentzian}
\end{equation}
Thus, we have applied the canonical-transformation method and reproduced systematically the 2D Schrödinger symmetry, the result obtained in a heuristic manner \cite{BenAchour:2023dgj}. 

We conclude %this derivation 
with three remarks.
First, the above symmetry holds for the specific choice of the the lapse function \eqref{N1}. Second, the Hamiltonian constraint is \textit{not} used to derive the symmetry. Third, the constraint itself appears as one of the generators, $\widetilde Q_+=H_g+\frac{c l_p }{2l_s^2}$. These are involved in understanding the physical meaning of the symmetry (see Secs.\ref{sec:interpretation} and \ref{sec:discussion}).

\subsection{(Anti)de-Sitter Schwarzschild metric}\label{sec:AdS}
Before closing this section, we here provides a physical aspect of this system by showing
%we here show 
that the solution of the Hamiltonian \eqref{H_g2} is uniquely given by the %(anti)de Sitter 
(A)dS Schwarzschild metric. %, which provides a physical aspect of this system. 
In the derivation, we impose the Hamiltonian constraint $H_g\approx0$ as late as possible, to make its role clear. 

First, the canonical equations provide a uniform linear motion in the configuration space $\{\tX^a\}$: 
\begin{align}\label{canonical_eq}
\dot{\tP_\pm}&=\{\tP_\pm,H_g\} =0 \Rightarrow \tP_\pm(\tau)=\rm{const.}\nn\\
\dot{\tX^\pm}&=\{\tX^\pm,H_g\} =-\frac{2}{\e c l_p}\tP_\mp \Rightarrow \tX^{\pm}(\tau)=-\frac{2}{\e c l_p}\tP_\mp \tau + \tB^{\pm}.
\end{align}
Here and in the remainder of this section, we regard $\tB^{\pm}$ and $\tP_\pm$ as constants $\beta^{\pm}$ and $\alpha_{\pm}$, respectively. 

We note from the metric \eqref{metric_X} that the areal radius of the 2-sphere, $r$, should be identified as 
\begin{equation}\label{r}
    r=\frac{l_s}{2}X^-.
\end{equation}
Therefore, from the canonical transformation \eqref{tX_L} and the trajectory \eqref{canonical_eq}, we can have $\frac{2}{l_s}r=X^-(\tau)=\tX^-(\tau)=-\frac{2}{\e c l_p}\tP_+ \tau + \tB^{-}$, leading to 
\begin{align}\label{tau}
    \tau = - \frac{\e \bar c}{\tP_+}\l(r-\frac{l_s}{2}\tB^-\r)
    \Rightarrow
    d\tau = - \frac{\e \bar c}{\tP_+} dr,
\end{align}
where we define 
\begin{equation}\label{bar_c}
    \bar c \equiv \frac{cl_p}{l_s}=\frac{l_0l_s}{l_p^2}.
\end{equation} 
Similarly, we have for $X^+$ 
\begin{align}\label{X+_L}
    X^+=\frac{\tP_-}{\tP_+}X^-+\frac{1}{\tP_+}(\tB^+\tP_+-  \tB^-\tP_-) + \frac{\e c^2 l_p^2}{12l_\Lambda^2}\frac{(X^-)^3}{\tP_+^2}
\end{align}

Combining these and the lapse \eqref{N1}, we can calculate 
%\begin{align}
%    &-\e \frac{X^-}{4X^+}d\tau^2 \nn\\
%    =&\frac{1}{\frac{\tP_+ \tP_-}{-\e \bar c^2}r-\frac{l_s}{2\e \bar c^2}\tP_+(\tP_+ \tB^+-\tP_-\tB^-)r-\frac{\Lambda}{3}r^2}dr^2
%\end{align}
\begin{align}
    -\e N^2 d\tau^2
    &=-\e \frac{X^-}{4X^+}\frac{\bar c^2}{\tP_+^2} dr^2\nn\\
    &=\frac{1}{\frac{4\tP_+ \tP_-}{-\e \bar c^2}-\frac{2GM}{r}-\frac{\Lambda}{3}r^2}dr^2\nn\\
    &\approx \frac{1}{1-\frac{2GM}{r}-\frac{\Lambda}{3}r^2}dr^2,
\end{align}
where we have defined 
\begin{align}\label{M_def}
    M\equiv \frac{\e l_s}{G \bar c^2}\tP_+(\tP_+ \tB^+-\tP_- \tB^-)
    =\frac{2}{G \bar c^3}\tP_+ \widetilde J_{+-},
\end{align}
and applied the Hamiltonian constraint condition in the last step: (from \eqref{H_g5})  
\begin{equation}\label{constraint_g}
    H_g\approx0\Rightarrow \tP_+ \tP_-\approx -\frac{\e \bar c^2}{4}.
\end{equation}
Similarly, we can have 
\begin{align}
    \e \frac{4X^+}{X^-}dt^2
    &=\e \frac{4X^+}{X^-}\frac{\tP_+^2}{\bar c^2}\frac{\bar c^2}{\tP_+^2}dt^2\nn\\
    &=-\l(\frac{4\tP_+ \tP_-}{-\e \bar c^2}-\frac{2GM}{r}-\frac{\Lambda}{3}r^2\r)dt_\infty^2\nn\\
    &\approx -\l(1-\frac{2GM}{r}-\frac{\Lambda}{3}r^2\r)dt_\infty^2,
\end{align}
where we have defined the time at $r=\infty$ as $dt_\infty \equiv \frac{\bar c}{\tP_+}dt$. 
Thus, after solving the canonical equations and imposing $H_g\approx0$, the metric \eqref{metric_X} becomes the (anti)de-Sitter Schwarzschild metric: 
\begin{align}\label{metric_SchL}
    ds^2\approx&-\l(1-\frac{2GM}{r}-\frac{\Lambda}{3}r^2\r)dt_\infty^2\nn\\
    &+\l(1-\frac{2GM}{r}-\frac{\Lambda}{3}r^2\r)^{-1}dr^2+r^2 d\Omega^2.
\end{align}

Here, the formula \eqref{M_def} plays the role of the energy of the system. Indeed, in the case of $\Lambda=0$ and $\e=-1$, it agrees with the Komar mass \cite{Wald:1984rg} (see Appendix \ref{A:M_Q}). $M$ commutes with $H_{g}$ \eqref{H_g5} and thus is a gauge-invariant observable. Note that if we introduce another gauge-invariant observable $T\equiv \frac{G\bar c^2}{\epsilon l_s \tP_+}$, which is associated with  $dt_\infty \equiv \frac{\bar c}{\tP_+}dt$, we can check that 
\begin{equation}\label{TM_rel}
    \{T,M\}=1,
\end{equation}
holds,\footnote{Similar relations appear in Refs.\cite{Kastrup:1993br, Kuchar:1994zk}.} and it would be interesting to study this system in terms of $(T,M)$ and consider the effect of the Schrödinger symmetry.
%%%%%%%%%%%%%%%%%%%%%%%%%%%%%%%%%%%%%%%%%%%%%%%%%%%%%%%%%%%%%%%%%%%%%
\section{Maxwell Field}
\label{sec:EM}
In the previous section, we have succeeded in applying the canonical-transformation method to the vacuum mini-superspace with cosmological constant $\Lambda$ and reproducing the known result, the 2D Schrödinger symmetry. In the following sections, we will apply the method to new cases where matter fields exist, increasing the dimension of the mini-superspace. In this section, we will consider Maxwell field $A_\mu$ with $\Lambda$ and show that the 3D Schrödinger symmetry appears.  

%In this section, we demonstrate the emergence of 3D Schrödinger symmetry in a static, spherically symmetric minisuperspace model coupled to an electromagnetic field \( A^\mu \) and a cosmological constant \( \Lambda \). The inclusion of the electromagnetic field increases the number of dimensions of the minisuperspace to three. As a result, a three-dimensional Schrödinger symmetry can be identified, and its two-dimensional subgroup corresponds to the symmetry observed in the vacuum case.

%%%%%%%%%%%%%%%%%%%%%%%%%%%%%%%
%\subsection{3D Schrödinger symmetry}
We follow the same steps as in the previous section to derive the symmetry. We first evaluate the Hamiltonian for the reduced Lagrangian of the action,  
\begin{align}
  S_{gE}[g_{\mu\nu},A_\mu]
   &=\frac{1}{16\pi G\hbar}\int_\mathcal{M}d^4x\sqrt{-g}(R-2\Lambda)\nn\\
   &~~~~~~~~~~-\frac{1}{4\hbar} \int_\MM d^4x\sqrt{-g}F_{\mu\nu}F^{\mu\nu},  
   \label{action_EL}
\end{align}
under the assumption $A_\mu(x)dx^\mu =A_0(\tau)dt$ in the spherically-symmetric static metric \eqref{metric_X}, where $F_{\mu\nu}\equiv \p_\mu A_\nu-\p_\nu A_\mu$.
Noting $F_{\mu\nu}F^{\mu\nu}=2g^{tt}g^{\tau\tau}\dot A_0^2$ and integrating time and 2-sphere, the Maxwell part $L_E$ is given through $S_E=\int d\tau L_E$ by 
\begin{align}\label{L_E}
L_E=\frac{c l_p}{4N}\sqrt{\frac{X^-}{X^+}}(X^-)^2(\dot X^A)^2,
\end{align}
where we define $X^A\equiv \sqrt{\frac{\pi}{\hbar}}l_p A_0=\sqrt{\pi G} A_0$. 
Introducing the canonical momentum for $X^A$ as 
\begin{equation}
    P_A:=\frac{\p L_E}{\p  \dot X_A}= \frac{c l_p}{2N}\sqrt{\frac{X^-}{X^+}}(X^-)^2 \dot X^A, 
\end{equation}
we have the Hamiltonian for the Maxwell part: 
\begin{align}
    H_E:=\dot X^A P_A - L_E = N\sqrt{\frac{X^+}{X^-}} \frac{P_A^2}{cl_p (X^-)^2}.
\end{align}
Combining this and the gravity one \eqref{H_g}, we gain the total Hamiltonian: 
\begin{align}\label{H_gE}
    H_{gE}&=H_g + H_E\nn\\
                &= N\sqrt{\frac{X^+}{X^-}}\l[-\frac{4}{\e c l_p}P_+P_- + \frac{P_A^2}{cl_p (X^-)^2}\r.\nn\\
                &~~~~~~~~~~~~~~~~~~~~~~~~\l. +\frac{cl_p}{l_\Lambda^2}(X^-)^2 -\frac{cl_p}{l_s^2}\r]\nn\\
                &\equiv  N \MH_{gE}.
\end{align}
Here, $\MH_{gE}$ is the Hamiltonian constraint, %which becomes $\MH_{gE}\approx 0$ on shell, 
and the canonical variables are $(X^+,X^-,X^A;P_+,P_-,P_A)$. % satisfy the relation \eqref{canonical}. 

In the chosen $N$ \eqref{N1}, the Hamiltonian \eqref{H_gE} becomes
\begin{align}\label{H_gE2}
    H_{gE}&=-\frac{2}{\e c l_p}P_+P_- + \frac{P_A^2}{2cl_p (X^-)^2} +\frac{cl_p}{2l_\Lambda^2}(X^-)^2 -\frac{cl_p}{2l_s^2}\\
    \label{H_gE3}
    &=\frac{1}{2}\MG^{ab}_E P_a P_b +\MV_\Lambda(X),
\end{align}
where $a,b = +, -, A$, and the supermetric is 3-dimensional: 
\begin{align}
\MG_{Eab}& =   cl_p 
\begin{pmatrix}
0 & -\frac{\epsilon}{2}& 0 \\
-\frac{\epsilon}{2} & 0 & 0\\
0 & 0 & (X^-)^2
\end{pmatrix},\nn\\
\MG_{E}^{ab}& =\frac{1}{cl_p}
\begin{pmatrix}
0 & -\frac{2}{\epsilon} & 0 \\
-\frac{2}{\epsilon} & 0 & 0\\
0 & 0 & \frac{1}{(X^-)^2}
\end{pmatrix},
\label{supermetric_gE}
\end{align}
This is again ``Lorentzian", $\det \MG_{E ab}<0$, and the lifted space is not conformally flat.

We next perform a canonical transformation to express the Hamiltonian $H_{gE}$ \eqref{H_gE2} as the free-particle form. We insert $\frac{1}{2cl_p}(-P_A^2+P_A^2)=0$ into $H_{gE}$ and write it as
\begin{align}\label{H_gE4}
    &H_{gE}\nn\\
    =&
    -\frac{2}{\e c l_p}P_+\l[P_- +\frac{\e P_A^2}{4P_+}\l(1- \frac{1}{(X^-)^2}\r) -\frac{\e c^2l_p^2}{4l_\Lambda^2}\frac{(X^-)^2}{P_+}\r]\nn\\
    &~~~~~~~~~~~~~~~~~~~~~~~~~~~~~~~~~~~~~~~+\frac{1}{2cl_p}P_A^2-\frac{cl_p}{2l_s^2}.
\end{align}
Noting $\{H_{gE},H_{gE} \}=\{P_+,H_{gE}\}=\{P_A,H_{gE}\}=0$, the bracket $[\cdots]$ also commutes with $H_{gE}$. Therefore, we set the new momenta as 
\begin{align}\label{tP_E}
    \tP_+&\equiv P_+,\nn\\
    \tP_-&\equiv P_- +\frac{\e P_A^2}{4P_+}\l(1- \frac{1}{(X^-)^2}\r) -\frac{\e c^2l_p^2}{4l_\Lambda^2}\frac{(X^-)^2}{P_+}, \nn\\
    \tP_A&\equiv P_A.
\end{align}
This can be reproduced by a generating function $F_{gE}(X,\tP)$, say, 
\begin{align}\label{F_E}
   F_{gE}(X,\tP)&=\int dX^+ P_+(X,\tP)+\int dX^- P_-(X,\tP)\nn\\
   &~~~~~~~~~~~~~~~~~~~~~~~~~~~~~~+\int dX^A P_A(X,\tP),\nn\\
    &=X^+ \tP^+ + X_- \tP^- + X^A \tP_A \nn\\
    &~~-\frac{\e \tP_A^2}{4\tP_+}\l(X^-+\frac{1}{X^-}\r)+ \frac{\e c^2l_p^2}{12l_\Lambda^2}\frac{(X^-)^3}{\tP_+}.
\end{align}
In turn, the new coordinates $\tX^a$ are obtained through $\tX^a=\frac{\p F_{gE}}{\p \tP_a}$ as 
\begin{align}\label{tX_E}
\tX^+&= X^+ + \frac{\e P_A^2}{4P_+^2}\l(X^-+\frac{1}{X^-}\r) - \frac{\e c^2l_p^2}{12l_\Lambda^2}\frac{(X^-)^3}{P_+^2},\nn\\
\tX^-&=X^-,\nn\\
\tX^A&=X^A-\frac{\e P_A}{2P_+}\l(X^-+\frac{1}{X^-}\r).
\end{align}
Indeed, these new variables $(\tX^a,\tP_a)$ satisfy the canonical relations \eqref{canonical_tilde}. We thus reach the desired form: 
\begin{align}\label{H_gE5}
    H_{gE}&=-\frac{2}{\e c l_p}\tP_+ \tP_- + \frac{1}{2cl_p}\tP_A^2 -\frac{cl_p}{2l_s^2}\\
    &=\frac{1}{2}\widetilde{\MG}^{ab}_E \tP_a \tP_b -\frac{cl_p}{2l_s^2}\nn
\end{align}
with the constant supermetric 
\begin{align}
\widetilde{\MG}_{Eab} =   cl_p 
\begin{pmatrix}
0 & -\frac{\epsilon}{2}& 0 \\
-\frac{\epsilon}{2} & 0 & 0\\
0 & 0 & 1
\end{pmatrix},~
\widetilde{\MG}_{E}^{ab} =\frac{1}{cl_p}
\begin{pmatrix}
0 & -\frac{2}{\epsilon} & 0 \\
-\frac{2}{\epsilon} & 0 & 0\\
0 & 0 & 1
\end{pmatrix}.
\label{supermetric_gE2}
\end{align}

Finally, we find the symmetry. According to the strategy of Sec.\ref{sec:strategy}, we set the generators as 
\begin{subequations} \label{Noethers_E}
\begin{gather}
\tP_a, ~~\tB^a := \tX^a - \widetilde{\MG}^{ab}_E \tP_b \tau,\\
\widetilde J_{ab} := \widetilde{\MG}_{Eac} \tX^c \tP_b- \widetilde{\MG}_{Ebc} \tX^c \tP_a\nn\\
~~~~~~=\widetilde{\MG}_{Eac} \tB^c \tP_b- \widetilde{\MG}_{Ebc} \tB^c \tP_a,\\
\widetilde Q_+:=\frac{1}{2}\widetilde{\MG}^{ab}_E \tP_a \tP_b=H_{gE}+\frac{cl_p}{2l_s^2} \\
\widetilde Q_0 := \frac{1}{2} \tB^a \tP_a, \\
\widetilde Q_- := \frac{1}{2} \widetilde{\MG}_{Eab} \tB^a \tB^b,
\end{gather}
\end{subequations}
where $(a,b,c)=(+,-,A)$. These constitute the 3D-Schrödinger algebra
\begin{equation}\label{3D_Sc_alg}
\mathfrak{sh}(3) = \left( \mathfrak{sl}(2,\mathbb{R}) \oplus \mathfrak{so}(2,1) \right) \oplus_s \left( \mathbb{R}^3 \oplus \mathbb{R}^3 \right),
\end{equation}
meaning that the system possesses the 3D-Schrödinger symmetry. 

This symmetry is a consequence of containing a mixture of electric and gravity components, say, $\widetilde{J}_{A\pm}=\frac{1}{2}cl_p(\e\tB^\mp\tP_A+\tB^A\tP_\pm)$. On the other hand, only the gravity sector $(\tB^\pm, \tP_\pm, \widetilde{J}_{+-})$ together with $(\widetilde{Q}_0,\widetilde{Q}_\pm)$ form the 2D-Schrödinger algebra as a sub-algebra of the 3D one, where $\tB^A$ and $\tP_A$ do not contribute. Therefore, taking the matter decoupling limit where $\tP_A\to0$ and $\tB_A\to0$, the 3D-Schrödinger algebra reduces to the 2D one. %\TS{This is strange to some extend. $\tilde{P}_-$ and $\tilde{X}^+$ is included $P_A$ and what is important is not $\tB^A\to0$ but $P_A\to0$ or $\tP_A\to0$.} 

%%%%%%%%%%%%%%%%%%%%%%%%%%%%%%%%%%
%\subsection{(Anti) de-Sitter Reissner-Nordström metric}
We conclude this section by checking that the on-shell metric of this system is uniquely given by the (anti-)de-Sitter Reissner-Nordström metric: 
\begin{align}\label{metric_RN}
    ds^2\approx&-\l(1-\frac{2GM}{r}+\frac{G\MQ^2}{4\pi r^2}-\frac{\Lambda}{3}r^2\r)dt_\infty^2\nn\\
    &+\l(1-\frac{2GM}{r}+\frac{G\MQ^2}{4\pi r^2}-\frac{\Lambda}{3}r^2\r)^{-1}dr^2+r^2 d\Omega^2,
\end{align}
where 
\begin{align}\label{Q_def}
    \MQ\equiv \sqrt{\frac{\pi}{G}} \frac{l_s}{\bar c}\tP_A,
\end{align}
is the electric charge (see Appendix \ref{A:RN} for the derivation).

\section{Multiple Massless Scalar Fields}\label{sec:scalar}
As another case with matter fields, we discuss the mini-superspace with $n$ massless scalar fields, whose solution corresponds to a generalized Janis-Newman-Winicour (JNW) solution \cite{Janis:1968zz} and its ``interior" \cite{Chiou:2008eg}, and we show that $(2+n)$ D-Schrödinger symmetry appears. An interesting point here is that the symmetry is obtained by using a different coordinate system $Y^a$ in the mini-superspace than the previous one $X^a$, together with a different choice of gauge $N$. %We discuss the implication. %This covariance of the symmetry in the mini-superspace should be a signal for a robustness of the symmetry. 

%%%%%%%%%%%%%%%%%%%%%%%%%%%%
\subsection{Y-coordinate in the mini-superspace}\label{sec:Y}
We consider a coupled system of gravity and $n$ free massless scalar fields $\phi_I=\phi_I(\tau)$ and construct a Hamiltonian suitable for the canonical-transformation method. Before going to the details, it is interesting to see what would happen if we included mass terms and cosmological constant. So, let us start with the action: 
\begin{align}\label{S_s}
S_{g\phi}[g_{\mu\nu},\phi_I]&=\frac{1}{ 16\pi G \hbar }\int d^4x\sqrt{-g}(R-2\Lambda)\nn\\ 
   &-\frac{1}{2\hbar}\sum_{I=1}^n\int d^4x\sqrt{-g}\l(g^{\mu\nu}\partial_{\mu}\phi_I\partial_{\nu}\phi_I+m_I^2 \phi_I^2\r).
\end{align}

First, if we use the ansatz metric \eqref{metric_X} %which we have considered so far, 
and evaluate the reduced Hamiltonian $H_{g\phi}$, we can see that it is difficult to apply the canonical-transformation method.% as in the previous cases.
\footnote{\label{foot:scalar_X}For the metric \eqref{metric_X} and $\phi_I=\phi_I(\tau)$, the action of scalar fields reduces to $S_\phi=\int d\tau L_\phi$ with 
\begin{equation}
    L_\phi = cl_pN\sqrt{\frac{X^+}{X^-}}(X^-)^2\sum_I\l(\frac{1}{\e N^2} (\dot X^I)^2 -m_I^2 (X^I)^2 \r),\nn
\end{equation}
where $X^I\equiv \sqrt{\frac{\pi}{\hbar}}l_p \phi_I$.
Defining the momenta $P_I\equiv \frac{\p L_\phi}{\p \dot X^I}$, we can calculate the Hamiltonian $H_\phi\equiv \sum_I P_I \dot X^I-L_\phi $ and add the gravity part $H_g$ \eqref{H_g} to reach the reduced Hamiltonian $H_{g\phi}=H_g+H_\phi$: 
\begin{align}
    H_{g\phi}&  
    =N\sqrt{\frac{X^+}{X^-}}\l[
    -\frac{4}{\e c l_p}P_+P_- -\frac{cl_p}{l_s^2}+\frac{cl_p}{l_\Lambda^2}(X^-)^2 \r.\nn\\
    &
    +\l.\sum_I\l(\frac{\e}{4cl_p X^+X^-}P_I^2+cl_pm_I^2(X^-)^2(X^I)^2\r)\r].\nn
\end{align}
Here, the mass terms lead to a non-trivial interaction between $X^-$ and $X^I$. Even if we choose the lapse \eqref{N1}, $N=\sqrt{\frac{X^-}{4X^+}}$, and set $m_I=0$ and $\Lambda=0$, the term of $P_I^2$ would remain mixing $X^+$ and $X^-$, and it would be difficult to find two quantities that commute with $H_{g\phi}$. }

One idea for overcoming this situation is to employ another coordinate system in the mini-superspace. We consider the $Y$-coordinate system related to the $X$-coordinate system as 
\begin{align}\label{YX}
Y^0:=\frac{1}{4}\log X^+X^-,~Y^1:=\frac{1}{4}\log\frac{X^+}{X^-},~Y^I:=\sqrt{\frac{\pi}{\hbar}}l_p \phi_I,
\end{align}
in which the spacetime metric \eqref{metric_X} is expressed as 
\begin{equation}\label{metrix_Y}
    ds^2=\e\l(-N^2 d\tau^2+4e^{4Y^1}dt^2\r)+\frac{l_s^2}{4}e^{4Y^0-4Y^1}d\Omega^2.
\end{equation}
We can use this and evaluate the action \eqref{S_s} by performing integration by parts and dropping boundary terms, to obtain the reduced Lagrangian $L_{g\phi}$ via $S_{g\phi}=\int d\tau L_{g\phi}$: 
\begin{align}
    L_{g\phi}&=Ne^{4Y^0-2Y^1}cl_p\l[
    \frac{1}{\e N^2}\l(-(\dot Y^0)^2+(\dot Y^1)^2+\sum_I(\dot Y^I)^2\r)\r. \nn\\
   &~~~~~~~~~~~~~\l. -\sum_I m_I^2(Y^I)^2-\frac{1}{l_\Lambda^2}+\frac{1}{l_s^2}e^{-4Y^0+4Y^1}\r].
\end{align}
Defining the momenta $\Pi_a\equiv \frac{\p L_{g\phi}}{\p \dot Y^a}$ (for $a=0,1,I$), we have the reduced Hamiltonian: 
\begin{align}\label{H_phi}
    H_{g\phi}&=N e^{-4Y^0+2Y^1}\l[
    \frac{1}{4\e cl_p}\l(-\Pi_0^2+\Pi_1^2+\sum_I\Pi_I^2\r)\r.\nn\\
    &\l.-\frac{cl_p}{l_s^2}e^{4Y^0}+cl_p e^{8Y^0-4Y^1}\l(\sum_I m_I^2 (Y^I)^2+\frac{1}{l_\Lambda^2}\r)\r].
\end{align}
The mass terms connect all the coordinates $Y^0,Y^1,Y^I$, and the cosmological constant links $Y^0$ and $Y^1$, which still makes the application of our method difficult. 

In the following, we assume
\begin{equation}
    m_I=0,~~~\Lambda=0
\end{equation}
and set the lapse as  
\begin{equation}\label{N2}
    N=2 e^{4Y^0-2Y^1}. 
\end{equation}
Then, we reach a simple form of the Hamiltonian: 
\begin{align}\label{H_phi2}
    H_{g\phi}&=\frac{1}{2\e cl_p}\l(-\Pi_0^2+\Pi_1^2+\sum_I\Pi_I^2\r)-\frac{2cl_p}{l_s^2}e^{4Y^0}\\
    &=\frac{1}{2\e cl_p}\eta^{ab}\Pi_a\Pi_b- \frac{2cl_p}{l_s^2}e^{4Y^0},\nn    
\end{align}
where %(diagonal of $\eta^{ab})\equiv (-1,1,\delta_{IJ})$ 
$\eta_{ab} \equiv \mathrm{diag}(-1,1,\cdots.,1)$ is a $(2+n)\times(2+n)$ ``Lorentzian" constant matrix. %Here, for convenience, we have chosen the gauge \eqref{N2} so that the kinematic term of the Hamiltonian \eqref{H_phi2} for $\e=-1$ takes the relativistic standard form $\frac{1}{2m}(-E^2 + P_i^2)$, where $P_0$ corresponds to ``energy $E$", although this is not important for the dynamics since it is determined essentially by $H_{g\phi}\approx0$ as discussed in Sec.\ref{sec:intro}.
This contains only one coordinate $Y^0$ so that we can implement the canonical-transformation program easily.
%%%%%%%%%%%%%%%%%%%%%%%
\subsection{$(2+n)$ D-Schrödinger symmetry}\label{sec:symmetry_s}
We now follow our strategy and show the appearance of the $(2+n)$ D-Schrödinger symmetry. 

We first change the canonical variables. 
In the Hamiltonian $H_{g\phi}$ \eqref{H_phi2}, $\Pi_1$ and $\Pi_I$ commute with $H_{g\phi}$, and therefore, 
the sum of the remaining two terms must commute with $H_{g\phi}$. This suggests as the new momenta
\begin{align}\label{tPi}
    \tPi_0\equiv \sqrt{\Pi_0^2+4\e \bar c^2 e^{4Y^0}},~\tPi_1\equiv \Pi_1,~\tPi_I\equiv \Pi_I.    
\end{align}
A generating function corresponding to these can be identified with 
\begin{align}\label{F_E}
   &F_{g\phi}(Y,\tPi)\nn\\
   %&=\int dY^0 P_0(Y,\tP)+\int dY^1 P_1(Y,\tP)+\sum_I \int dY^I P_I(Y,\tP)\nn\\
   &=\int dY^0 \sqrt{\tPi_0^2-4\e \bar c^2 e^{4Y^0}}+\int dY^1 \tPi_1+\sum_I \int dY^I \tPi_I\nn\\
    &=\frac{1}{2} \sqrt{\tPi_0^2-4\e \bar c^2 e^{4Y^0}}+\frac{\tPi_0}{4} \log \l(\frac{\e\tPi_0-\e\sqrt{\tPi_0^2-4\e \bar c^2 e^{4Y^0}}}{\tPi_0+\sqrt{\tPi_0^2-4\e \bar c^2 e^{4Y^0}}}\r)\nn\\
    &~~~~~~~~+Y^1 \tPi^1 +\sum_I Y_I \tPi^I.
\end{align}
In turn, this generates the new coordinates
\begin{align}\label{tY}
    \tY^0&=\frac{1}{4}\log \left(\frac{\e \Pi_0-\e\sqrt{\Pi_0^2-4\epsilon\bar{c}^2e^{4Y^0}}}{\Pi_0+\sqrt{\Pi_0^2-4\epsilon\bar{c}^2e^{4Y^0}}}\right),\nn\\
    &~~~~~~~~~\tY^1=Y^1,~~\tY^I=Y^I.
\end{align}
Indeed, these satisfy the canonical relation:
%and the canonical relation holds indeed:
\begin{equation}
    \{\tY^a,\tPi_b\}=\delta^a_b,~\{\tY^a,\tY^b\}=0=\{\tPi_a,\tPi_b\}.
\end{equation}
Thus, the Hamiltonian \eqref{H_phi2} is expressed in terms of $(\tY^a,\tPi_a)$ as 
\begin{align}\label{H_phi3}
    H_{g\phi}&=\frac{1}{2(-\e) cl_p}\l(-\tPi_0^2+\tPi_1^2+\sum_I\tPi_I^2\r)\nn\\
    &=\frac{1}{2}\MG^{ab}_\phi\tPi_a\tPi_b,    
\end{align}
with the (2+n)-dimensional constant supermetric:
\begin{align}
    \MG_{\phi ab} =  (-\e) cl_p \eta_{ab},~~\MG_{\phi}^{ab} =\frac{1}{(-\e)cl_p}\eta^{ab}.
\end{align}
%where $\eta_{ab} = \mathrm{diag}(-1,1,\cdots.,1)$.
%\begin{align}
%\MG_{\phi ab} =  (-\e) cl_p 
%\begin{pmatrix}
%-1  & 0 & 0 \\
%0 & 1 & 0\\
%0 & 0 & \delta_{IJ}
%\end{pmatrix},~
%\MG_{\phi}^{ab} =\frac{1}{(-\e)cl_p}
%\begin{pmatrix}
%-1 & 0 & 0 \\
%0 & 1 & 0\\
%0 & 0 & \delta_{IJ}
%\end{pmatrix}.
%\label{supermetric_phi}
%\end{align}
Here, for convenience, we multiplied $-1$ in the right hand side of \eqref{H_phi3} so that the Hamiltonian \eqref{H_phi3} for $\e=-1$ corresponds to the Hamiltonian constraint of a relativistic massless particle, $\MH_{\rm{rel}}=-E^2 + P_i^2$, where $\tPi_0$ plays a role of ``energy $E$", although this is not important for dynamics since it is determined essentially by $H_{g\phi}\approx0$ as discussed in Sec.\ref{sec:intro}. 

Now that the Hamiltonian \eqref{H_phi3} has the same form as the free-particle Hamiltonian \eqref{hamiltonian_free}, we can replace $(x^i,p_i,g^{ij})$ with $(\tY^a,\tPi_a,\MG^{ab}_\phi)$ and apply the argument in Sec.\ref{sec:particle} to conclude that the spherically-symmetric static system coupled with $n$ free massless scalar fields has the $(2+n)$ D-Schrödinger symmetry: 
\begin{equation}\label{scalar_sch_sym}
\mathrm{Sh}(2+n) = \left( \mathrm{SL}(2,\mathbb{R}) \times \mathrm{SO}(1,1+n) \right) \ltimes \left( \mathbb{R}^{2+n} \times \mathbb{R}^{2+n} \right).
\end{equation}

The $(2+n)$ D-Schrödinger algebra $\mathfrak{sh}(2+n)$ consists of three sectors. The gravity sector has only gravity components $(\tPi_0,\tPi_1,\tB^0,\tB^1,\tJ_{01})$, forming together with $\tQ_{\pm}, \tQ_0$ a sub-algebra $\mathfrak{sh}(2)$; the scalar-field sector is composed by $(\tPi_I,\tB^I,\tJ_{IJ})$, leading together with $\tQ_{\pm}, \tQ_0$ to a sub-algebra $\mathfrak{sh}(n)$; and the gravity-scalar sector contains the mixed components $\tJ_{0I},\tJ_{1I}$. 
Therefore, in the matter decoupling limit where $\tPi_I\to0$ and $\tB^I\to0$, only the gravity sector remains, and the algebra reduces to the 2D-Schrödinger algebra $\mathfrak{sh}(2)$.  %\TS{This is also strange to some extend.} \YY{Please modify as you want.}

Note that in the matter decoupling limit, this model reduces to the vacuum model with $\Lambda=0$ in Sec.\ref{sec:Schwarzschild}. The argument above and the one in Sec.\ref{sec:2D_Sch} show that the algebra $\mathfrak{sh}(2)$ holds true both for $Y$-coordinates with the lapse \eqref{N2} and for $X$-coordinates with another lapse \eqref{N1}. Therefore, the 2D-Schrödinger symmetry in the same vacuum model can be realized in different choices of mini-superspace coordinates and lapses. This result is expected to provide, albeit partially, evidence of the covariance of the symmetry (i.e., independence of the choice of mini-superspace coordinates or lapse).

%The remarkable point here is that this algebra $\mathfrak{sh}(2)$ is expressed in terms of $Y$-coordinates with the lapse choice \eqref{N2} while the one obtained in Sec.\ref{sec:2D_Sch} is written in terms of $X$-coordinates under a different lapse \eqref{N1}.  Therefore, the 2D-Schrödinger symmetry can be realized in different choices of mini-superspace coordinates and lapses. This means the covariance of the Schrödinger symmetry in the mini-superspace, which signals a robustness of the symmetry (at least, for pure-gravity cases).

%Note that in the matter decoupling limit, both Model I (without cosmological constant) and Model II reduce to the same model, the vacuum spherically symmetric static model. They lead to the 2D Schrödinger symmetry in terms of different lapse functions and different mini-superspace coordinates. This is expected to support (at least, partially) the covariance of the 2D Schrödinger symmetry in the mini-superspace.

%%%%%%%%%%%%%%%%%%%%%%%
\subsection{Generalized JNW metric and its ``interior"}\label{JNW}
To see the physics of this system, we solve the Hamiltonian \eqref{H_phi3}. %and derive the generalized Janis-Newman-Winicour (JNW) metric with $n$ massless scalar fields and its ``interior" metric. 
Here, we provide the results (see Appendix \ref{A:JNW} for the derivation).

The metric is given by 
\begin{align}\label{metric_JNW}
    ds^2&=(-\e)\l(-\l[\frac{(-\epsilon)(R-r_-)}{R+r_+}\r]^{\frac{1}{\mu}}dt_\infty^2\r.\nn\\
    &~~~~~~~\l.+\l[\frac{R+r_+}{(-\epsilon)(R-r_-)} \r]^{\frac{1}{\mu}}dR^2\r)+r(R)^2d\Omega^2,
\end{align}
with 
\begin{align}\label{r(R)}
    r(R)^2=(R+r_+)^{1+\frac{1}{\mu }}
        [(-\epsilon)(R-r_-)]^{1-\frac{1}{\mu}}.
\end{align}
Here, we introduce a new radial coordinate $R$ by 
\begin{equation}
   e^{4\tY^0}=:\frac{R+r_+}{(-\epsilon)(R-r_-)},
\label{R_def}
\end{equation}
and we define
\begin{align}\label{defs}
    \mu\equiv\frac{\tPi_0}{\tPi_1},&~r_{\pm}\equiv \frac{r_0}{2}(\mu\pm1),~r_0\equiv \frac{l_s \tPi_1}{2\bar c}e^{-2\l(\frac{1}{\mu}\tB^0+\tB^1\r)},\nn\\
    dt_\infty\equiv& 2 e^{2\l(\frac{1}{\mu}\tB^0+\tB^1\r)}dt=\frac{l_s \tPi_1}{\bar c r_0}dt.
\end{align} 
We can check $\{\mu,H_{g\phi}\}=\{r_0,H_{g\phi} \}=0$ for \eqref{H_phi3}, meaning that $\mu$ and $r_0$ are gauge invariant. 
Note that we have not used the Hamiltonian constraint $H_{g\phi}\approx 0$ yet. 

The scalar fields $\phi_I =\frac{1}{\sqrt{\pi G}} \tY^I$ are described by 
\begin{align}\label{phi_JNW}
    \phi_I(R)=\frac{\MA_I}{\mu}\log \frac{(-\epsilon)(R-r_-)}{R+r_+}+\phi_I^{(0)},
\end{align}
where 
\begin{align}\label{defs2}
    \MA_I \equiv \frac{1}{\sqrt{16\pi G}} 
    \frac{\tPi_I}{\tPi_1},~~
      \phi_I^{(0)}\equiv \frac{1}{\sqrt{\pi G}}
    \l( \frac{\tPi_I}{\tPi_0} \tB^0 + \tB^I \r).
\end{align}

Finally, we impose the Hamiltonian constraint \eqref{H_phi3}:  
\begin{align}\label{mu_JNW}
    H_{g\phi}\approx 0 &\Rightarrow -\tPi_0^2+\tPi_1^2+\sum_I\tPi_I^2\approx 0.\nn\\
    &\Rightarrow \mu^2 \equiv\l( \frac{\tPi_0}{\tPi_1}\r)^2 \approx 1+ 16\pi G \sum_I \MA_I^2.
\end{align}
This provides a constraint to the physical value of $\mu$:
\begin{equation}\label{mu_value}
    \mu\geq 1,
\end{equation}
where $\mu=1$ holds only for $\MA_I=0$.  

In this solution, the energy-momentum tensor $T_{\mu\nu}=\sum_I(\p_\mu \phi_I \p_\nu \phi_I-\frac{1}{2}g_{\mu\nu}g^{\alpha\beta}\p_\alpha \phi_I \p_\beta \phi_I)$ satisfies 
\begin{equation}\label{TTTT}
    -T^t{}_t=T^R{}_R=-T^\theta{}_\theta=-T^\phi{}_\phi, 
\end{equation}
with 
\begin{align}\label{E_density}
    -T^t{}_t&=\frac{1}{2}g^{\tau\tau}\sum_I\dot \phi_I^2\nn\\
        &=\frac{r_0^2\sum_I \MA_I^2}{2(-\epsilon)}(R+r_+)^{-2-\frac{1}{\mu}}[(-\epsilon)(R-r_-)]^{-2+\frac{1}{\mu}}.
\end{align}
%where we have used the configuration \eqref{metric_JNW} and \eqref{phi_JNW} and the $\tau$-$R$ relation in Eq.\eqref{expressions_R}. 

Let us now investigate the physical properties of this solution for several cases. 

%%%%%%%%%%%%%%%%%%%
\subsubsection*{\rm{(i)} $\mu=1$ for $R\neq0$: Schwarzschild metric} 
In this case, the constraint \eqref{mu_JNW} means the vanishing amplitude of scalar fields: $\MA_I=0$. From \eqref{E_density}, then, the energy-momentum tensor becomes zero, leading to a vacuum configuration. Here, we have $r_-=0$ and $r_+=r_0$ (from \eqref{defs}), and the relation \eqref{r(R)} provides $r=R+r_0$ for $R\neq0$. Thus, in the case of $\mu=1$ and $R\neq 0$, the metric \eqref{metric_JNW} reduces to the Schwarzschild metric both for $\epsilon=-1$ and $\epsilon=+1$: 
\begin{equation}\label{Sch_r0}
    ds^2=-\l(1-\frac{r_0}{r}\r)dt_\infty^2+\l(1-\frac{r_0}{r}\r)^{-1}dr^2+r^2 d\Omega^2.
\end{equation}
Here, the ADM energy $M$ is given by  
\begin{equation}\label{M_def2}
    M=\frac{r_0}{2G}=\frac{l_s \tPi_1}{4G \bar c}e^{-2\l(\tB^0+\tB^1\r)},
\end{equation}
where we set $\mu=1$ in the definition of $r_0$ \eqref{defs}.
Although this expression is different from that of \eqref{M_def}, we can check that $\frac{r_0}{2G}$ agrees with the Komar mass even in the case of $\mu\neq1$ (see Appendix \ref{A:M_Q}). $r_0$ is gauge invariant (as shown above), and so this expression of $M$ is. Note that introducing a gauge-invariant quantity $T\equiv 2\frac{G\bar c}{l_s}e^{2\l(\frac{1}{\mu}\tB^0+\tB^1\r)}$, we can show that the canonical relation \eqref{TM_rel} holds for this expression of $T$ and $M$ too (in a general value of $\mu$).

%%%%%%%%%%%%%%%%%%%%
\subsubsection*{\rm{(ii)} $\mu>1$ and $\epsilon=-1$: a generalized JNW solution}
This describes the configuration for $R>r_-$ that is  given (from \eqref{r(R)}, \eqref{metric_JNW}, and  \eqref{phi_JNW}) by 
\begin{align}\label{metric_JNW_out}
    ds^2=-\l[\frac{R-r_-}{R+r_+}\r]^{\frac{1}{\mu}}dt_\infty^2+\l[\frac{R+r_+}{R-r_-} \r]^{\frac{1}{\mu}}dR^2+r(R)^2d\Omega^2,
\end{align}
with 
\begin{align}\label{r_phi_out}
    r(R)^2&=(R+r_+)^{1+\frac{1}{\mu }}
        (R-r_-)^{1-\frac{1}{\mu}},\nn\\
        \phi_I(R)&=\frac{\MA_I}{\mu}\log \frac{R-r_-}{R+r_+}+\phi_I^{(0)},
\end{align}
where $\mu$ satisfies the constraint \eqref{mu_JNW}. This is the JNW metric \cite{Janis:1968zz} that is generalized to $n$ scalar fields. 

\begin{figure}[h]
  \centering
 \includegraphics[scale=0.32]{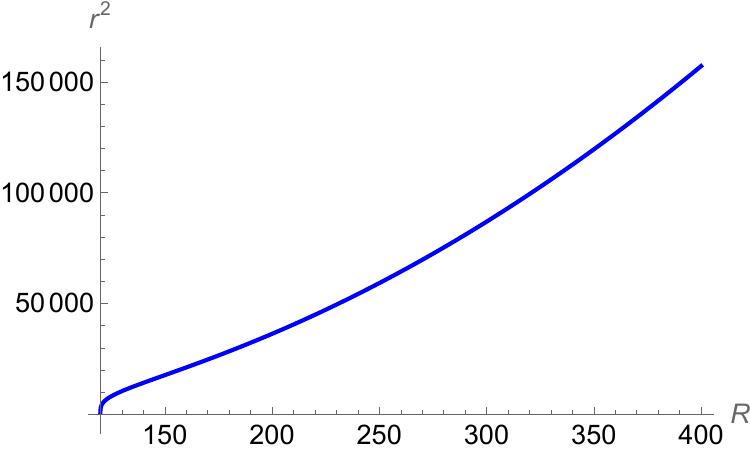}
 \includegraphics[scale=0.32]{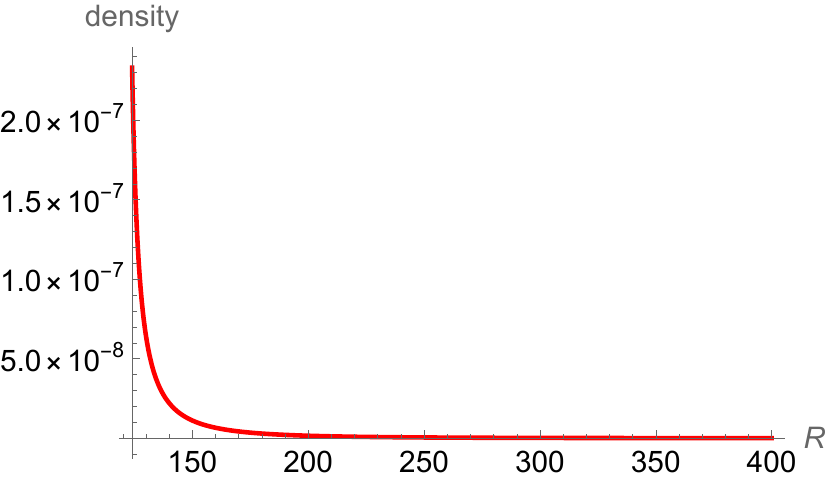}
\caption{For the case of $\epsilon=-1$, the areal radius $r(R)^2$ (Left) and energy density $-T^t{}_t(R)$ (Right). $\mu=1.4$ and $r_0=100$.}
\label{f:JNW_out} 
\end{figure}

We see the physics briefly \cite{Janis:1968zz}. The energy density \eqref{E_density} is non-zero due to $\MA_I\neq0$ (from \eqref{mu_JNW} and $\mu>1$), approaching to zero for $R\to \infty$ and to positive infinity for $R\to r_-$ (see Right of Fig.\ref{f:JNW_out}). 
On the other hand, the areal radius $r(R)$ \eqref{r_phi_out} becomes large for a large $R$ and vanishes for $R\to r_-$ (see Left of Fig.\ref{f:JNW_out}). Therefore, this represents an asymptotically-flat non-vacuum spacetime with a naked singularity at the center, $r=0$. It is not an isotropic physical fluid: $T^R_R\neq T^\theta_\theta=T^\phi_\phi$ and $T^\theta_\theta<0$. Note that one should be careful when taking the limit of $\mu\to1$ at $R=r_-$ \cite{Janis:1968zz}.

%$r_-$ becomes zero for $\mu=1$, but writing $r(R)^2$ as 
%\begin{align}
%    r(R)^2&= \l[R+\frac{r_0}{2}(\mu+1)\r]^{1+\frac{1}{\mu}}\l[R-\frac{r_0}{2}(\mu-1)\r]^{1-\frac{1}{\mu}}\nn\\
%    &\equiv \l[R+\frac{r_0}{2}(\mu+1)\r]^{1+\frac{1}{\mu}}J(R,\mu),
%\end{align}
%the function $J(R,\mu)$ is undefined at $R=0$ for $\mu\to1$.
%%%%%%%%%%%%%%%%%%%%
\subsubsection*{\rm{(iii)} $\mu>1$ and $\epsilon=+1$: a closed universe}
This case corresponds to the domain $-r_+<R<r_-$, and the configuration is given by
\begin{align}\label{metric_JNW_in}
    ds^2=-\l[\frac{r_++R}{r_--R} \r]^{\frac{1}{\mu}}dR^2+\l[\frac{r_--R}{r_++R}\r]^{\frac{1}{\mu}}dt_\infty^2+r(R)^2d\Omega^2,
\end{align}
with 
\begin{align}\label{r_phi_in}
    r(R)^2&=(r_++R)^{1+\frac{1}{\mu }}
        (r_--R)^{1-\frac{1}{\mu}},\nn\\
        \phi_I(R)&=\frac{\MA_I}{\mu}\log \frac{r_--R}{r_++R}+\phi_I^{(0)}.
\end{align}

\begin{figure}[h]
  \centering
 \includegraphics[scale=0.32]{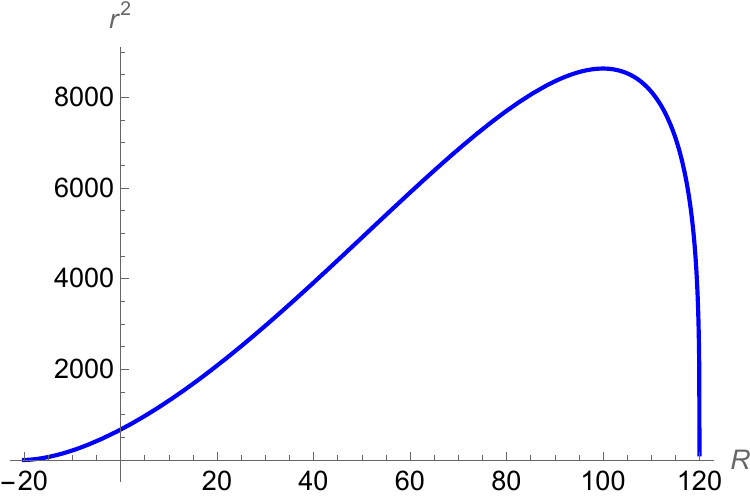}
 \includegraphics[scale=0.32]{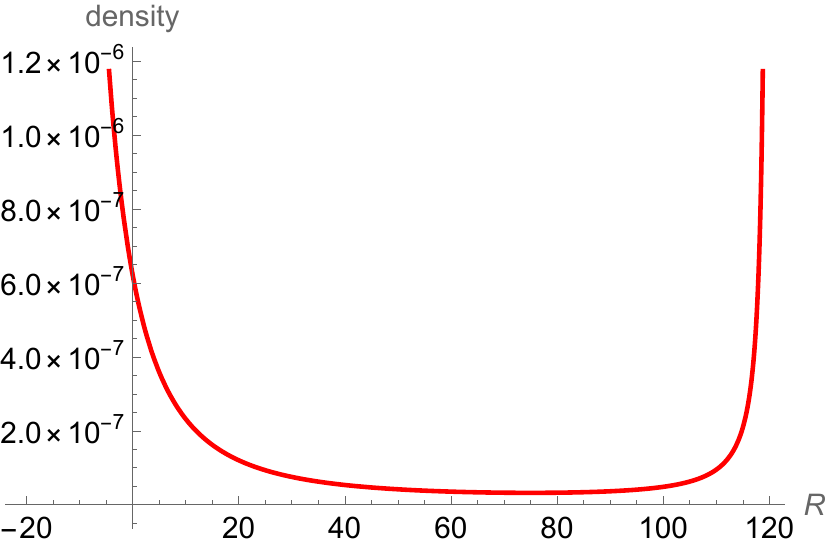}
\caption{For the case of $\epsilon=+1$, the areal radius $r(R)^2$ (Left) and energy density $-T^R{}_R(R)$ (Right). $\mu=1.4$ and $r_0=100$. Here, $R$ plays a role of the time coordinate.}
\label{f:JNW_in} 
\end{figure}

We examine the meaning of this configuration. In the metric \eqref{metric_JNW_in}, $R$ is timelike and $t_\infty$ spacelike, and it should be natural to interpret $-T^R{}_R$ as the energy-density $\rho$ and $T^t_t$ as the radial pressure. From \eqref{TTTT}, then $\rho=p$ holds, where $p=T^t{}_t=T^\theta{}_\theta=T^\phi{}_\phi$. Therefore, this is a causal-limit isotropic fluid \cite{Zeldovich:1961sbr}. %while the configuration is not homogeneous but spherically symmetric. 
Next, we see from \eqref{r_phi_in} that $r(R)^2$ vanishes at $r=-r_+$ and $r=r_-$ as shown in Left of Fig.\ref{f:JNW_in}. We also check from \eqref{TTTT} and \eqref{E_density} that the energy-density $\rho=-T^R{}_R|_{\epsilon=1}$ diverges at $r=-r_+$ and $r=r_-$ as shown in Right of Fig.\ref{f:JNW_in}.  
Furthermore, the spacetime volume is finite: 
\begin{align}\label{V4}
    \mathcal{V}_{\rm 4D}&=\int d^4x \sqrt{-g}\nn\\
    &=4\pi \Delta t_\infty \int^{r_-}_{-r_+}dR~r(R)^2\nn\\
    %&=\Delta t_\infty \frac{2\pi^2}{3}r_0^3\frac{\mu^2-1}{\sin\frac{\pi}{\mu}}\nn\\
    %&=\frac{2\pi^2}{3}l_p^2 r_0^2 \tPi_1 \frac{\mu^2-1}{\sin\frac{\pi}{\mu}}\nn\\
    &=\frac{\pi^2l_p^6}{6l_0^2}\tPi_1^3e^{-4\l(\frac{1}{\mu}\tB^0+\tB^1\r)}\frac{\mu^2-1}{\sin\frac{\pi}{\mu}},
\end{align}
where $\Delta t_\infty=\frac{l_0l_s\tPi_1}{\bar c r_0}=\frac{l_p^2\tPi_1}{r_0}$ and the definitions \eqref{defs} have been used. Note that it is still non-zero after taking $\mu\to1$, $\lim_{\mu\to1}\mathcal{V}_{\rm 4D}=\frac{\pi l_p^6}{3 l_0^2}\tPi_1^3 e^{-4\l(\frac{1}{\mu}\tB^0+\tB^1\r)}$, but the meaning is unclear since the finiteness comes from the singular behavior for $\mu\to1$ in the last factor of \eqref{V4}.

Thus, this configuration represents a Kantowski-Sachs-type closed universe for a causal-limit isotropic fluid and has the initial and final singularities at $R=-r_+,~r_-$ corresponding to a big bang and big crunch, respectively. 
Indeed, setting $n=1$ and using the scalar field $\phi_{I=1}$ as a ``clock" (instead of $R$), the metric \eqref{metric_JNW_in} agrees with the one obtained in Ref.\cite{Chiou:2008eg}. The new point we find is that the closed universe and the JNW spacetime can be unified into the single expression \eqref{metric_JNW}: for the domain $-r_+<R<r_-$, it represents the closed universe while for $r_-<R<\infty$ the JNW spacetime. In this sense, the closed universe corresponds to the ``interior" of the JNW metric, although the two spacetimes are disconnected at $R=r_-$ where the singularity exists in both sides.

Let us conclude this section by summarizing the results. For a coupled system of gravity and $n$ massless scalar fields, we have employed the different choice of canonical variables and lapse from the ones used in the previous cases, and established the $(2+n)$D Schrödinger symmetry. In the vacuum case, this leads to a partial evidence for the covariance of the symmetry in the mini-superspace. Then, we have obtained the generalized JNW solution and the closed universe solution as its ``interior". They are characterized by the mass parameter $r_0$ and the parameter $\mu$ associated with %related to 
the amplitude of the scalar fields. In terms of the canonical variables, the solutions are determined by the values of the generators of the symmetry:  $(\tPi_0,\tPi_1,\tB^0,\tB^1)$. Therefore, the symmetry transformation can change the values. We will discuss this point in the next section.

%%%%%%%%%%%%%%%%%%%%%%%%%%%%%%%%%%%%%%%%%%%%%%%%%%%%%%%%%%%%%%%%%%%
\section{Interpretation of the Symmetries under Hamiltonian Constraints}
\label{sec:interpretation}
In the previous sections, we have identified Schrödinger symmetries in the mini-superspace models. These systems differ from ordinary non-relativistic systems due to the presence of Hamiltonian constraints; in particular, 
they significantly complicate the interpretation of the symmetries. 

We first summarize what we have done so far. For the mini-superspace models, we have employed the canonical-transformation method and expressed the Hamiltonian constraints $H\approx0$ (for certain choices of $N$) as 
\begin{align}\label{H_again}
    H=\frac{1}{2}\MG^{ab}P_aP_b+\lambda\equiv H_0+\lambda 
\end{align}
to reach Schrödinger symmetries. Here, $\MG^{ab}$ is a constant invertible matrix, $\lambda$ is a constant characterizing the dynamics, and 
$P_a$ is used to denote $\tP_a$ or $\tPi_a$ for simplicity.  

%\textcolor{red}{The action corresponding to \eqref{H_again} is}
%\begin{align}\label{S_again}
%    S=\int d\tau\frac{1}{2}\MG_{ab}\frac{dX^a}{d\tau}\frac{dX^b}{d\tau}.
%\end{align}
%\textcolor{red}{However, in this action, the lapse function $N$ is fixed at the level of the action. Consequently, while the Hamiltonian equations are derived from its variation, they lack information regarding the constraints. The Noether charge we derived can be regarded as the generator of a symmetry that leaves this action—which ignores the constraints—invariant. Yet, since physical dynamics must properly take constraints into account, the physical significance of this symmetry is not trivial. Indeed, not all of these symmetries naively leave the action invariant without any gauge fixing.}

\begin{figure}[h]
  \centering
 \includegraphics[scale=0.28]{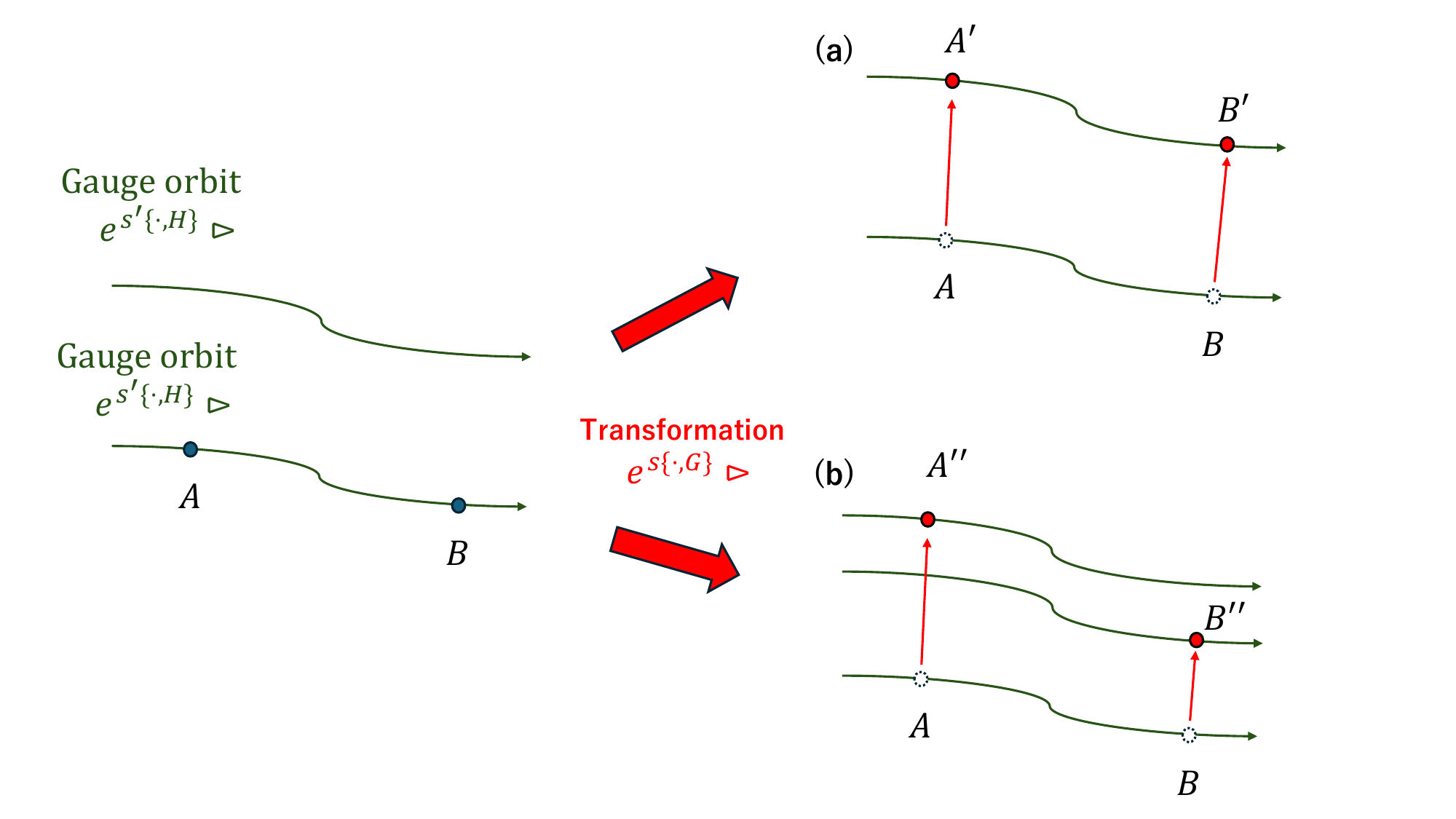}
\caption{Schematic representation of a gauge-independent transformation mapping one gauge orbit to another in the phase space. Two physically equivalent points, $A$ and $B$, lying on a single gauge orbit generated by $e^{s'\{\cdot,H\}}\triangleright$, are mapped by the transformation $e^{s\{\cdot,G\}}\triangleright$. (a) The mapped points $A'$ and $B'$ reside on the same gauge orbit, remaining physically equivalent after the transformation. Such a transformation is gauge-independent. (b) In contrast, the mapped points $A''$ and $B''$ land on distinct gauge orbits. Here, the originally equivalent points $A$ and $B$ are transformed into physically distinct points. In this case, the transformation by $G$ is gauge-dependent.}
\label{f:PointIII} 
\end{figure}

Now, before studying the physical meaning of the symmetries, we list the points we should keep in mind when considering systems with gauge constraints in general. Consider a symmetry generator $G$ satisfying the condition \eqref{symmetry_cond}. Then, the following questions arise.
\begin{itemize}
\item[(I)] When a state satisfying the constraints is transformed by $G$, does the transformed state satisfy the constraints again?

\item[(II)] Does applying a transformation $G$ to a gauge-invariant observable $\MO$ after using the constraints yield the same result as using the constraints after transforming $\MO$ by $G$? (Here, $\MO$ commutes with the constraints.)

\item[(III)] 
States connected by gauge transformations are physically equivalent. When transformed by $G$, are they still physically equivalent?
\end{itemize}
Let us add a little more. For (I),  the physical states are restricted on a hypersurface in the phase space determined by the constraints, and some transformation $G$ 
may move such states outside the surface, rendering them no longer physical ones. For (II), using the constraints, one can construct different (but weakly equivalent) expressions of $\MO$. Applying the transformation $G$ to each generically can yield different results (even if %$H=0 is$ 
the constraints are used afterward).
For example, in the case of the vacuum model (Sec.\ref{sec:Schwarzschild}), consider the transformation of 
the ADM mass $M$ by $G=\tB^-$. The definition \eqref{M_def} contains $\tP_-$, and the mass changes by the transformation. However, if we use the constraint \eqref{H_g5} and express $\tP_-$ in $\tP_+$ first, then the mass does not change. For (III), physically equivalent states are represented by a single gauge orbit. If a transformation $G$ moves a gauge orbit to another one, that transformation is gauge-independent. In other words, the transformation of a gauge-invariant observable $\MO$ by such a $G$ is gauge-invariant. See Fig.\ref{f:PointIII}.

The points (I)--(III) are satisfied if a symmetry generator $G$ commutes with the gauge constraints, as discussed below. 
% in an example.\TS{(← I feel we should not depend on "examples" about this point.)\;\;\;} 
%\textcolor{red}{Specifically, if and only if the generator $G$ satisfies $\{G, H\} \approx 0$ (where $H$ denotes the constraint), the states remain on the hypersurface $H \approx 0$. In addition, for such a weakly commuting generator $G$, the transformation it generates, $e^{s\{\cdot,G\}}\triangleright$, preserves weak equalities: if $A \approx B$, then $e^{s\{\cdot,G\}}\triangleright A \approx e^{s\{\cdot,G\}}\triangleright B$. Furthermore, the finite transformation generated by $G$ weakly commutes with the gauge transformation generated by the constraint $H$, yielding $e^{s\{\cdot,G\}}\triangleright e^{t\{\cdot,H\}}\triangleright \approx e^{t\{\cdot,H\}}\triangleright e^{s\{\cdot,G\}}\triangleright$. Here, $e^{s\{\cdot,G\}}\triangleright$ and $e^{t\{\cdot,H\}}\triangleright$ are the finite transformations by $G$ and $H$, respectively. (These operators are defined in \eqref{fin_trans}.) We will see this for an example.}
%\textcolor{red}{Indeed, the transformation induced by such a generator maps the solutions into the ones for the original theory.} \textcolor{red}{In other words, such a transformation} 
% the transformation induced by such a generator 
%does not alter the theory; that is, it acts as a symmetry transformation that leaves the action (without fixing $N$) invariant (see Appendix \ref{A:Action} for details).
Indeed, the transformation induced by such generators maps a solution to another in the original theory (dynamical symmetry) and does not change the theory itself. (In the action level, it acts as a symmetry transformation that leaves the action (without fixing $N$) invariant. See Appendix \ref{A:Action}.)  
%In other words, such a transformation} 
% the transformation induced by such a generator 
%does not alter the theory; that is, it acts as a symmetry transformation that leaves the action (without fixing $N$) invariant (see Appendix \ref{A:Action} for details).
On the other hand, for symmetry generators $G$ that do not commute with the gauge constraints, the three points require careful attention. This is the main topic we will discuss below.

Let us focus on our system \eqref{H_again} again. The three symmetry generators $B^a$, $Q_0$, and $Q_-$ do not commute with the Hamiltonian constraint. The interpretation of these symmetries remains unclear, and there can be various possibilities. One interpretation of transformations by $Q_0$ and $Q_-$ involves incorporating the transformation of the lapse function $N$ \cite{Achour2019}. In contrast, to the best of our knowledge, no such interpretation exists for $B^a$. 

Here, we propose another possible interpretation, which can also be applied to $B^a$:
\textit{such non-commuting symmetries transform the theory itself, and the transformed configurations are physical solutions of this new theory}. First, we will develop a general argument for this view (Sec. \ref{sec:Noether}), and then explicitly demonstrate it with examples (Sec. \ref{sec:solution_trans} and Sec.\ref{sec:theory_trans}).

%%%%%%%%%%%%%%%%%%%%%%%%%%%%%%%%%%%%%%%%%
\subsection{Symmetry generating a new theory}\label{sec:Noether}
%We argue 
We explain an interpretation that the symmetries by non-commuting generators $G$ with the Hamiltonian constraint $H$ transform the theory to a new theory. 

We start with reviewing Noether's theorem \textit{without} considering $H\approx 0$. 
We first check the role of the symmetry condition \eqref{symmetry_cond}. Under an infinitesimal canonical transformation generated by a function $G(x, p, t)$, the transformed canonical variables $(x', p')$ and the transformed Hamiltonian $H'(x', p', t)$ are given by
\begin{subequations}\label{XPcon}
\begin{gather}
x'^i=x^i+\epsilon\{x^i,G\},\\
p'_i=p_i+\epsilon\{p_i,G\},\\
H'(x', p', t)=H(x,p)+\epsilon\frac{\partial G}{\partial t},
\end{gather}
\end{subequations}
where $H(x,p)$ is the original Hamiltonian and $\epsilon$ is a small parameter.
If $G$ satisfies the symmetry condition \eqref{symmetry_cond}, it follows from \eqref{XPcon} that 
\begin{align}\label{KH}
H'(x,p)=H(x,p). 
\end{align}
That is, the Hamiltonian preserves its functional form. Therefore, symmetry transformations leave the theory invariant.
In particular, symmetry transformations can be interpreted as transformations of the initial conditions of the solution trajectory.

%We next 
We see this explicitly for  finite canonical transformations of $X^a$ by a generator $G$ of the Schrödinger symmetry in the Hamiltonian \eqref{H_again}. 
Noting $X^a=B^a+\MG^{ab}P_b \tau$ from \eqref{b_free}, %\eqref{solution_free}, 
it can be calculated as 
\begin{align}\label{X'}
    X'^a&=e^{s\{\cdot,G\}}\triangleright X^a\nn\\
    &=e^{s\{\cdot,G\}}\triangleright B^a+\MG^{ab}(e^{s\{\cdot,G\}}\triangleright P_b)\tau,
\end{align}
where a canonical transformation of a function $f$ by a generator $G$ for a finite parameter $s$ is denoted as 
\begin{align}\label{fin_trans}
    e^{s\{\cdot,G\}}\triangleright f := f+s\{f,G\}+\frac{1}{2!}s^2 \{\{f,G\},G \} \nn\\
    +\frac{1}{3!}s^3 \{\{\{f,G\},G \},G\}+ \cdots.
\end{align}
Here, $e^{s\{\cdot,G\}}\triangleright B^a$ and $e^{s\{\cdot,G\}}\triangleright P_b$ are expressed only in terms of the generators of the Schrödinger symmetry, since the symmetry algebra is closed. Note also that the generators can be built from $P_a$ and $B^a$ (see Sec.\ref{sec:particle}).

%Now, for 
For a solution to the canonical equation, we have a uniform linear motion in the configuration space $\{X^a\}$ (as seen in \eqref{solution_free}):
\begin{align}\label{X_sol}
    X^a(\tau)=\beta^a+\MG^{ab}\alpha_b \tau,
\end{align}
which corresponds to replacing the generators $(P_a,B^a)$ with integration constants $(\alpha_a,\beta^a)$ as in Sec.\ref{sec:particle}. Then, the transformed trajectory is given from \eqref{X'} by 
\begin{align}\label{X'_tau}
     X'(\tau)&=(e^{s\{\cdot,G\}}\triangleright B^a)|_{P_a\to\alpha_a,B^a\to\beta^a}\nn\\
    &~~~~~~~~~+\MG^{ab}(e^{s\{\cdot,G\}}\triangleright P_b)|_{P_a\to\alpha_a,B^a\to\beta^a}\tau\nn\\
    &\equiv \beta'^a + \MG^{ab}\alpha'_a \tau,
\end{align}
where $(\alpha'_a,\beta'^a)$ are another constants according to the argument above. 
%\TS{($\alpha'_a\equiv(e^{s\{\cdot,G\}}\triangleright P_b)|_{P_a\to\alpha_a,B^a\to\beta^a}$,$\beta'^a\equiv(e^{s\{\cdot,G\}}\triangleright B^a)|_{P_a\to\alpha_a,B^a\to\beta^a}$)}
Thus, a uniform linear motion (a solution to the canonical equation) $X^a(\tau)$ is mapped to another one $X'^a(\tau)$ by the symmetry transformations. This is the meaning of the dynamical symmetry in the cases without a constraint $H\approx 0$.

Now, let us consider the effect of $H\approx 0$. Suppose that the original trajectory $X^a(\tau)$ \eqref{X_sol} satisfies it: 
\begin{align}\label{H_al}
    H|_{P_a\to\alpha_a}=\frac{1}{2}\MG^{ab}\alpha_a \alpha_b +\lambda=0.
\end{align}
Then, $X^a(\tau)$ presents a physical solution of the theory with the Hamiltonian \eqref{H_again}. Noting that $\MG^{ab}$ is Lorentzian, the constraint \eqref{H_al} can be interpreted as a ``dispersion relation" between ``energy" $\alpha_0$ and ``momentum" $\alpha_{\rm other}$ for the characteristic ``mass" $\lambda$. 

Due to this constraint, all the components of $\alpha_a$ are not independent for physical states. On the other hand, under the gauge transformation generated by the Hamiltonian \eqref{H_again}, $\alpha_a$ remains invariant, while $\beta^a$ is gauge-dependent and carries unphysical degrees of freedom. This clear separation into $\alpha_a$ and $\beta^a$ is a key advantage of working with the Hamiltonian of the form \eqref{H_again}.

Next, let us examine whether the transformed trajectory $X'^a(\tau)$ \eqref{X'_tau} satisfies the Hamiltonian constraint, $H|_{P_a\to\alpha_a'}=0$, or not. %How about $X'^a(\tau)$ \eqref{X'}? Does $H|_{P_a\to\alpha_a'}=0$ hold? To examine this, first 
First, we can check, by explicit calculation, 
\begin{align}\label{H'_P'}
    e^{s\{\cdot,G\}}\triangleright H_0=\frac{1}{2}\MG^{ab}(e^{s\{\cdot,G\}}\triangleright P_a)(e^{s\{\cdot,G\}}\triangleright P_b). 
\end{align}
Noting $(e^{s\{\cdot,G\}}\triangleright P_a)|_{P_a\to \alpha_a, B^a\to \beta^a}\equiv\alpha'_a$, therefore, in order to see $H|_{P_a\to\alpha_a'}=0$, we just have to show $(e^{s\{\cdot,G\}}\triangleright H)|_{P_a\to \alpha_a, B^a\to \beta^a}=0$. 
Here, the expression
\begin{align}
    e^{s\{\cdot,G\}}\triangleright H=\lambda + H_0 + s \{H_0,G\}+\frac{1}{2}s^2 \{\{H_0,G\},G\}+\cdots.
\end{align}
means from \eqref{H'_P'} that
\begin{align}\label{H_case1}
    \{H_0,G\}=0~{\rm and}~ H|_{P_a\to\alpha_a}=0 \Longrightarrow H|_{P_a\to\alpha'_a}=0.
\end{align}
For this case, therefore, the transformed solution $X'(\tau)$ is a new physical solution within the original theory. (We will see this explicitly in Sec.\ref{sec:solution_trans}.) This is the reason why point (I) in the last subsection is satisfied for symmetry generators that commute with $H$. 

Similarly, when $\{G,H_0\}=0$, points (II) and (III) hold. For point (II), since $G$ doest not change $H$, even if we rewrite a quantity $X$ using $H=0$ into another expression before applying the transformation $G$, the result will match that obtained by applying $G$ to the original expression of $X$. 
For point (III), $G$ and $H$ commute, and so do $e^{s\{\cdot,G\}}\triangleright$ and $e^{s'\{\cdot,H\}}\triangleright$, meaning that a gauge orbit is mapped to anothoer one by $e^{s\{\cdot,G\}}\triangleright$. 

However, if $\{H_0,G\}\neq0$, the transformed trajectory $X'(\tau)$  does not satisfy $H|_{P_a\to \alpha'_a}=0$ and thus does not become a physical solution to the original theory with $H$. To interpret such a transformation as a kind of symmetry, we consider the following. For a physical solution with $(\alpha_a, \beta^a)$ to the original theory with $H$ (that is, $H|_{P_a\to\alpha_a}=0$), we use a generator $G$ of Schrödinger symmetry and introduce a new characteristic constant by\footnote{Here, the value of $s$ is chosen in a specific manner. See below for examples.} 
\begin{align}\label{lam_new}
    \lambda_{\rm New}:= \lambda+(H_0-e^{s\{\cdot,G\}}\triangleright H_0)|_{P_a\to\alpha_a,B^a\to\beta^a},
\end{align}
to define a new Hamiltonian or new dispersion relation: 
\begin{equation}\label{H_New}
    H_{\rm New}:= H_0+\lambda_{\rm New}.
\end{equation}
Then, for a trajectory $X'(\tau)$ generated by $G$, we can use the property \eqref{H'_P'} and calculate 
\begin{align}
    H_{\rm New}|_{P_a\to\alpha'_a,B^a\to\beta'^a}&=\lambda_{\rm New}+ (e^{s\{\cdot,G\}}\triangleright H_0)|_{P_a\to\alpha_a,B^a\to\beta^a}\nn\\
    &= \lambda+H_0|_{P_a\to\alpha_a}\nn\\
    &=0.
\label{Hlamc}
\end{align}
This shows that the transformed trajectory $X'(\tau)$ is a physical solution to the $new$ theory $H_{\rm New}$.\footnote{\label{foot:theory} A theory is specified by a Hamiltonian constraint $H\approx 0$ and canonical-transformation relations $\tX^a(X,P), \tP_a(X,P)$, which enable us to translate the uniform linear motion in $\{\tX^a\}$ into the spacetime metric $g_{\mu\nu}$ in terms of $X^a$. Here, the new theory has a different constraint but the same relations.} Note that the new Hamiltonian also holds the Schrödinger symmetry, since $\lambda_{\rm new}$ is constant and the analysis we have done so far is applicable to $H_{\rm New}$. Thus, this procedure generates a new theory preserving the Schrödinger symmetry and transforms a physical solution of the original theory to a physical solution of the new theory.\footnote{A similar idea—symmetry that generates new theories—is studied in other approaches \cite{BenAchour:2020xif}.} (We will demonstrate this explicitly in Sec.\ref{sec:theory_trans}.) 

%From a perspective of this symmetry transformation that 
Adopting this symmetry transformation rule, which
changes $\alpha_a$, $\beta^a$, and $\lambda$ simultaneously, let us consider points (I), (II) and (III). 
Point (I) is achieved by adopting the idea of moving the constraint surface itself while preserving the way of transformation of a point in the phase space (as in \eqref{Hlamc}). 
% For Point (II), $H=H_0+\lambda$ remains invariant according to this generalized transformation, and no ambiguity appears in transforming a quantity $X$. 

For Point (II), since the generalized transformation keeps the Hamiltonian constraint as in Eq.\eqref{Hlamc}, different but weakly-equivalent expressions of $\MO$ transforms by $G$ in the same manner.  
%the value of the Hamiltonian $H=H_0+\lambda$ remains invariant under this generalized transformation; that is, $H|_{P_a\to\alpha_a}=H|_{P_a\to\alpha'_a,\lambda\to\lambda_{\rm New}}$. This implies that the weak equality $H\approx0$ is preserved under this generalized transformation. Consequently, the transformation of a physical quantity $\MO$ becomes independent of the different representations made by the constraint $H\approx0$. 
%To see this briefly, let us consider an example in the vacuum case (Sec.\ref{sec:Schwarzschild}). For the AdS-Schwarzschild solution characterized by a set $(\alpha_+,\alpha_-;\beta^+,\beta^-)$ satisfying $H_g|_{\tP_a\to\alpha_a}=0$, that is, from Eq.\eqref{H_g5}, 
%\begin{align}\label{Hconst_alpha}
%-\frac{2}{\epsilon cl_p}\alpha_+\alpha_-+\lambda=0,
%\end{align}
For example, consider the (A)dS-Schwarzschild solution characterized by a set $(\alpha_+,\alpha_-;\beta^+,\beta^-)$ satisfying $H_g|_{\tP_a\to\alpha_a}=0$, that is, $-\frac{2}{\epsilon cl_p}\alpha_+\alpha_-+\lambda=0$, where $\lambda=-\frac{cl_p}{2l_s^2}$ from Eq.\eqref{H_g5}. By using this, the mass $M$ \eqref{M_def} can be expressed in various ways, say,  
\begin{align}%\label{M_def_1}
M&= \frac{\epsilon l_s}{G \bar c^2}\alpha_+(\alpha_+ \beta^+-\alpha_-\beta^-)\equiv M_1\nonumber\\
&=\frac{\epsilon l_s}{G \bar c^2}(\alpha_+^2 \beta^+-\frac{\epsilon cl_p}{2}\lambda\beta^-)\equiv M_2.\nn
\end{align}
Then, apply the transformation $e^{s\{\cdot,\tB^-\}}\triangleright$ (for a finite value $s$) to $M_1$, $M_2$, \textit{and} $H_g$. According to \eqref{lam_new}, $\alpha_-$ and $\lambda$ transform as
\begin{align}%\label{Trans_rules}
\alpha_-\rightarrow\alpha_-'=\alpha_--s,~~
\lambda \rightarrow\lambda_{\rm{New}}=\lambda-\frac{2s}{\epsilon c l_p}\alpha_+,\nn
\end{align} 
and $\alpha_+, \beta^+,\beta^-$ do not change. 
These satisfy %which, of course,  satisfies 
$H_g^{\rm New}|_{\tP_a\to \alpha'_a}=0$ as in Eq.\eqref{Hlamc}. This leads to 
\begin{align}
M_1\rightarrow M_1'&=\frac{\epsilon l_s}{G \bar c^2}\alpha_+\left\{\alpha_+ \beta^+-(\alpha_--s)\beta^-\right\}\nonumber\\
&=M_1+\frac{\epsilon l_s}{G \bar c^2}\alpha_+\beta^-s,\nonumber\\
M_2 \rightarrow M_2'&=\frac{\epsilon l_s}{G \bar c^2}\left\{\alpha_+^2 \beta^+-\frac{\epsilon cl_p}{2}\left(\lambda-\frac{2s}{\epsilon c l_p}\alpha_+\right)\beta^-\right\}\nonumber\\&=M_2+\frac{\epsilon l_s}{G \bar c^2}\alpha_+\beta^-s.\nonumber
\end{align}
Therefore, $M_1'=M_2'$ holds, and the way of the transformation does not depend on the expression of $M$. Note that if $\lambda$ were not transformed simultaneously, we would have $M'_1\neq M'_2$.

We finally discuss point (III). We here have to be careful with transformation of not only a gauge-invariant observable $\MO$ but also $\lambda$, since $\lambda$ transforms as in Eq.\eqref{lam_new} and determines the theory. From Eq.\eqref{lam_new} and the fact that $H$ does not involve $B^a$, the transformation of $\lambda$ induced by generators that depend at most on linear forms of $B^a$ is gauge independent. Actually, for a function $F(P)$ and a generator $G$, 
\begin{align}
   \{F(P),G\}|_{P_a\to \alpha_a, B^a\to \beta^a}=\left(-\frac{\partial F}{\partial P_a}\frac{\partial G}{\partial B^a}\right)|_{P_a\to \alpha_a, B^a\to \beta^a}
\end{align}
is independent of $\beta^a$ if $\frac{\partial G}{\partial B^a}$ does not include $B^a$, and inductively, terms of any order in $s$ in \eqref{lam_new} become independent of $\beta^a$. Note that one of the Schrödinger symmetry generators, $Q_-$, contains a quadratic term in $B^a$, and the resulting new theory is gauge-dependent.\footnote{\label{foot:gaugedependence}The presence of $\beta^a$ does not necessarily imply gauge dependence. There exist physical quantities that involve $\beta^a$ but are nonetheless gauge-independent. For example, $J_{ab}$.}

On the other hand, the transformation of gauge-invariant observables $\MO$ can be gauge dependent. For instance, the transformation of $\MO=J_{ab}$ by $B^a$ is gauge dependent, since their commutation relation is linear on $B^a$. (This corresponds to the case (b) in Fig.\ref{f:PointIII}.) Once we determine the value of $\beta^a$, the transformation is uniquely determined. There is room for discussion regarding the role of the gauge in this view of symmetry.

Thus, the above perspective can be considered as one interpretation of Noether's theorem in a theory with the Hamiltonian constraint $H\approx 0$ and symmetry generators $G$ that do not commute with $H$. 

It is worth noting that transformations commuting with $H$ can be regarded, from this perspective, as a special case where $\lambda$ remains unchanged. Furthermore, in this interpretation, all symmetry transformations satisfying the condition \eqref{symmetry_cond} preserve the property \eqref{KH} of leaving the time evolution operator $H_0$ invariant for certain choices of $N$.\footnote{\label{foot:action and H} %\textcolor{red}{The invariance of the time evolution operator under a specific choice of lapse corresponds to leaving the action with gauge fixing \eqref{S_again_fixing} invariant. See Appendix \ref{A:Action} for details.} 
The invariance of $H_0$ for a specific choice of $N$ corresponds to leaving the gauge-fixed action \eqref{S_again_fixing} invariant. See Appendix \ref{A:Action}.} %under this choice of $N$. 
Therefore, this interpretation of the $\lambda$ transformation can be viewed as a generalization of the standard dynamical symmetry. 
%as an extension of the symmetry perspective—where initial conditions are changed without altering the time-evolution operator $H_0$—to the transformation of $\lambda$.

%%%%%%%%%%%%%%%%%%%%%%%%%%%%%%%%%%%%%%%%%
\subsection{$\{G,H\}=0$: Solution transformations}\label{sec:solution_trans}
We use the mini-superspace models discussed so far and demonstrate that generators $G$ satisfying %with 
$\{G,H\}=0$ map a physical solution to another in the same theory with $H$, where both satisfy $H\approx 0$. Among the generators of Schrödinger symmetry, $\tP_a$ and $\tJ_{ab}$ are of this type. We will pick up particularly interesting cases. 

%%%%%%%%%%%%%%%%%%%%%
\subsubsection*{Ex.1: $\widetilde J_{-A}$ in the Maxwell-field mini-superspace}
We consider the Maxwell-field mini-superspace with the Hamiltonian constraint \eqref{H_gE5},
\begin{align}
    H_{gE}&=-\frac{2}{\e c l_p}\tP_+ \tP_- + \frac{1}{2cl_p}\tP_A^2 -\frac{cl_p}{2l_s^2},\nn
\end{align}
and study transformation of a (A)dS-RN solution \eqref{metric_RN} by  $\widetilde J_{-A}$. In particular, we %let us 
map it to a (A)dS Schwarzschild solution \eqref{metric_SchL} with a different value of $M$. 

We start by using the 3D Schrödinger algebra \eqref{3D_Sc_alg} and calculating the finite canonical transformation %of interest 
by $\widetilde J_{-A}=\widetilde\MG_{-+}^E\tB^+\tP_A-\widetilde\MG_{AA}^E\tB^A\tP_-=\frac{(-\epsilon)c l_p}{2}\tB^+\tP_A-cl_p \tB^A \tP_-$: 
\begin{align}\label{trans_A}
    \tP'_+&=e^{s\{\cdot,\widetilde J_{-A}\}}\triangleright \tP_+=\tP_+ +\frac{\epsilon cl_ps}{2}\tP_A +\frac{\epsilon c^2l_p^2s^2}{4}\tP_-,\nn\\
    \tP'_-&=e^{s\{\cdot,\widetilde J_{-A}\}}\triangleright \tP_-=\tP_-,\nn\\
    \tP'_A&=e^{s\{\cdot,\widetilde J_{-A}\}}\triangleright \tP_A=\tP_A+ cl_ps \tP_-,\nn\\
    \tB'^+&=e^{s\{\cdot,\widetilde J_{-A}\}}\triangleright \tB^+=\tB^+,\nn\\
    \tB'^-&=e^{s\{\cdot,\widetilde J_{-A}\}}\triangleright \tB^-=\tB^--cl_ps\tB^A+\frac{\epsilon c^2l_p^2s^2}{4}\tB^+,\nn\\
    \tB'^A&=e^{s\{\cdot,\widetilde J_{-A}\}}\triangleright \tB^A=\tB^A-\frac{\epsilon cl_ps}{2}\tB^+. 
\end{align}

Suppose that the original (A)dS-RN solution is characterized by a set of the integration constants $(\alpha_+,\alpha_-,\alpha_A;\beta^+,\beta^-,\beta^A)$ satisfying the Hamiltonian constraint \eqref{H_gE5}: $H_{gE}|_{\tP_a\to\alpha_a}=0$. 

Now, we set the transformation parameter $s$ such that the electric charge \eqref{Q_def} vanishes: $\MQ'=\sqrt{\frac{\pi}{G}}\frac{l_s}{\bar c}\alpha'_A=0$, where $\alpha'_A=P'_A|_{\tP_a\to\alpha_a,\tB^a\to\beta^a,s\to s_*}=0$ with
\begin{equation}\label{s_A}
    s=-\frac{1}{cl_p}\frac{\alpha_A}{\alpha_-}\equiv s_*.
\end{equation}
Then, the other charges transform from \eqref{trans_A} as follows:
\begin{align}\label{trans_A_sol}
    \alpha'_+&=\tP'_+|_{\tP_a\to\alpha_a,\tB^a\to\beta^a,s\to s_*}=\alpha_++\frac{(-\epsilon)\alpha_A^2}{4\alpha_-},\nn\\
    \alpha'_-&=\tP'_-|_{\tP_a\to\alpha_a,\tB^a\to\beta^a,s\to s_*}=\alpha_-,\nn\\
    \beta'^+&=\tB'^+|_{\tP_a\to\alpha_a,\tB^a\to\beta^a,s\to s_*}=\beta^+,\nn\\
    \beta'^-&=\tB'^-|_{\tP_a\to\alpha_a,\tB^a\to\beta^a,s\to s_*}=\beta^-+\frac{\alpha_A}{\alpha_-}\beta^A+\frac{\epsilon\alpha_A^2}{4\alpha_-^2}\beta^+,\nn\\
    \beta'^A&=\tB'^A|_{\tP_a\to\alpha_a,\tB^a\to\beta^a,s\to s_*}=\beta^A+\frac{\epsilon \alpha_A}{2\alpha_-}\beta^+.
\end{align}
According to these, the value of the mass \eqref{M_def} becomes 
%These change the value of mass $M$ \eqref{M_def}: 
$M'=\frac{\epsilon l_s}{G\bar c^2}\alpha'_+(\alpha'_+\beta'^+-\alpha_-'\beta'^-)$.
On the other hand, the new charges still satisfy the original Hamiltonian constraint: 
\begin{align}
    H_{gE}|_{\tP_a\to\alpha'_a}&=
    -\frac{2}{\e c l_p}\alpha'_+ \alpha'_- + \frac{1}{2cl_p}\alpha_A^{'2} -\frac{cl_p}{2l_s^2}\nn\\
    &= -\frac{2}{\e c l_p}\l(\alpha_++\frac{(-\epsilon)\alpha_A^2}{4\alpha_-}\r)\alpha_- -\frac{cl_p}{2l_s^2}\nn\\
    &=H_{gE}|_{\tP_a\to \alpha_a}=0.
\end{align}
Therefore, we reach the AdS-Schwarzschild solution with mass $M'$. This shows that the symmetry transformation by $\widetilde J_{A-}$ for $s=s_*$ maps the (A)dS-RN solution with $(M,\MQ)$ to the (A)dS-Schwarzschild solution with $M'$ in the same Hamiltonian constraint $H_{gE}\approx0$.

%%%%%%%%%%%%%%%%%%
\subsubsection*{Ex.2: $\tJ_{1\phi}$ in the scalar-field mini-superspace model.}
We next provide an example in the scalar-field mini-superspace model \eqref{H_phi3}. For simplicity, we consider the case of $n=1$ here: 
\begin{align}\label{H_phi_single}
    H_{g\phi}&=\frac{1}{2(-\e) cl_p}\l(-\tPi_0^2+\tPi_1^2+\tPi_\phi^2\r).    
\end{align}
Let us apply the symmetry transformation by $\tJ_{1\phi}$ and map the JNW solution (and its interior) to the Schwarzschild solution. 

First, the finite canonical transformation of the relevant generators is given by, through the Schrödinger algebra associated with \eqref{scalar_sch_sym},
\begin{align}\label{trans_scalar}
    \tPi'_1&=e^{s\{\cdot,\widetilde J_{1\phi}\}}\triangleright \tPi_1=\tPi_1\cos{\Theta}-\tPi_\phi \sin{\Theta},\nn\\
    \tPi'_\phi&=e^{s\{\cdot,\widetilde J_{1\phi}\}}\triangleright \tPi_\phi=\tPi_\phi\cos{\Theta}+\tPi_1 \sin{\Theta},\nn\\
    \tB'^1&=e^{s\{\cdot,\widetilde J_{1\phi}\}}\triangleright \tB^1=\tB^1\cos{\Theta}+\tB^\phi \sin{\Theta},\nn\\
    \tB'^\phi&=e^{s\{\cdot,\widetilde J_{1\phi}\}}\triangleright \tB^\phi=\tB^\phi\cos{\Theta}-\tB^1 \sin{\Theta},
\end{align}
where $\Theta\equiv (-\epsilon)cl_p s$. 

We assume that the original solution metric \eqref{metric_JNW} with the condition \eqref{mu_JNW} is specified by a set of the constants $(\alpha_0,\alpha_1,\alpha_\phi;\beta^0,\beta^1,\beta^\phi)$, where the Hamiltonian constraint \eqref{H_phi_single} holds: $H_{g\phi}|_{\tPi_a\to\alpha_a}=0$. 

We then choose the parameter $s=\frac{\Theta}{(-\epsilon)cl_p}$ such that $\alpha'_\phi=\tPi'_\phi|_{\tPi_a\to\alpha_a,\tB^a\to\beta^a,s\to s_*}=0$. That is,  
\begin{align}\label{Theta_*}
    \Theta_*=(-\epsilon)cl_p s_*=\arctan\l({-\frac{\alpha_\phi}{\alpha_1}}\r).
\end{align}
Applying this to $\alpha'_a=\tPi'_a|_{\tPi_a\to\alpha_a,\tB^a\to \beta^a,s\to s_*}$ and using $\cos^2{\Theta_*}=\frac{1}{1+\alpha_\phi^2/\alpha_1^2}$, we can calculate
\begin{align}
    &-\alpha'^2_0+\alpha'^2_1+\alpha'^2_\phi\nn\\
    =&-\alpha_0^2+ \alpha_1^2 \cos^2{\Theta_*}+\alpha_\phi^2 \sin^2{\Theta_*}-2\alpha_1 \alpha_\phi \cos{\Theta_*}\sin{\Theta_*}\nn\\
    =&\cos^2{\Theta_*}\l[-\frac{\alpha_0^2}{\cos^2{\Theta_*}}+\alpha_1^2+\alpha_\phi^2\tan^2{\Theta_*}-2\alpha_1\alpha_\phi \tan{\Theta_*}\r]\nn\\
    =&\cos^2{\Theta_*}\l[-\alpha_0^2\l(1+\frac{\alpha_\phi^2}{\alpha_1^2}\r)+\alpha_1^2+\alpha_\phi^2\frac{\alpha_\phi^2}{\alpha_1^2}-2\alpha_1\alpha_\phi \frac{-\alpha_\phi}{\alpha_1}\r]\nn\\
    =&\cos^2{\Theta_*}\l[-\alpha_0^2+\alpha_1^2+\alpha_\phi^2+\l(\frac{\alpha_\phi}{\alpha_1}\r)^2(-\alpha_0^2+\alpha_\phi^2)+\alpha_\phi^2\r]\nn\\
    =&0,
\end{align}
where in the last step we have used $H_{g\phi}|_{\tPi_a\to\alpha_a}=0$. Therefore, the transformed configuration $Y'^a(\tau)=\beta'^a+\MG^{ab}_\phi\alpha'_b\tau$ satisfies the original Hamiltonian constraint. Here, the $\mu$ parameter \eqref{defs} changes as $\mu=\frac{\alpha_0}{\alpha_1}>1\to\frac{\alpha'_0}{\alpha'_1}=1$ because of $\alpha'_\phi=0$. Therefore, this is the Schwarzschild solution with a different value of mass \eqref{M_def2}, $M'=\frac{l_s \alpha'_1}{4G \bar c}e^{-2\l(\beta'^0+\beta'^1\r)}$.

%%%%%%%%%%%%%%%%%%%%%%%%%%%%%%%%%%%%%%%%
\subsection{$\{G,H\}\neq0$: Theory transformations}\label{sec:theory_trans}
Finally, we present two examples where a theory itself changes under transformations by a symmetry generator $G$ that does not commute with the Hamiltonian constraint $H$. $\tB^a$, $\widetilde{Q}_0$, and  $\widetilde{Q}_-$ belong to this type. We then discuss what the resulting theory is.%what kind of theory it is. 

%%%%%%%%%%%%
\subsubsection*{Ex.3: $\tB^+$ in the Maxwell-field mini-superspace model.}
We discuss the symmetry transformation by the Galilei boost $\tB^+$ in the Maxwell-field mini-superspace model \eqref{H_gE5}, where the constant $\lambda$ in \eqref{H_again} corresponds to $\frac{-cl_p}{2l_s^2}$. %First, 
Suppose again that the original (A)dS-RN solution is specified by a set $(\alpha_+,\alpha_-,\alpha_A; \beta^+, \beta^-, \beta^A)$ that satisfies $H_{gE}|_{\tP_a\to\alpha_a}=0$. We define the new parameter \eqref{lam_new} by 
\begin{align}\label{lam_new_E}
\lambda_{\rm New}:&= \frac{-cl_p}{2l_s^2}+(H_0-e^{s\{\cdot,\tB^+\}}\triangleright H_0)|_{\tP_a\to\alpha_a,\tB^a\to\beta^a,s\to s_*}\nn\\
&=\frac{-cl_p}{2l_s^2}+\frac{2\alpha_-}{(-\epsilon)cl_p}s_*\equiv -\frac{cl_p}{2l_s^2}(1+k_*),
\end{align}
for a certain value $s_*$, where we have employed the 3D Schrödinger algebra \eqref{3D_Sc_alg} and defined $k_*\equiv \frac{4\epsilon\alpha_-}{\bar c^2}s_*$. This gives the new dispersion relation:
\begin{align}\label{H_gE_New}
    H^{\rm New}_{gE}=\frac{2}{(-\e) c l_p}\tP_+ \tP_- + \frac{1}{2cl_p}\tP_A^2 -\frac{cl_p}{2l_s^2}(1+k_*).%+\frac{2\alpha_-}{(-\epsilon)cl_p}s_*.
\end{align}

At the same time, the configuration $X^a(\tau)=\beta^a+\MG^{ab}_E\alpha_b \tau$ is transformed 
to the new one $X'^a(\tau)=\beta'^a+\MG^{ab}_E\alpha'_b \tau$, where 
\begin{align}\label{alpha'_Ex3}
    \alpha'_+&=\l(e^{s\{\cdot,\tB^+\}\}}\triangleright \tP_+\r)_{\tP_a\to\alpha_a,\tB^a\to\beta^a,s\to s_*}=\alpha_+-s*\nn\\
    \alpha'_-&=\l(e^{s\{\cdot,\tB^+\}\}}\triangleright \tP_-\r)_{\tP_a\to\alpha_a,\tB^a\to\beta^a,s\to s_*}=\alpha_-\nn\\
    \alpha'_A&=\l(e^{s\{\cdot,\tB^+\}\}}\triangleright \tP_A\r)_{\tP_a\to\alpha_a,\tB^a\to\beta^a,s\to s_*}=\alpha_A\nn\\
    \beta'^a&=\l(e^{s\{\cdot,\tB^+\}\}}\triangleright \tB^a\r)_{\tP_a\to\alpha_a,\tB^a\to\beta^a,s\to s_*}=\beta^a. 
\end{align}
Then, $\alpha'_a$ satisfy not the original Hamiltonian constraint but the new one:
\begin{align}\label{H_E_New_constraint}
    &H^{\rm New}_{gE}|_{\tP_a\to\alpha'_a}\nn\\
    &=\frac{2}{(-\e) c l_p}(\alpha_+-s*)\alpha_- + \frac{1}{2cl_p}\alpha_A^2 -\frac{cl_p}{2l_s^2}+\frac{2\alpha_-}{(-\epsilon)cl_p}s_*\nn\\
    &=0,
\end{align}
because of $H_{gE}|_{\tP_a\to\alpha_a}=0$. Therefore, this symmetry transformation generates the new theory with $H_{gE}^{\rm New}$ and maps the solution of the original theory $H_{gE}$ into another solution of the new one $H_{gE}^{\rm New}$. 

Now, two questions arise: (i) What kind of spacetime is the obtained configuration? and (ii) What is the new theory $H_{gE}^{\rm New}$? 

We first note from footnote \ref{foot:theory} that the original relations $\tX^a(X,P),\tP_a(X,P)$ are kept in the new theory, and the same geometric interpretation holds except for the parts involving the Hamiltonian constraint. Then, the calculation for $g_{rr}$ \eqref{tautau_E} is modified as 
\begin{align}
    -\e N^2 d\tau^2
    &=\frac{1}{\frac{1}{\bar c^2} \l( -\frac{4}{\e} \alpha'_+ \alpha'_- +\alpha'^2_A \r)-\frac{2GM'}{r}+\frac{G\MQ^2}{4\pi r^2}-\frac{\Lambda}{3}r^2}dr^2\nn\\
    &\approx \frac{1}{1+k_*-\frac{2GM'}{r}+\frac{G\MQ^2}{4\pi r^2}-\frac{\Lambda}{3}r^2}dr^2.
\end{align}
Here, $H^{\rm New}_{gE}|_{\tP_a\to\alpha'_a}\approx 0$ is applied, and the value of the mass \eqref{M_def} changes from $M$ to $M'=\frac{\e l_s}{G \bar c^2}\alpha'_+(\alpha'_+ \beta^+-\alpha_- \beta^-)$ while the electric charge $\MQ$ doesn't. We can also have a similar modification for $g_{tt}$ (see \eqref{tt_E}). Therefore, we reach a different metric from the (A)dS-RN metric:
\begin{align}\label{metric_RN_k}
    ds^2\approx&-\l(1+k_*-\frac{2GM'}{r}+\frac{G\MQ^2}{4\pi r^2}-\frac{\Lambda}{3}r^2\r)dt_\infty^2\nn\\
    &+\l(1+k_*-\frac{2GM'}{r}+\frac{G\MQ^2}{4\pi r^2}-\frac{\Lambda}{3}r^2\r)^{-1}dr^2+r^2 d\Omega^2.
\end{align}

Even for the case of $\Lambda=0$, it is not asymptotically flat. A new term due to $k_*\neq0$ appears in the energy density and radial pressure: 
\begin{align}
    G^t{}_t&=G^r{}_r=-\Lambda-\frac{G\MQ^2}{4\pi r^4}+\frac{k_*}{r^2},\nn\\G^\theta{}_\theta&=G^\phi{}_\phi=-\Lambda+\frac{G\MQ^2}{4\pi r^4},\nn\\
    R&=4\Lambda-\frac{2k_*}{r^2}.
\end{align}
This is a solution to the new theory $H^{\rm New}_{gE}$.

Next, to identify what the new theory is, let us review how we have obtained it. We first note from \eqref{lam_new_E} that the shift $-\frac{cl_p}{2l_s^2}\to-\frac{cl_p}{2l_s^2}(1+k_*)$ corresponds to generating the new theory \eqref{H_gE_New}. We then remember that the term $-\frac{cl_p}{2l_s^2}$ in the original theory \eqref{H_gE5} comes from the second term, $\frac{cl_p}{l_s^2}$, in the reduced gravity Langrangian \eqref{L_g}. Therefore, we can see that a 2D covariant term \begin{align}\label{S_new_E}
    \Delta S^{\rm New}_{gE}&=\frac{1}{16\pi G\hbar}\int d^4x \sqrt{-g}\frac{8\pi k_*}{\MA}\nn\\
    &=\int d^2x \sqrt{-g_{2d}}\frac{k_*}{2l_p^2}
\end{align}
makes the shift effect. Here, $\MA=4\pi r^2$, and $\sqrt{-g_{2d}}$ is the area element for the 2D normal part to the spatial sphere. Thus, the new metric \eqref{metric_RN_k} solves the equation of motion for the Einstein-gravity coupled with Maxwell field plus the 4D cosmological constant $\Lambda$ and the \textit{2D cosmological constant} $\Lambda_{2d}\equiv-\frac{k_*}{4l_p^2}$.\footnote{A similar argument was made by a different approach \cite{BenAchour:2020xif}}. Although more studies on this new theory would be needed, at least, it keeps the 3D Schrödinger symmetry (as seen at the end of Sec.\ref{sec:Noether}). %, which should provide an interesting aspect of our interpretation that symmetries generates new theories. 
This theory-symmetry method would be one procedure for generating a gravity theory with Schrödinger symmetry.

%%%%%%%%%%%%
\subsubsection*{Ex.4: $\tB^\phi$ in the scalar-field mini-superspace model.}
We now consider the transformation by $\tB^\phi$ in the single scalar-field mini-superspace model \eqref{H_phi_single}, and discuss the new theory and configuration. 

In particular, for the original solution metric \eqref{metric_JNW} with constants $(\alpha_0,\alpha_1,\alpha_\phi;\beta^0,\beta^1,\beta^\phi)$ satisfying the Hamiltonian constraint $H_{g\phi}\approx 0$, we study the finite transformation by $\tB^\phi$ for $s=\alpha_\phi\equiv s_*$. For this parameter we evaluate $\lambda_{\rm New}$ \eqref{lam_new} and find 
\begin{align}\label{H_phi_new}
    H_{g\phi}^{\rm New}=\frac{1}{2(-\e) cl_p}\l(-\tPi_0^2+\tPi_1^2+\tPi_\phi^2\r)+\frac{1}{2(-\epsilon)cl_p}\alpha_\phi^2   
\end{align}
Because of $\lambda_{\rm New}={\rm const.}\neq 0$, this theory is not Einstein gravity plus a scalar field but keeps the 3D Schrödinger symmetry. 

By a similar calculation to \eqref{alpha'_Ex3}, we can check that $\alpha'_0=\alpha_0$, $\alpha'_1=\alpha_1$, and $\alpha'_\phi=0$, meaning that the scalar field $\phi$ is not excited now. Nevertheless, this new configuration satisfies the new Hamiltonian constraint:
\begin{align}\label{H_phi_new=0}
    H_{g\phi}^{\rm New}|_{\tPi_a\to \alpha'_a}&=\frac{1}{2(-\e) cl_p}\l(-\alpha'^2_0+\alpha'^2_1+\alpha'^2_\phi+\alpha_\phi^2\r)\nn\\
    &=\frac{1}{2(-\e) cl_p}\l(-\alpha^2_0+\alpha^2_1+0+\alpha^2_\phi\r)=0.
\end{align}
Therefore, the new configuration is given by the JNW-type metric \eqref{metric_JNW} with $\mu^2=\l(\frac{\alpha'_0}{\alpha'_1}\r)^2\approx 1+\l(\frac{\alpha_\phi}{\alpha_1}\r)^2$, where the new constraint \eqref{H_phi_new=0} is applied. It does not excite the scalar field $\phi$ but has the same metric as the original configuration.  

What is this new theory? One simple interpretation is to consider it as the Einstein gravity coupled with the scalar field $\phi$ in the background of another scalar field $\psi$ with the same excitation $\alpha_\phi$ as the original scalar-field configuration. More precisely, considering the $Y^\phi$ solution \eqref{YI_sol} and the relation $\tY^\phi=Y^\phi=\sqrt{\frac{\pi}{\hbar}}l_p \phi$, we set the background field as  
\begin{align}
    \psi\to \psi_{\rm back}(\tau)=\sqrt{\frac{\hbar}{\pi}}\frac{\alpha_\phi}{(-\epsilon)cl_p^2}\tau+\psi_0,
\end{align}
where $\psi_0$ is an arbitrary constant. 
Then, we add 
\begin{align}\label{S_psi}
S_{\psi}^{\rm (back)}[g_{\mu\nu}]=
%S_{\psi}[g_{\mu\nu},\psi_{\rm back}]=
-\frac{1}{2\hbar}\int d^4x\sqrt{-g}g^{\mu\nu}\partial_\mu \psi_{\rm back}(\tau) \partial_\nu \psi_{\rm back}(\tau)
\end{align}
to the Einstein-Hilbert action coupled with single massless scalar field $\phi_{I=1}=\phi$ \eqref{S_s} where $\Lambda=0$, $m_I=0$, and $I=1$. This is the new theory. In other words, this can be regarded as an effective theory obtained by introducing another field $\psi$ with the same excitation as the excitation of field $\phi$ in the original configuration and then integrating it out. The new configuration with the JNW metric is a solution of this new theory, due to the background field $\psi_{\rm back}$ which replaces the effect of the original field $\phi$. Better understanding the physical meaning of this would require further research.

%%%%%%%%%%%%%%%%%%%%%%%%%%%%%%%%%%%%%%%%%%%%%%%
\section{Conclusions and Prospects}
\label{sec:discussion}
Symmetry should be a useful concept in exploring the principles of quantum gravity. The gauge symmetry of spacetime is general covariance, and the Hamiltonian constraint plays the central role in the dynamics of gravity. Then, what is the physical symmetry in the configuration space of gravity?
Hamiltonian constraints define an infinite-dimensional configuration space, superspace. 
Imposing geometric symmetries (such as homogeneity) on spacetime yields a finite-dimensional configuration space, mini-superspace. This reduction procedure can be viewed as a kind of fluid limit for (quantum) gravitational configurations. Generically, when many microscopic degrees of freedom are coarse-grained into a fluid description, symmetries like volume-preserving diffeomorphism invariance often emerge, independent of the details of the constituents. Therefore, it is expected that some common symmetry may emerge in various mini-superspaces, potentially carrying the universal symmetry of quantum gravity. Indeed, 2D Schrödinger symmetry was obtained in the homogeneous isotropic universe model and the spherically symmetric static vacuum model. In this paper, to investigate its robustness, we analyzed spherically symmetric static mini-superspace models coupled with matter fields at a classical level. Due to spherical symmetry and staticity, these become one-dimensional dynamical systems with Hamiltonian constraint $H = N\MH \approx 0$, where the radial coordinate is the evolution parameter.

To find the symmetries of such systems, we developed the method utilizing canonical transformations. This method has the advantage of being directly calculable within the standard canonical formalism but works only for certain types of theories; it is complementary to the ED-lift method, which is applicable to any potential term systematically but is considered in the lifted space.
%This method not only applies to a broader class of models than conventional approaches but also possesses the advantage of being systematically treated within the standard framework of canonical theory. 
First, applying the canonical transformation method to the spherically symmetric static mini-superspace model with Maxwell field and cosmological constant, we demonstrated that the 3D Schrödinger symmetry emerges for a certain choice of $N$ and the $X$ coordinate system in mini-superspace. Here, the dimension of the mini-superspace increases by one for Maxwell field $A_0$, and the solution is given by the (A)dS-RN solution. Next, we analyzed the spherically symmetric static mini-superspace model coupled to $n$ massless scalar fields. Using a different $N$ and coordinate system $Y$ than before, we showed that the $(n+2)$D Schrödinger symmetry arises. Solving this model yielded the JNW solution and its ``interior" solution in a unified manner. 

Taking the matter decoupling limit and setting the cosmological constant to zero, both models become the same one, the vacuum mini-superspace model. Then, the 2D Schrödinger symmetry is obtained for different lapses and different mini-superspace coordinate systems. This fact can be interpreted as partially reflecting the covariance of Schrödinger symmetry in the mini-superspace.

Thus, the result that Schrödinger symmetry is obtained for mini-superspaces with different dimensions and different types of matter fields partially supports the robustness of the emergence of Schrödinger symmetry in the fluid limit.

We finally explored the physical meaning of Schrödinger symmetry under the Hamiltonian constraint $H\approx 0$. Generators $G$ commuting with $H$ transform a solution satisfying $H\approx0$ to another one (dynamical symmetry). On the other hand, this interpretation does not hold for generators $G$ that do not commute with $H$. Then, we proposed one possible interpretation: such a symmetry transforms the theory itself, and the configuration obtained by the transformation is a solution of the new theory. Indeed, this provides interesting views to such symmetries in our models as shown in explicit examples.
%We explained this with examples. While this interpretation remains open to debate, 
This interpretation provides a first step toward understanding the meaning of symmetry under Hamiltonian constraints. 

Let us now consider several future prospects.

The primary concern is whether the  Schrödinger symmetries are gauge-dependent. The result mentioned above—the appearance of the 2D symmetry in two different lapse and mini-superspace coordinate systems within the single model—suggests that this symmetry is not gauge-dependent. One approach toward a more complete understanding is to consider our scalar-field model, use the scalar field as a clock  (deparametrization) \cite{Oriti:2023yjj}, and perform a symmetry analysis in terms of purely physical degrees of freedom. %We would like to attempt this in the future.

%First, 
Next, it is crucial to investigate the robustness of Schrödinger symmetry in a broader context. Our current discussion is limited to cases reducible to the free-particle Hamiltonian without potential via canonical transformations. However, potentials of special forms, such as $V(x)=\frac{k}{x^2}$, also preserve Schrödinger symmetry \cite{Boyer:1974qt,Nikitin:2017gtu,Duval:2024eod}. Therefore, it is important to consider such more general cases and investigate whether other matter fields, such as massive scalar field or Dirac field, possess Schrödinger symmetry, or reconsider whether the scalar-field model in terms of the $X$ coordinate in footnote \ref{foot:scalar_X} leads to the symmetry. Conversely, it would be interesting to assume such special potentials and examine the spacetime structure derived from them or consider the corresponding modified gravity theories.

A related problem is the exploration for dynamics that eliminates singularities. One approach is to construct an effective theory inspired by DeWitt's idea \cite{DeWitt:1967yk}. One would examine what behavior trajectories on the mini-superspace should exhibit to avoid singularities, and search for potentials that can realize it while preserving Schrödinger symmetry. The corresponding theory should contain effective dynamics for resolving singularities.

Also, a deeper understanding of the interpretation of symmetries that generate theories is needed. For this, it should be useful to consider an extended phase space formulation where time (and its conjugate, energy) is incorporated as canonical variables \cite{Fahn:2018ahm,struckmeier2005hamiltonian}. This could allow the symmetry condition \eqref{symmetry_cond} to be written in the form of an extended Poisson bracket, $\{G,H\}_*=0$, enabling a more systematic investigation of the symmetries in systems with $H=0$, including generators that do not commute with $H$.

Furthermore, there are various possibilities for quantizing the mini-superspace models using Schrödinger symmetry. One approach is to study quantization techniques specific to particular potentials \cite{Essin:2006sic} and to consider how they apply to the models. Another method adopts the idea of group quantization \cite{Isham:1988bx} to quantize the Schrödinger group. Here, the dimension of our mini-superspace models is larger than 2, and it would be more non-trivial to construct the Casimirs and representations than the previous study \cite{Sartini:2021ktb}.  

In particular, quantization of the scalar-field mini-superspace model is interesting. The interior region of the horizon in the Schwarzschild metric corresponds—via swapping two coordinates $t$ and $r$— to a Kantowski-Sachs-type cosmological spacetime, where no matter field exists. 
From this correspondence, quantum black-hole models are expected to exhibit a strong link with quantum cosmology models \cite{Modesto:2005zm}. On the other hand, a scalar field can be used as a ``clock” in relational formalism to quantize models \cite{Blyth:1975is,Halliwell:1984eu,Domagala:2010bm,Ashtekar:2011ni}. One of our interesting results in the scalar-field mini-superspace model is that, the JNW solution and the Kantowski-Sachs-type spacetime (with the causal-limit fluid) are unified into the single metric \eqref{metric_JNW}, meaning that the latter corresponds to the ``interior” of the former. In particular, both possess the Schrödinger symmetry. Therefore, the quantization of this model based on the symmetry and this interior-exterior relationship should provide a promising avenue for connecting quantum black hole models and quantum universe models.

%Here, the nonlinearity inherent to gravity dynamics should be noted. 
In any quantization method, the non-linearity inherent to gravity dynamics must be carefully considered. One feature of our canonical transformation method is that it reduces the system's motion to a uniform linear motion on the mini-superspace, where the Hamiltonian constraint $H=0$ governs the dispersion relation. Here, the linearity of the solution—and at the quantum level, the principle of superposition—holds true, making it possible to superpose states with different values of charges. However, translating it into geometric language requires us to use the relationship $\tX(X,P), \tP(X,P)$ between the original canonical variables $(X,P)$ and the new ones $(\tX,\tP)$, from which the non-linearity appears. This point will likely require careful attention during quantization.

Finally, it is crucial to examine the Schrödinger symmetry found in our mini-superspace models from the perspective of a hydrodynamic approximation derived from a full theory of quantum gravity. Ref.\cite{Oriti:2024elx} focuses on the Schrödinger symmetry in the FLRW mini-superspace model and discusses the potential for removing cosmological singularities and realizing accelerated expansion through corrections to the quantum fluid equations that preserve the symmetry \cite{Marchetti:2025jze}. It would be interesting to check whether a similar mechanism works in the spherically-symmetric static mini-superspace models. %, the role of the hydrodynamic approximation and symmetry could be more non-trivial due to the presence of structures specific to inhomogeneities, such as event horizons and central singularities. Therefore, testing the framework in such systems could 
This test could serve as a crucial benchmark for assessing the universality of the hydrodynamic picture in quantum gravity. On the other hand, Ref.\cite{BenAchour:2023dgj} pointed out the existence of Schrödinger symmetry in the spherically-symmetric static vacuum mini-superspace and suggested that nonlinear terms preserving the symmetry could serve as quantum corrections. However, the form of such corrections may depend on the choice of the lapse function. Indeed, the symmetry they identified corresponds to only one of the two possible choices of the lapse function and mini-superspace coordinate we found. It would be intriguing to consider different lapses and coordinates and investigate whether there are differences in the physical effects arising from the correction terms. Furthermore, we showed that Schrödinger symmetry persists
even when adding matter fields. If the emergence of Schrödinger symmetry in such a mini-superspace is rigorously linked to the validity of the hydrodynamic approximation in a full quantum gravity theory incorporating matter fields, the findings of this study would have significant implications.

%%%%%%%%%%%%%%%%%%%%%%%%%%%%%%%
\stoptoc
\section*{Acknowledgments}
\resumetoc
We would like to thank Jibril Ben Achour, Che-Yu Chen, Kristina Giesel, Philipp Höhn, Christy Kelly, Etera Livine, Daniele Oriti, Francesco Sartini, and Shinji Tsujikawa for fruitful and inspiring discussions. We benefited from discussions at the workshop ``\textit{Concepts of Quantum and Spacetime} (KEK, Japan, March 2026)”. T.S. is supported by Waseda University Special Research Projects (Nos.~2025C-493) and the 15th Early Bird Program of Waseda University.
Y.Y. appreciated the hospitality of the members of the Institute for Theoretical Physics III at the Friedrich-Alexander-University Erlangen-Nuremberg, where he had a fruitful discussion and completed this work. Y.Y. was partially supported by Japan Society for the Promotion of Science (No.21K13929).

%%%%%%%%%%%%%%%%%%%%%%%%%%%%%%%%%
\appendix

%%%%%%%%%%%%%%%%%%%%%%%%%%%%%%%%%%%%%%%%%%%
% \section{ED method}\label{A:ED}
\section{Eisenhart-Duval Lift Method}\label{A:ED}
We provide a short review of the Eisenhart-Duval (ED) lift method for obtaining Schrödinger symmetry \cite{cccc1220-d9a2-3162-a3c5-a8782bbc8c8b,BenAchour:2022fif}, and discuss what happens if we apply it to our models.

One of the standard approaches to systematically identifying conserved quantities of a dynamical system is to seek Killing vectors \cite{Wald:1984rg}.
Consider a system defined on an $n$-dimensional configuration space with coordinates $x^i$, governed by the Hamiltonian:
\begin{equation}
   H = \frac{1}{2} g^{ij}(x) p_i p_j + V(x).
\end{equation}
(Note that it is quadratic in momentum \cite{Dantas:2026hvz}.) 
If there exists
%a spatial 
a Killing vector $\xi^i(x)$ satisfying the Killing equation,
\begin{equation}
\mathcal{L}_\xi g_{ij}=0, % =\nabla_i \xi_j + \nabla_j \xi_i = 0
\end{equation} 
and leaving the potential invariant ($\xi^i \partial_i V = 0$), then we can obtain a quantity $Q = \xi^i p_i$ that is conserved on shell: $\frac{d}{dt}Q = 0.$
%generates a conserved quantity $Q = \xi^i p_i$, provided that the potential is invariant along the vector field ($\xi^i \partial_i V = 0$):
%\begin{equation}
%\frac{d}{dt}\left(\xi^i p_i\right) = 0.
%\end{equation}
Note that such conserved quantities are  linear in momentum. Therefore, this approach is insufficient to derive  Schr\"odinger symmetry, since the Schr\"odinger algebra includes generators that are explicitly time-dependent and quadratic in momenta.

%However, conserved quantities derived from these standard spatial Killing vectors are manifestly linear in the momenta $p_i$ and do not explicitly contain the time parameter $t$. When attempting to derive the Schr\"odinger symmetry, this becomes a significant drawback, as the Schr\"odinger algebra essentially includes generators (such as Galilean boosts and dilatations) that are explicitly time-dependent and quadratic in momenta.

To overcome this limitation, one can use a powerful geometric framework known as the \textit{Eisenhart-Duval (ED) lift method}. % is employed. 
In the ED-lift method, the original $n$-dimensional configuration space is extended (``lifted") to a $(n+2)$-dimensional space by introducing two auxiliary coordinates ($u$ and $w$) and the ED metric:
%$u$, which plays the role of the time parameter $t$ in the original system, and a supplementary coordinate $w$. The geometry of this extended space is governed by the ED metric:
\begin{equation}
ds^2_{\mathrm{ED}} = 2dudw - 2V(x)du^2 + g_{ij}(x)dx^i dx^j.
\end{equation}

In this $(n+2)$-dimensional space, the dynamics of the original system are elegantly mapped to null geodesics. The Hamiltonian of the lifted space is given by
\begin{equation}
\mathcal{H}_{\mathrm{ED}} = \frac{1}{2} g^{MN}_{\mathrm{ED}} P_M P_N = P_u P_w + V(x)P_w^2 + \frac{1}{2} g^{ij}(x)p_i p_j,
\end{equation}
where $P_M = (P_u, P_w, p_i)$ are the conjugate momenta to $X^M=(u, w, x^i)$. Since the metric is independent of $w$, the corresponding momentum $P_w$ is a constant of motion.
Consider a geodesic line in this space. Setting $P_w = 1$ and restricting the geodesic onto the null hypersurface $\mathcal{H}_{\mathrm{ED}} = 0$ immediately yields $P_u = -H$, identifying $P_u$ with the original Hamiltonian and $u$ with the conjugate time coordinate $t$. Therefore, a physical solution is represented by a null geodesics with $P_w=1$.

Now that %Because the 
physical trajectories correspond to null geodesics, %the symmetry analysis is no longer restricted to exact Killing vectors. Instead, Conformal Killing Vectors 
conformal Killing vectors (CKVs) of the lifted metric provide conserved quantities. A CKV $\Xi^M(X)$ %in the ED space 
is determined by %the conformal Killing equation
\begin{equation}
\label{CKVequation}
\mathcal{L}_\Xi g^{\mathrm{ED}}_{MN} = 
%\nabla_M \Xi_N + \nabla_N \Xi_M = 
2 \omega g^{\mathrm{ED}}_{MN},
\end{equation}
where $\omega=\omega(X)$ is a scalar function. The phase space function $\mathcal{Q} = \Xi^M P_M$ associated with this CKV is conserved along the null geodesics. This %is 
can be confirmed by explicitly 
evaluating its Poisson bracket with the lifted Hamiltonian, which vanishes on-shell:
\begin{equation}
\frac{d \mathcal{Q}}{d\lambda} = \{\mathcal{Q}, \mathcal{H}_{\mathrm{ED}}\} = \omega \mathcal{H}_{\mathrm{ED}} \approx 0 \quad (\text{on } \mathcal{H}_{\mathrm{ED}} = 0)
\end{equation}

By evaluating %this lifted charge $\mathcal{Q}$ 
$\mathcal{Q}$ for a physical solution %on the null hypersurface  
(i.e., substituting $P_w=1$, $P_u=-H$, and $u=t$), we obtain the %actual
conserved quantities of the original $n$-dimensional system, unless $\mathcal{Q}$ contains $w$. In particular, $\mathcal{Q}$ can depend on $X^u=u=t$ and contain $P_u = -H$, expressing time-dependent and higher-order momentum conserved charges. Therefore, this procedure can be used to find Schr\"odinger symmetry.

%Crucially, because the CKV component $\Xi^u$ multiplies $P_u = -H$ (which is quadratic in $p_i$), and the components $\Xi^M$ can explicitly depend on $u=t$, the ED method systematically derives the explicitly time-dependent and higher-order momentum conserved charges required to establish the full Schr\"odinger symmetry.

Despite its conceptual elegance, the ED-lift method presents significant practical challenges when applied to general systems. The core of the issue lies in the geometric properties of the lifted space. If the $(n+2)$-dimensional lifted space is conformally flat, there exist coordinate systems where the metric becomes simple: %takes the form:
\begin{equation}
g^{\mathrm{ED}}_{MN}(X) = \Omega^2(X) \eta_{MN},
\end{equation}
where $\eta_{MN}$ is the flat Minkowski metric and $\Omega(X)$ is a scalar.  %conformal factor. 
Then, it is straightforward to solve  the conformal Killing equations \eqref{CKVequation} and determine the CKVs. 
%Thanks to this simple, determining the CKVs is remarkably straightforward. In such cases, the ED method is exceptionally powerful, allowing for a systematic extraction of the full dynamical symmetry algebra.

However, if the lifted space is \textit{not} conformally flat, one is forced to directly solve the equations \eqref{CKVequation}. This requires tackling a highly complex system of coupled partial differential equations, %involving the Christoffel symbols, which is mathematically daunting and restricts 
which restricts the systematic discovery of symmetries. 
%It is important to note a crucial caveat:
Note that even if the original $n$-dimensional configuration space with metric $g_{ij}(x)$ is flat (or conformally flat), the presence of a non-trivial potential $V(x)$ can warp the $(n+2)$-dimensional lifted space, %explicitly 
destroying its conformal flatness.

% Despite its conceptual elegance, the ED lift method has practical limitations when applied systematically to general systems. The core of the issue lies in the geometric properties of the lifted space. In an $N$-dimensional manifold (for $N \ge 3$), the maximum number of independent Conformal Killing Vectors (CKVs) is given by $\frac{1}{2}(N+1)(N+2)$, which is attained if and only if the space is conformally flat. Therefore, when the ED lifted space is conformally flat, the method is exceptionally powerful, allowing for a systematic extraction of the full dynamical symmetry algebra.

% However, if the lifted space is not conformally flat, systematically finding the CKVs by solving the conformal Killing equations becomes highly restrictive and mathematically challenging. It is important to note a crucial caveat: even if the original $n$-dimensional configuration space with metric $g_{ij}(x)$ is perfectly flat (or conformally flat), the presence of a non-trivial potential $V(x)$ can warp the $(n+2)$-dimensional lifted space, explicitly destroying its conformal flatness.

This is precisely the obstacle encountered in the systems discussed in this paper. Consider the Hamiltonian of the spherically-symmetric static space with a cosmological constant $\Lambda$, given by Eq.\eqref{H_g3}.
%Eq.\eqref{H_g2} and equivalently rewritten as Eq.~\eqref{H_g3}. 
The original 2D mini-superspace metric $\mathcal{G}_{ab}$ given in Eq.\eqref{metric_L} is flat.
%manifestly flat, and hence conformally flat. 
By applying the ED-lift method, the 4-dimensional lifted metric takes the form: %would take the form:
\begin{equation}
ds^2_{\mathrm{ED}} = 2dudw - 2\mathcal{V}_\Lambda(X)du^2 + \mathcal{G}_{ab}dX^a dX^b,
\end{equation}
where the %spacetime 
coordinates are $(u, w, X^+, X^-)$. %Computing the curvature of this 4D lifted spacetime reveals that the Ricci tensor identically vanishes ($R_{\mu\nu} = 0$), implying that the Weyl tensor $C_{\mu\nu\rho\sigma}$ precisely coincides with the Riemann tensor. 
Due to the $X^-$-dependence of the potential $\mathcal{V}_\Lambda(X)$, the Weyl tensor acquires non-vanishing components: %, specifically:
\begin{equation}
C_{u-u-} = -\frac{c l_p}{l_\Lambda^2} \neq 0,
\end{equation}
where the index $-$ corresponds to the $X^-$ coordinate. In general, a manifold of dimension four or higher is conformally flat if and only if its Weyl tensor strictly vanishes. Therefore, the lifted space is no longer conformally flat. % as long as the cosmological constant is present ($\Lambda \neq 0$). 
Consequently, this geometric deformation hinders the direct and systematic application of the ED-lift method to find the full set of Schr\"odinger symmetries although the method may be able to find a part of it.
% \cite{BenAchour:2022fif}.

%A completely 
An analogous situation arises in the scalar-field mini-superspace model.
In the Hamiltonian \eqref{H_phi2},
%Considering the Hamiltonian given in Eq.~\eqref{H_phi2}, although 
the original mini-superspace metric $\eta_{ab}$ is flat while the potential term exhibits an exponential dependence on the coordinate $Y^0$. %, namely $V(Y) \propto e^{4Y^0}$. 
Just as the above case, therefore, the lifted space is not conformally flat. %in the cosmological constant case, the conformal flatness in the lifted space is explicitly broken by this potential, rendering the ED method equally ineffective for systematically extracting the symmetries of this system.

While the ED-lift method is straightforward and elegant, and in principle applicable to cases where momentum terms are quadratic \cite{Dantas:2026hvz}, it is not always practical in every situation, as demonstrated in the examples above. It may therefore be necessary to consider alternative methods tailored to specific cases. Consequently, in this paper, we develop a method based on canonical transformations rather than point transformations to search for symmetries.

%While the ED method remains a conceptually useful and elegant tool, these geometric limitations in the presence of specific potentials necessitate an alternative approach. For this reason, rather than relying on the ED lift, we establish the symmetries in this paper by systematically constructing the symmetry generators via canonical transformations directly in the phase space.

\section{Mass and Electric Charge}\label{A:M_Q}
We check the consistency for the formulae of mass $M$, \eqref{M_def} and \eqref{M_def2},  and electric charge $\MQ$ \eqref{Q_def}. 

\subsubsection*{Mass}
The Komar mass formula can evaluate the ADM energy of an asymptotically-flat stationary spacetime in a covariant way \cite{Wald:1984rg}: 
\begin{equation}\label{Komar}
    M_{\rm Komar}\equiv -\frac{1}{8\pi G} \int_{S_\infty} dS_{\mu\nu} \nabla^\mu t^\nu,
\end{equation}
where $t^\mu$ is the timelike Killing vector satisfying $t^2\xrightarrow[r\to\infty]{}-1$, and $dS_{\mu\nu}$ is the area element for a 2-sphere $S_\infty$ in an asymptotically flat vacuum region. 
For a spherically-symmetric static spacetime with the metric \eqref{metric_X}, we can calculate it as 
\begin{align}
    M_{\rm Komar}&= -\frac{1}{8\pi G}\int_{S_\infty} \N^{t_\infty}t^{\tau} dS_{t_\infty \tau}\times2 \nn\\
    &= -\frac{1}{4\pi G}\int_{S_\infty} g^{t_\infty t_\infty}\N_{t_\infty}t^{\tau} \sqrt{-g}d\theta d\phi \nn\\
    &=\frac{1}{8\pi G}4\pi r^2 \frac{1}{\sqrt{-g_{2d}}}\p_\tau (- g_{t_\infty t_\infty})\label{Komar_2},
    %&= \frac{\e l_s^2 \tP_+}{4G\bar c}\l(X^+ \dot X^--X^-\dot X^+ \r),
\end{align}
where $g_{2d}$ is the determinant of the 2D metric normal to 2-sphere, and 
$dt_\infty=\frac{1}{k} dt$ is the time coordinate at $r=\infty$ with a constant $k$. 

First, we prove that the mass \eqref{M_def} for $\Lambda=0$ and $\e=-1$ agrees with the Komar formula \eqref{Komar_2}. 
We start by expressing the metric \eqref{metric_X} in terms of $t_\infty$: 
\begin{align}
   ds^2=\epsilon\frac{4X^+(\tau)}{X^-(\tau)}\frac{\tP_+^2}{\bar c^2}dt^2_\infty
   -\e \frac{X^-(\tau)}{4X^+(\tau)}d\tau^2+\frac{l_s^2}{4}{X^-(\tau)}^2d\Omega^2,
   \label{metric_Komar}
\end{align}
where we used the gauge $N=\sqrt{\frac{X^-}{4X^+}}$ and $dt_\infty \equiv \frac{\bar c}{\tP_+} dt$. 
We can then calculate the Komar mass \eqref{Komar_2} as 
\begin{align}
    M_{\rm Komar}&= \frac{\e l_s^2 \tP_+}{2G\bar c}\l(X^+ \dot X^--X^-\dot X^+ \r)\nn\\
    &=\frac{l_s}{G\bar c^2}P_+(B^- P_-- B^+ P_+).
\end{align}
Here, noting that $(\tX^a,\tP_a)=(X^a,P_a)$ holds in the case of $\Lambda=0$, we have applied the canonical equations \eqref{canonical_eq}: $\dot X^{\pm}=-\frac{2}{\e c l_p} P_\mp={\rm const.}$ and $X^\pm=-\frac{2}{\e c l_p}P_\mp \tau + B^\pm$. This agrees with the definition of $M$ \eqref{M_def} for $\Lambda=0$ and $\e=-1$. %Note here that the Komar formula is applied to an asymptotically flat vacuum region, where $\e=-1$ holds.

Next, we show that in the Y-coordinate, the Komar formula \eqref{Komar_2} becomes the simpler formula \eqref{M_def2}. In terms of $dt_\infty= \frac{l_s \tPi_1}{\bar c r_0}dt\equiv \frac{1}{k} dt$ and the lapse \eqref{N2}, the Y-metric \eqref{metrix_Y} is expressed as
\begin{equation}\label{metrix_Y2}
    ds^2=4\e\l(-e^{8Y^0-4Y^1} d\tau^2+k^2e^{4Y^1}dt_\infty^2\r)+\frac{l_s^2}{4}e^{4Y^0-4Y^1}d\Omega^2.
\end{equation}
The formula \eqref{Komar_2} can be evaluated as  
\begin{align}
    M_{\rm Komar}&=\frac{(-\e)}{2G}l_s^2 k \dot Y^1\nn\\
    &=\frac{(-\e)}{2G}l_s^2 \frac{\bar c r_0} {l_s \tPi_1}\frac{\tPi_1}{(-\e) c l_p}\nn\\
    &=\frac{r_0}{2G},
\end{align}
where we used the definition of $k$ and the trajectory \eqref{Y1_sol}. This is consistent with the definition \eqref{M_def2}.

%%%%%%%%%%%%%%%
\subsubsection*{Electric charge}
The standard electric charge is given by 
\begin{align}\label{Q}
    \MQ_{\rm{elecric}}\equiv \int_\Sigma d\Sigma_{\mu}j^\mu =\frac{1}{2} \int_S dS_{\mu\nu} F^{\mu\nu}.
\end{align}
Here, $j^\mu$ is electric current, $d\Sigma_\mu$ is the volume element of a hypersurface $\Sigma$, and we applied Maxwell's equations $\nabla_\nu F^{\mu\nu}=j^\mu$ and Stokes' theorem. 

In the metric \eqref{metric_Komar}, we can evaluate this as 
\begin{align}\label{Q1}
    \MQ_{\rm{electric}}&=\frac{1}{2}\int_S dS_{t_{\infty}\tau} F^{t_\infty \tau} \times 2 \nn\\
    &=\int_S d\theta d\phi \sqrt{-g}g^{t_\infty t_\infty} g^{\tau \tau} F_{t_\infty \tau} \nn\\
    &=4\pi r^2 \frac{1}{\sqrt{-g_{2d}}}\dot A_{t_\infty}\nn\\
    &=\sqrt{\frac{\pi}{G}}l_s^2 (X^-)^2 \dot X^A,
\end{align}
where we recalled $X^A\equiv \sqrt{\pi G}A_t$. 
Using the relation \eqref{tX_E} of $\tX^A$ and $X^A$ and the canonical equation, we have 
\begin{align}
    \dot X^A&=\dot \tX^A+\frac{\e \tP_A}{2\tP_+}\dot \tX^-\l(1-\frac{1}{(X^-)^2}\r)\nn\\
    &=\frac{\tP_A}{cl_p}+\frac{\e \tP_A}{2\tP_+}\frac{-2}{\e cl_p}\tP_+\l(1-\frac{1}{(X^-)^2}\r)\nn\\
    &=\frac{1}{cl_p^2 (X^-)^2}\tP_A.
\end{align}
Substituting this into \eqref{Q1}, we obtain 
\begin{equation}
    \MQ_{\rm{electric}}=\sqrt{\frac{\pi}{G}}\frac{l_s}{\bar c}\tP_A,
\end{equation}
which matches the definition of $\MQ$ \eqref{Q_def}. 

%%%%%%%%%%%%%%%%%%%%%%%%%%%%%%%
\section{Derivation of (A)dS-RN metric}\label{A:RN}
We take the same steps as in Sec.\ref{sec:AdS} and show that the solution to the Hamiltonian \eqref{H_gE5} is given uniquely by the (Anti) de-Sitter Reissner-Nordström metric. 

We first solve the canonical equations \eqref{canonical_eq} where $H_g$ is replaced with $H_{gE}$ \eqref{H_gE5}. We have the relation \eqref{tau} between the parameter $\tau$ and the areal radius $r$. As in \eqref{X+_L}, we can calculate 
\begin{align}\label{X+_E}
    X^+&=\frac{\tP_-}{\tP_+}X^-+\frac{1}{\tP_+}(\tB^+\tP_+-  \tB^-\tP_-)\nn\\ 
    &-\frac{\e \tP_A^2}{4\tP_+^2}\l(X^-+\frac{1}{X^-}\r)    
    + \frac{\e c^2 l_p^2}{12l_\Lambda^2}\frac{(X^-)^3}{\tP_+^2}.
\end{align}

We use these ingredients and the lapse \eqref{N1} and obtain
\begin{align}\label{tautau_E}
    -\e N^2 d\tau^2
    &=-\e \frac{X^-}{4X^+}\frac{\bar c^2}{\tP_+^2} dr^2\nn\\
    &=\frac{1}{\frac{1}{\bar c^2} \l( -\frac{4}{\e} \tP_+ \tP_ - +\tP_A^2 \r)-\frac{2GM}{r}+\frac{G\MQ^2}{4\pi r^2}-\frac{\Lambda}{3}r^2}dr^2,\nn\\
    &\approx \frac{1}{1-\frac{2GM}{r}+\frac{G\MQ^2}{4\pi r^2}-\frac{\Lambda}{3}r^2}dr^2.
\end{align}
In the second line, we applied the mass formula \eqref{M_def} and introduced the electric charge formula \eqref{Q_def}:  
\begin{align}%\label{Q_def2}
    \MQ\equiv \sqrt{\frac{\pi}{G}} \frac{l_s}{\bar c}\tP_A\nn,
\end{align}
which agrees with the standard definition % of electric charge 
(see Appendix \ref{A:M_Q}) and commutes with $H_{gE}$. In the last line, we employed the constraint (from \eqref{H_gE5}): 
\begin{equation}
    H_{gE}\approx 0 \Rightarrow -\frac{4}{\e} \tP_+ \tP_-+ \tP_A^2 \approx \bar c^2.
\end{equation}
Similarly, we have 
\begin{align}\label{tt_E}
    \e \frac{4X^+}{X^-}dt^2
    &=\e \frac{4X^+}{X^-}\frac{\tP_+^2}{\bar c^2}\frac{\bar c^2}{\tP_+^2}dt^2\nn\\
    %&=-\l(\frac{1}{\bar c^2} \l( -\frac{4}{\e} \tP_+ \tP_+ +\tP_A^2 \r)-\frac{2GM}{r}+\frac{G\MQ^2}{4\pi r^2}-\frac{\Lambda}{3}r^2\r)dt_\infty^2\nn\\
    &\approx -\l(1-\frac{2GM}{r}+\frac{G\MQ^2}{4\pi r^2}-\frac{\Lambda}{3}r^2\r)dt_\infty^2,
\end{align}
where we used $dt_\infty \equiv \frac{\bar c}{\tP_+}dt$.  
Thus, the solution is given uniquely by the (A)dS Reissner-Nordström metric: 
\begin{align}\label{metric_RN_A}
    ds^2\approx&-\l(1-\frac{2GM}{r}+\frac{G\MQ^2}{4\pi r^2}-\frac{\Lambda}{3}r^2\r)dt_\infty^2\nn\\
    &+\l(1-\frac{2GM}{r}+\frac{G\MQ^2}{4\pi r^2}-\frac{\Lambda}{3}r^2\r)^{-1}dr^2+r^2 d\Omega^2.
\end{align}

%%%%%%%%%%%%%%%%%%%%%%%%%%%%%%
\section{Derivation of JNW metric}\label{A:JNW}
We derive the solution configuration given in Sec.\ref{JNW}. We first solve the canonical equations for $H_{g\phi}$ \eqref{H_phi3}:
\begin{align}
    \dot \tPi_a=\{\tPi_a,H_{g\phi} \}=0&\Rightarrow \tPi_a(\tau)={\rm const}.\nn\\
    \dot\tY^a=\{\tY^a,H_{g\phi}\}=\frac{\eta^{ab}\tPi_b}{(-\epsilon)cl_p}&\Rightarrow \tY^a(\tau)=\frac{\eta^{ab}\tPi_b}{(-\epsilon)cl_p}\tau+\tB^a,
\end{align}
where $(\tPi_a,\tB^a)$ are considered as constants characterizing the trajectory again. 
More explicitly, we have 
\begin{subequations}
    \begin{gather}\label{Y0_sol}
    \tY^0(\tau)=\frac{\tPi_0}{\epsilon cl_p}\tau+\tB^0\Rightarrow\tau=\frac{\e c l_p}{\tPi_0}(\tY^0-\tB^0)\\
    \label{Y1_sol}
    Y^1(\tau)=\tY^1(\tau)=\frac{\tPi_1}{(-\epsilon)cl_p}\tau+\tB^1=-\frac{\tPi_1}{\tPi_0}\tY^0+\frac{\tPi_1}{\tPi_0}\tB^0+\tB^1,\\
    \label{YI_sol}
     Y^I(\tau)=\tY^I(\tau)=\frac{\tPi_I}{(-\epsilon)cl_p}\tau+\tB^I=-\frac{\tPi_I}{\tPi_0}\tY^0+\frac{\tPi_I}{\tPi_0}\tB^0+\tB^I.
\end{gather}
\end{subequations}
On the other hand, $Y^0$ and $\tY^0$ are related by the first equation in \eqref{tY}. We can solve it for $Y^0$ as 
% \begin{equation}\label{YY}
%     e^{4Y^0}=\frac{\tPi_0^2}{\bar c^2}\frac{ e^{4\tY^0}}{(e^{4\tY^0}+\epsilon )^2}.
% \end{equation}
\begin{align}\label{YY}
    e^{4Y^0}&=\frac{\tPi_0^2}{\bar c^2}\frac{ e^{4\tY^0}}{(e^{4\tY^0}+\epsilon )^2}\\
    &=\begin{cases}
    \frac{\tPi_0^2}{4\bar c^2}\frac{1}{(\mathrm{cosh}(2\tY^0))^2} & (\text{for } \epsilon = 1) \nonumber\\
    \frac{\tPi_0^2}{4\bar c^2}\frac{1}{(\mathrm{sinh}(2\tY^0))^2} & (\text{for } \epsilon = -1) \nonumber
\end{cases}
\end{align}

Now, we introduce a new radial coordinate $R$ by \eqref{R_def}:
\begin{equation}
   e^{4\tY^0}=:\frac{R+r_+}{(-\epsilon)(R-r_-)}
\label{R_def_A}
\end{equation}
for $R\neq r_-$, where $r_\pm$ are constants to be determined, and $\epsilon=-1$ corresponds to $R>r_-$ and $\epsilon=1$ to $R<r_-$. 
We want to express the metric \eqref{metrix_Y} in terms of $R$ (instead of $\tau$). We start with evaluating the areal radius $r$: 
\begin{align}\label{r(R)_A}
    r^2&=\frac{l_s^2}{4}e^{4Y^0-4Y^1}\nn\\
    &=\frac{l_s^2\tPi_0^2}{4\bar c^2}\frac{ e^{4\tY^0}}{(e^{4\tY^0}+\epsilon )^2}e^{4\frac{\tPi_1}{\tPi_0}\tY^0-4\l(\frac{\tPi_1}{\tPi_0}\tB^0+\tB^1\r)}\nn\\
    &=\frac{l_s^2\tPi_0^2}{4\bar c^2}e^{-4\l(\frac{\tPi_1}{\tPi_0}\tB^0+\tB^1\r)}\frac{(R+r_+)^{1+\frac{\tPi_1}{\tPi_0}}[(-\epsilon)(R-r_-)]^{1-\frac{\tPi_0}{\tPi_1}}}{(r_+ + r_-)^2}\nn\\
    &=(R+r_+)^{1+\frac{1}{\mu }}
        [(-\epsilon)(R-r_-)]^{1-\frac{1}{\mu}},
\end{align}
which provides \eqref{r(R)}. Here, in the second line, we used the solution \eqref{Y1_sol} and the relation \eqref{YY}; in the third line, we applied the definition \eqref{R_def_A}; and in the last line, we introduced \eqref{defs}:
\begin{align}\label{defs_A}
    \mu\equiv\frac{\tPi_0}{\tPi_1},~r_{\pm}\equiv \frac{r_0}{2}(\mu\pm1),~r_0\equiv \frac{l_s \tPi_1}{2\bar c}e^{-2\l(\frac{1}{\mu}\tB^0+\tB^1\r)}.
\end{align}
%Note that $\mu$ and $r_0$ commute with $H_{g\phi}$ \eqref{H_phi3} and thus are gauge invariant. 
%Note that we can check $\{\mu,H_{g\phi}\}=\{r_0,H_{g\phi} \}=0$ for \eqref{H_phi3}, meaning that $\mu$ and $r_0$ are gauge invariant. 
Similarly, we can calculate the $tt$-component as 
\begin{align}\label{tt_R}
    4\epsilon e^{4Y^1}dt^2&=4e^{-4\frac{\tPi_1}{\tPi_0}\tY^0+4\l(\frac{\tPi_1}{\tPi_0}\tB^0+\tB^1\r)}dt^2\nn\\
    &=\epsilon\l(\frac{(-\epsilon)(R-r_-)}{R+r_+}\r)^{\frac{1}{\mu}}dt_\infty^2,
\end{align}
where we defined
$dt_\infty\equiv 2 e^{2\l(\frac{1}{\mu}\tB^0+\tB^1\r)}dt=\frac{l_s \tPi_1}{\bar c r_0}dt$. 
Furthermore, we can check from the definitions \eqref{R_def_A} and the trajectory \eqref{Y0_sol} and \eqref{Y1_sol} that 
\begin{align}\label{expressions_R}
    d\tau &= \frac{cl_p (r_++r_-)}{4\tPi_0}\frac{1}{(-\e)(R-r_-)(R+r_+)}dR, \nn\\
    e^{4Y^0}&=\frac{\tPi_0^2}{\bar c^2 (r_++r_-)^2}(-\e)(R-r_-)(R+r_+),\nn\\
    e^{-4Y^1}&=\l(\frac{R+r_+}{(-\e)(R-r_-)}\r)^{\frac{\tPi_1}{\tPi_0}}e^{-4\l(\frac{\tPi_1}{\tPi_0}\tB^0+\tB^1\r)}
\end{align}
holds, enabling us to express the $\tau\tau$-component as 
\begin{align}
    -4\epsilon e^{8Y^0-4Y^1}d\tau^2=(-\e)\l(\frac{R+r_+}{(-\epsilon)(R-r_-)} \r)^{\frac{1}{\mu}}dR^2,
\end{align}
where we used the definitions \eqref{defs_A}. Thus, we reach \eqref{metric_JNW}:
\begin{align}\label{metric_JNW_A}
    ds^2&=(-\e)\l(-\l[\frac{(-\epsilon)(R-r_-)}{R+r_+}\r]^{\frac{1}{\mu}}dt_\infty^2\r.\nn\\
    &~~~~~~~\l.+\l[\frac{R+r_+}{(-\epsilon)(R-r_-)} \r]^{\frac{1}{\mu}}dR^2\r)+r(R)^2d\Omega^2,
\end{align}
where $r(R)$ is given by \eqref{r(R)_A}. Here, 
%Note that we have not used 
the Hamiltonian constraint $H_{g\phi}\approx 0$ has not been set yet. 

Next, we evaluate the scalar fields $\phi_I =\frac{1}{\sqrt{\pi G}} \tY^I$: 
\begin{align}\label{phi_JNW_A}
    \phi_I(R)&=\frac{1}{\sqrt{\pi G}}\l(-\frac{\tPi_I}{\tPi_0}\tY^0+\frac{\tPi_I}{\tPi_0}\tB^0+\tB^I\r)\nn\\
&=\frac{1}{\sqrt{\pi G}}\l(\frac{\tPi_I}{4\tPi_0}\log \frac{(-\epsilon)(R-r_-)}{R+r_+}+\frac{\tPi_I}{\tPi_0}\tB^0+\tB^I\r)\nn\\
&=\frac{\MA_I}{\mu}\log \frac{(-\epsilon)(R-r_-)}{R+r_+}+\phi_I^{(0)},
\end{align}
which is \eqref{phi_JNW}.
Here, in the first line we applied the trajectory \eqref{YI_sol}; in the second line we used the definition \eqref{R_def_A}; and in the last line we defined \eqref{defs2}:
\begin{align}\label{defs2_A}
    \MA_I \equiv \frac{1}{\sqrt{16\pi G}} 
    \frac{\tPi_I}{\tPi_1},~~
      \phi_I^{(0)}\equiv \frac{1}{\sqrt{\pi G}}
    \l( \frac{\tPi_I}{\tPi_0} \tB^0 + \tB^I \r).
\end{align}

%Finally, we impose the Hamiltonian constraint \eqref{H_phi3}:  
%\begin{align}\label{mu_JNW}
%    H_{g\phi}\approx 0 &\Rightarrow -\tPi_0^2+\tPi_1^2+\sum_I\tPi_I^2\approx 0.\nn\\
%    &\Rightarrow \mu^2 \equiv\l( \frac{\tPi_0}{\tPi_1}\r)^2 \approx 1+ 16\pi G \sum_I \MA_I^2.
%\end{align}
%This provides a constraint to the physical value of $\mu$:
%\begin{equation}\label{mu_value}
%    \mu\geq 1,
%\end{equation}
%where $\mu=1$ holds only for $\MA_I=0$.  

For energy-momentum tensor $T_{\mu\nu}=\sum_I(\p_\mu \phi_I \p_\nu \phi_I-\frac{1}{2}g_{\mu\nu}g^{\alpha\beta}\p_\alpha \phi_I \p_\beta \phi_I)$, we can check that the relation \eqref{TTTT}
\begin{equation}%\label{TTTT_A}
    -T^t{}_t=T^R{}_R=-T^\theta{}_\theta=-T^\phi{}_\phi\nn
\end{equation}
holds with the condition \eqref{E_density}:
\begin{align}%\label{E_density_A}
    -T^t{}_t&=\frac{1}{2}g^{\tau\tau}\sum_I\dot \phi_I^2\nn\\
        &=\frac{r_0^2\sum_I \MA_I^2}{2(-\epsilon)}(R+r_+)^{-2-\frac{1}{\mu}}[(-\epsilon)(R-r_-)]^{-2+\frac{1}{\mu}}\nn.
\end{align}
Here, we have used the configuration \eqref{metric_JNW_A} and \eqref{phi_JNW_A} and the $\tau$-$R$ relation in Eq.\eqref{expressions_R}.

%%%%%%%%%%%%%%%%%%%%%%%%%%%%%%%
\section{Lagrangian Interpretation of Symmetry under Hamiltonian constraints}\label{A:Action}
In Sec.\ref{sec:interpretation}, we examined in detail the symmetries in the canonical formalism with Hamiltonian constraints, particularly providing an interpretation that certain symmetry transformations generate new theories. In this Appendix, we will study systems with Hamiltonian constraints in the action principle, explore the relationship between the canonical and Lagrangian descriptions, and show that the interpretation of generating new theories by symmetry transformations also holds in the action principles. 

The mini-superspace models in this paper can generally be described by the following action: 
\begin{align}\label{S_again}
S[X,\MN]=\int d\tau \MN\left[\frac{1}{2\MN^2}\MG_{ab}(X)\frac{dX^a}{d\tau}\frac{dX^b}{d\tau}-\mathcal{V}(X)\right].
\end{align}
Here, since the lapse $N(\tau)$ in the spacetime metric \eqref{metric_X} is an arbitrary function of $\tau$ and may depend on $X^a(\tau)$, we have absorbed a function $f(X)$ that may appear alongside $N$, and introduced a new independent variable $\MN(\tau)\equiv N(\tau)f(X(\tau))$. Indeed, say, the gravity action \eqref{L_g} and the scalar-field action in footnote \ref{foot:scalar_X} take this form. The variation of $\MN$ yields 
\begin{align}\label{S_const}
    \frac{\delta S}{\delta \MN}=-\frac{1}{2\MN^2}\MG_{ab}\frac{dX^a}{d\tau}\frac{dX^b}{d\tau}-\mathcal{V}=0,
\end{align}
which corresponds to the Hamiltonian constraint $\MH\approx0$. Therefore, the action \eqref{S_again} contains the effect of the constraint. On the other hand, the variation of $X^a$ provides the equation of motion for $X^a$ (geodesic equation in the mini-superspace), which is equivalent to the canonical equation for the Hamiltonian $H=\MN \MH$. Solving this together with the constraint \eqref{S_const} determines a physical solution $X^a(\tau)$. Therefore, all the equations are derived from the action principle of the non gauge-fixed action \eqref{S_again}, which determines the full dynamics and defines the theory. In this sense, we call it the \textit{full action}.

Alternatively, we can fix the lapse as $\MN(X)=1$ at the action level and consider the gauge-fixed action:
\begin{align}\label{S_again_fixing}
S_{\mathrm{fix}}[X]=\int d\tilde{\tau}\left(\frac{1}{2}\mathcal{G}_{ab}(X)\frac{dX^a}{d\tilde{\tau}}\frac{dX^b}{d\tilde{\tau}}-\mathcal{V}(X)\right),
\end{align}
where  $\tilde{\tau}$ denotes the time coordinate in this gauge. In the action principle of this gauge-fixed action, the constraint \eqref{S_const} cannot be derived while the $X$ equation of motion can in this gauge. Therefore, considering only the gauge-fixed action $S_{\rm fix}$ corresponds to analyzing the $X$ equation of motion, equivalent to the canonical equations for $H=\MH$, without imposing $\MH\approx0$. % the Hamiltonian constraint. 
In order to obtain a physical solution, we need to impose the constraint \eqref{S_const} (for $\MN=1$) \textit{by hand} and solve it with the $X$ equations. %This is what we have done within the canonical formalism in the main text. Indeed, say, the gravity Hamiltonian \eqref{H_g3} is obtained by performing the Legendre transformation on $S_{\rm fix}$.\footnote{\label{foot:gauge fixing} \TS{We note that in the Hamiltonian formalism, selecting a specific lapse function does not amount to complete gauge fixing; correct physical results are obtained as long as the weak equality $H \approx 0$ is imposed. Indeed, the choice of lapse determines the infinitesimal increment $d\tau$ rather than the $\tau$ itself. Consequently, merely imposing $H \approx 0$ leaves a residual gauge redundancy. While one could eliminate this redundancy by fixing a reference time $\tau_0$ as a constant of motion, from the perspective of Schrödinger symmetry, it is more natural to characterize the system by the initial values of $(\tilde{X}^a, \tilde{P}_a)$. In this framework, a non-trivial gauge degree of freedom persists among the initial values of $\tilde{X}^a$. This is the reason we employ the term “choice of lapse" rather than “gauge fixing" in our Hamiltonian formulation.}\YY{Choosing $\MN$ fixes it completely in the Lagrangian formalism? If so, why?}}

Now, we examine the transformations generated by Noether charges in the full action and the gauge-fixed action. In this paper (the canonical formalism), Noether charges are defined by the condition \eqref{symmetry_cond}, which leaves the gauge-fixed action invariant. Indeed, this condition %for the Hamiltonian $H$ derived from a gauge-fixed action $S_{\rm fix}$, which 
guarantees the invariance of $H$ for such transformations (see Eq.\eqref{KH}), and the canonical form of the gauge-fixed action $S_{\rm fix}^{\rm can}[X,P]=\int d\tilde{\tau}(P_a \dot X^a-H)$ is invariant, since for general canonical transformations, the symplectic structure $P_a dX^a$ is invariant. %Therefore, the canonical form of the gauge-fixed action $S_{\rm fix}^{can}[X,P]=\int d\tilde{\tau}(P_a \dot X^a-H)$ is invariant, up to a total derivative, for the symmetry transformations (see below for an example). 
However, this does not mean the invariance of the full action $S[X,\MN]$. This is natural, since the gauge-fixed action does not have the effect of the Hamiltonian constraint. In fact, not all such transformations leave the full action invariant (see below). %Note here that a physical solution must satisfy the Hamiltonian constrain, which is encoded into the the full action. Therefore, certain symmetry transformations are not dynamical symmetries that map a physical solution to another.
Therefore, some transformations generated by Noether charges satisfying \eqref{symmetry_cond} can change the theory. 

%Next, we examine the physical interpretation of the transformations generated by the Noether charges defined in Eq. \eqref{symmetry_cond}. Since the Hamiltonian is derived from the gauge-fixed action, these transformations can be regarded as symmetries that leave the gauge-fixed action invariant up to a total derivative. This is consistent with the property that transformations made by Noether charges leave the Hamiltonian (the generator of time evolution) invariant. Indeed, the gauge-fixed action directly corresponds to the dynamical equations of motion or Hamilton's equations. However, the physical solution must be derived from the action without gauge fixing. Crucially, not all of these transformations leave this action invariant. This explains why, from the perspective of the action, certain transformations are not regarded as physical symmetries that map one physical solution to another.

To illustrate these points concretely, let us consider a simple case where $\MG_{ab}$ is a constant invertible matrix and $\mathcal{V}(X)={\rm const.}=\lambda$:
\begin{align}\label{S_again_free}
S[X,\MN]=\int d\tau \MN \left(\frac{1}{2\MN^2}\mathcal{G}_{ab}\frac{dX^a}{d\tau}\frac{dX^b}{d\tau}-\lambda\right).
\end{align}
If we fix $\MN=1$, this becomes 
\begin{align}\label{S_again_fixing_free}
S_{\mathrm{fix}}[X]=\int d\tilde{\tau}\left(\frac{1}{2}\mathcal{G}_{ab}\frac{dX^a}{d\tilde{\tau}}\frac{dX^b}{d\tilde{\tau}}-\lambda\right).
\end{align}
Note that this form of the action is derived from the Hamiltonian \eqref{H_again} via the Legendre transformation.%\footnote{\label{foot:Canonical trans}\TS{Note that we derived the Hamiltonian in the free-particle form, where the Schrödinger symmetry is manifest, via a canonical transformation. The Lagrangian obtained by Legendre-transforming this Hamiltonian involves variables that mix the original coordinates and momenta, thereby obscuring the aesthetic simplicity typically associated with the Lagrangian picture. Nevertheless, this Lagrangian and the resulting action yield the correct physical dynamics.}}

As an example, we consider the Galilean boost transformation given by
\begin{align}\label{Btrans}
X^a \rightarrow X'^a=X^a - v^a \eta,
\end{align}
where $v^a$ is constant and $\eta=\eta(\tau):= \int^\tau d\tau' \MN(\tau')$. For the gauge fixing $\MN=1$, this becomes 
\begin{align}\label{Btrans_fix}
X^a \rightarrow X'^a=X^a - v^a \tilde{\tau}.
\end{align}
Under this transformation, the gauge-fixed action \eqref{S_again_fixing_free} transforms as $S_{\mathrm{fix}}\to S'_{\mathrm{fix}}=S_{\mathrm{fix}}[X']$: 
\begin{align}\label{S_again_fixing_transformed}
S'_{\mathrm{fix}}&=\int d\tilde{\tau}\left(\frac{1}{2}\mathcal{G}_{ab}\left(\frac{dX^a}{d\tilde{\tau}}-v^a\right)\left(\frac{dX^b}{d\tilde{\tau}}-v^b\right)-\lambda\right)\nonumber\nonumber\\
&=\int d\tilde{\tau}\Bigg(\frac{1}{2}\mathcal{G}_{ab}\frac{dX^a}{d\tilde{\tau}}\frac{dX^b}{d\tilde{\tau}}-\lambda\nonumber\\
&\;\;\;\;\;\;\;\;\;\;\;\;+\frac{d}{d\tilde{\tau}}\left(-\mathcal{G}_{ab}v^aX^b\right)+\frac{1}{2}\mathcal{G}_{ab}v^av^b\Bigg)\nn\\
&=S_{\rm fix}+{\rm (surface~terms)}
\end{align}
Here, the last term in the second line is constant and expressed as $\frac{d}{d\tilde{\tau}}\l(\frac{1}{2}\mathcal{G}_{ab}v^av^b \tilde{\tau}\r)$. Therefore, the transformation \eqref{Btrans_fix} leaves %the gauge-fixed action 
$S_{\rm fix}$ invariant up to the total derivative. %Note that this does not change the Hamiltonian constraint since $S_{\rm fix}$ does not provide it.

%Next, let us consider the Galilean boost transformation given by
%\begin{align}\label{Btrans}
%X^a \rightarrow X'^a=X^a - v^a \tilde{N}\tau =X^a - v^a \tilde{\tau},
%\end{align}
%where $v^a$ is constant. Under this transformation, the gauge-fixed action transforms as
%\begin{align}\label{S_again_fixing_transformed}
%S'_{\mathrm{fix}} &=\int d\tilde{\tau}\left\{\frac{1}{2}\mathcal{G}_{ab}\left(\frac{dX^a}{d\tilde{\tau}}-v^a\right)\left(\frac{dX^b}{d\tilde{\tau}}-v^b\right)-\lambda\right\}\nonumber\nonumber\\
%&=\int d\tilde{\tau}\Bigg\{\frac{1}{2}\mathcal{G}_{ab}\frac{dX^a}{d\tilde{\tau}}\frac{dX^b}{d\tilde{\tau}}\nonumber\\
%&\;\;\;\;\;\;\;\;\;\;\;\;-\frac{d}{d\tilde{\tau}}\left(\mathcal{G}_{ab}v^aX^b\right)+\frac{1}{2}\mathcal{G}_{ab}v^av^b-\lambda\Bigg\}.
%\end{align}
%Since the term $\frac{1}{2}\mathcal{G}_{ab}v^av^b$ is just constant, this transformation leaves the gauge-fixed action invariant up to a total derivative.

On the other hand, the full action \eqref{S_again_free} changes for the boost \eqref{Btrans} as $S\to S'=S[X', \MN]$: 
\begin{align}\label{S_again_transformed}
S'&=\int d\tau \MN \left(\frac{1}{2\MN^2}\mathcal{G}_{ab}\left(\frac{dX^a}{d\tau}-v^a\MN\right)\left(\frac{dX^b}{d\tau}-v^b\MN\right)-\lambda\right)\nonumber\\
&=\int d\tau \Bigg(\MN \left[\frac{1}{2\MN^2}\mathcal{G}_{ab}\frac{dX^a}{d\tau}\frac{dX^b}{d\tau}-\lambda+\frac{1}{2}\mathcal{G}_{ab}v^av^b\right]\nonumber\\
&~~~~~~~~~~~~~~~~~~~~~~~~~~~~
-\frac{d}{d\tau}\left(\mathcal{G}_{ab}v^aX^b\right)\Bigg)\nn\\
&=\int d\tau \MN \left[\frac{1}{2\MN^2}\mathcal{G}_{ab}\frac{dX^a}{d\tau}\frac{dX^b}{d\tau}-\lambda+\frac{1}{2}\mathcal{G}_{ab}v^av^b\right]\nn\\
&~~~~~~~~~~~~~~~~~~~~~~~~~~~~~~~~~+{\rm (surface~term)}.
\end{align}
Unlike the gauge-fixed case, the term $\frac{1}{2}\MN\mathcal{G}_{ab}v^av^b$ cannot be ignored due to the presence of $\MN$. Consequently, the Hamiltonian constraint is modified as
\begin{align}\label{Hconst_trans}
\frac{\delta S'}{\delta \MN}=-\frac{1}{2\MN^2}\mathcal{G}_{ab}\frac{dX^a}{d\tau}\frac{dX^b}{d\tau}-\lambda+\frac{1}{2}\mathcal{G}_{ab}v^av^b=0,
\end{align}
which leads to the new Hamiltonian constraint \eqref{H_New}. Thus, this symmetry transformation changes the full action, modifying the constraint. Note that the modification is just a constant shift, $\lambda\to \lam- \frac{1}{2}\mathcal{G}_{ab}v^av^b$, which does not affect the variation of $X^a$. This corresponds to the fact that the time evolution operator $H_0$ remains invariant in the canonical formalism (see Sec.\ref{sec:Noether}).

%On the other hand, the action without gauge fixing transforms as
%\begin{align}\label{S_again_transformed}
%S'&=\int d\tau \tilde{N}\left\{\frac{1}{2\tilde{N}^2}\mathcal{G}_{ab}\left(\frac{dX^a}{d\tau}-\tilde{N}v^a\right)\left(\frac{dX^b}{d\tau}-\tilde{N}v^b\right)-\lambda\right\}\nonumber\\
%&=\int d\tau \Bigg[\tilde{N}\left(\frac{1}{2\tilde{N}^2}\mathcal{G}_{ab}\frac{dX^a}{d\tau}\frac{dX^b}{d\tau}+\frac{1}{2}\mathcal{G}_{ab}v^av^b-\lambda\right)\nonumber\\
%&\;\;\;\;\;\;\;\;\;\;\;\;\;\;\;\;\;\;\;\;\;\;\;\;\;\;\;\;\;\;\;\;\;\;\;\;\;\;\;\;\;\;\;\;\;\;\;\;\;\;\;\;\;\;\;\;\;\;\;\;\;\;
%-\frac{d}{d\tau}\left(\mathcal{G}_{ab}v^aX^b\right)\Bigg].
%\end{align}
%Unlike the gauge-fixed case, the term $\frac{\tilde{N}}{2}\mathcal{G}_{ab}v^av^b$ cannot be ignored here due to the presence of the lapse function $\tilde{N}$. Consequently, the Hamiltonian constraint is modified as
%\begin{align}\label{Hconst_trans}
%\frac{\delta S'}{\delta \tilde{N}}=-\frac{1}{2\tilde{N}^2}\mathcal{G}_{ab}\frac{dX^a}{d\tau}\frac{dX^b}{d\tau}+\frac{1}{2}\mathcal{G}_{ab}v^av^b-\lambda=0.
%\end{align}
%Thus, these symmetry transformations do not leave the action without gauge fixing invariant; instead, they correspond to modifying the theory itself. On the other hand, this constant term does not affect the variation with respect to $X^a$. This corresponds to the fact that the time evolution operator remains invariant in the canonical formalism. 

Note that there exists another interpretation of such transformations: if $\MN$ is also transformed properly, the full action can be invariant, leading to dynamical symmetries. Indeed, this idea works for the generators $Q_+$, $Q_0$, and $Q_-$ corresponding to conformal transformations \cite{Achour2019}. For the Galilean boost $B^a$, however, to the best of our knowledge, there exists no such interpretation.

%Here, we should comment that there exists an alternative interpretation for transformations: one involves simultaneously transforming the lapse function $\tilde{N}$. Indeed, for the generators $Q_+$, $Q_0$, and $Q_-$ corresponding to conformal transformations, such an interpretation allows them to be regarded as symmetries that preserve the action \cite{Achour2019}. However, regarding the transformation generated by the Galilean boost $B^a$, to the best of our knowledge, there exists no such interpretation that renders the action without gauge fixing invariant by simultaneously transforming $\tilde{N}$.

Thus, the interpretation of generating new theories by symmetry transformations are consistent both in the canonical and Lagrangian formalisms. We here stress that the advantage of the canonical-formalism analysis in Sec.\ref{sec:interpretation} lies in its detailed examination of of the subtle points associated with gauge constraints (point I, II, III).

\stoptoc
\bibliographystyle{bib-style}
\bibliography{bibs_sano}

\end{document}